\documentclass[journal]{IEEEtran}
\IEEEoverridecommandlockouts
\usepackage{amsmath}
\usepackage{amsfonts}
\usepackage{bbding}
\usepackage{amssymb}
\usepackage{array}
\usepackage{subfigure}
\usepackage{multirow}
\usepackage{pmboxdraw}

\usepackage{graphicx}
\usepackage[named]{algo}
\usepackage{algorithmic}
\usepackage{bm}
\usepackage{psfrag}
\usepackage{stfloats}
\usepackage[compress]{cite}
\makeatletter
\renewcommand{\citepunct}{,\penalty\@m\hskip.13emplus.1emminus.1em}
\renewcommand{\citedash}{\hbox{--}\penalty\@m}
\makeatother
\usepackage{setspace}
\usepackage{color}
\usepackage{xcolor}
\allowdisplaybreaks

\usepackage{amsthm}
\usepackage{stfloats}

\newtheorem{theorem}{Theorem}

\usepackage{hyperref}

\begin{document}
\title{A Tutorial on Ultra-Reliable and Low-Latency Communications in 6G: Integrating Domain Knowledge into Deep Learning}
\author{Changyang~She,~\IEEEmembership{Member,~IEEE,} Chengjian~Sun,~\IEEEmembership{Student Member,~IEEE,}
Zhouyou~Gu,
Yonghui~Li,~\IEEEmembership{Fellow,~IEEE,} Chenyang~Yang,~\IEEEmembership{Senior Member,~IEEE,} H.~Vincent~Poor~\IEEEmembership{Fellow,~IEEE,} and Branka~Vucetic~\IEEEmembership{Fellow,~IEEE}
	
    \thanks{C. She, Z. Gu, Y. Li, and B. Vucetic are with the School of Electrical and Information Engineering, University of Sydney, Sydney, NSW 2006, Australia (e-mail: shechangyang@gmail.com, \{zhouyou.gu, yonghui.li, branka.vucetic\}@sydney.edu.au). }
	
    \thanks{C. Sun and C. Yang are with the School of Electronics and Information Engineering, Beihang University, Beijing 100191, China (email:\{sunchengjian,cyyang\}@buaa.edu.cn).}

    \thanks{H. V. Poor is with the Department of Electrical Engineering at Princeton University, Princeton, NJ 08544, USA. (e-mail: poor@princeton.edu).}

}

\maketitle
\begin{abstract}
As one of the key communication scenarios in the 5th and also the 6th generation (6G) of mobile communication networks, ultra-reliable and low-latency communications (URLLC) will be central for the development of various emerging mission-critical applications. State-of-the-art mobile communication systems do not fulfill the end-to-end delay and overall reliability requirements of URLLC. In particular, a holistic framework that takes into account latency, reliability, availability, scalability, and decision making under uncertainty is lacking. Driven by recent breakthroughs in deep neural networks, deep learning algorithms have been considered as promising ways of developing enabling technologies for URLLC in future 6G networks. This tutorial illustrates how domain knowledge (models, analytical tools, and optimization frameworks) of communications and networking can be integrated into different kinds of deep learning algorithms for URLLC. We first provide some background of URLLC and review promising network architectures and deep learning frameworks for 6G. To better illustrate how to improve learning algorithms with domain knowledge, we revisit model-based analytical tools and cross-layer optimization frameworks for URLLC. Following that, we examine the potential of applying supervised/unsupervised deep learning and deep reinforcement learning in URLLC and summarize related open problems. Finally, we provide simulation and experimental results to validate the effectiveness of different learning algorithms and discuss future directions.
\end{abstract}
\vspace{0.2cm}

\section{Introduction}
As one of the new communication scenarios in the 5th Generation (5G) mobile communications, Ultra-Reliable and Low-Latency Communications (URLLC) are crucial for enabling a wide range of emerging applications \cite{3GPP2017Scenarios}, including industry automation, intelligent transportation, telemedicine, Tactile Internet, and Virtual/Augmented Reality (VR/AR) \cite{Philipp2017Latency,Factory2015Yilmaz,Meryem2016Tactile,Imran2018How,Mischa2018Towards}. Since these applications fall within the scope of Industrial Internet-of-Things (IoT) \cite{GEIIOT}, Germany Industrie 4.0 \cite{German}, and Made in China 2025 \cite{MadeInChina}, URLLC has the potential to improve our everyday life and spur significant economic growth in the future.

\begin{figure}[tbp]
        \centering
        \begin{minipage}[t]{0.45\textwidth}
        \includegraphics[width=1\textwidth]{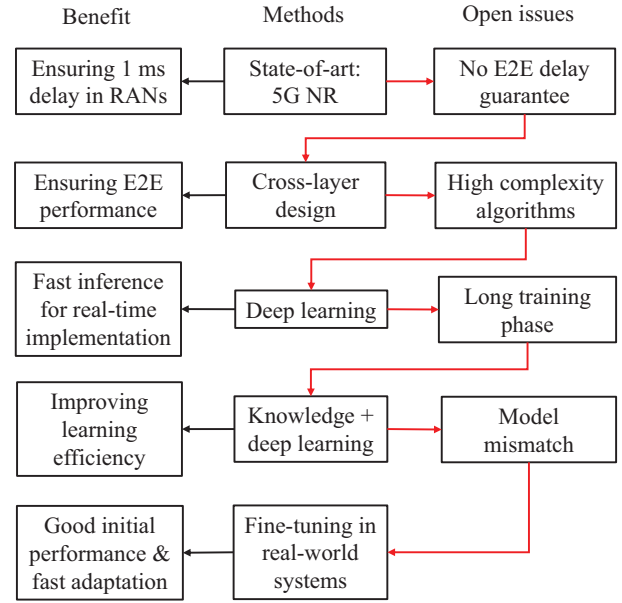}
        \end{minipage}
        \caption{Potential methods to achieve URLLC \cite{Changyang2020Deep,Dora2019Deep,Dora2020Deep}.}
        \label{fig:motivation}
        \vspace{-0.2cm}
\end{figure}

\subsection{Motivation}
According to the requirements in 5G standards \cite{3GPP2017Scenarios}, to support emerging mission-critical applications, the End-to-End (E2E) delay cannot exceed 1 ms and the packet loss probability should be $10^{-5}\sim10^{-7}$. Compared with the existing cellular networks, the delay and reliability require significant improvements by at least two orders of magnitude for 5G networks. This capability gap cannot be fully resolved by the 5G New Radio (NR), i.e., the physical-layer technology for 5G \cite{3GPP2017Agree}, even though the transmission delay in Radio Access Networks (RANs) achieves the $1$ ms target. Transmission delay contributes only a small fraction of the E2E delay, as the stochastic delays in upper networking layers, such as queuing delay, processing delay, and access delay, are key bottlenecks for achieving URLLC. Beyond 5G systems or so-called Sixth Generation (6G) systems should guarantee the E2E delay bound with high reliability. These stringent requirements create unprecedented research challenges to wireless network design. The open issues and potential methods for addressing them are illustrated in Fig. \ref{fig:motivation}. In the sequel, we describe the motivation for applying these methods in URLLC.

\subsubsection{Cross-layer Design} Existing design methods dividing divide communication networks into multiple layers according to the Open Systems Interconnection model \cite{jiang2018low}. Communication technologies in each layer are often developed without considering the impacts on other layers, despite the fact that the interactions across different layers are known to significantly impact on the E2E delay and reliability. Most existing approaches do not reflect such interactions; this leads to suboptimal solutions and thus we are yet to be able to meet the stringent requirements of URLLC. To guarantee the E2E delay and the reliability of the communication system, we need accurate and analytically tractable cross-layer models to reflect the interactions across different layers.

\subsubsection{Deep Learning} With 5G NR, the radio resources are allocated in each Transmission Time Interval (TTI) with duration of $0.125 \sim 1$~ms \cite{3GPP2017Agree}. To implement optimization algorithms in 5G systems, the processing delay should be less than the duration of one TTI. Since the cross-layer models are complex, related optimization problems are non-convex in general (See some examples on cross-layer optimization in \cite{Shengfeng2016Convexity,Cross2018she,Hu2018TWC,tang2019service,guo2019resource}.). The optimization algorithms in these papers require high computing overhead, and hence can hardly be implemented in 5G systems.

According to several survey papers \cite{zhang2019deep,zappone2019wireless,xiong2019deep}, deep learning has significant potential to address the above issue in beyond 5G/6G networks.
The basic idea is to approximate the optimal policy with a deep neural network (DNN). After the training phase, a near-optimal solution of an optimization problem can be obtained from the output of the DNN in each TTI. Essentially, by using deep learning, we are trading off the online processing time with the computing resource for off-line training.

\subsubsection{Integrating Knowledge into Learning Algorithms} Although deep learning algorithms have shown significant potential, the application of deep learning in URLLC is not straightforward. As shown in \cite{Changyang2020Deep}, deep learning algorithms converge slowly in the training phase and need a large number of training samples to evaluate or improve the E2E delay and reliability. If some knowledge of the environment is available, such as the estimated packet loss probability of a certain decision, the system can exploit this knowledge to improve the learning efficiency \cite{gu2016continuous}. Domain knowledge of communications and networking including models, analytical tools, and optimization frameworks have been extensively studied in the existing literature \cite{WirelessCom,Mor2013Queue}. How to exploit them to improve deep learning algorithms for URLLC has drawn significant attention as well, including in \cite{sun2019PIMRC,he2019model,zappone2019wireless}.

\subsubsection{Fine-tuning in Real-world Systems} Communication environments in wireless networks are non-stationary in general. Theoretical models used in off-line training may not match this non-stationary nature of practical networks. As a result, a DNN trained off-line cannot guarantee the Quality-of-Service (QoS) constraints of URLLC. Such an issue is referred to as the model-mismatch problem in \cite{Dora2020Deep}. To handle the model mismatch, wireless networks should be intelligent to adjust themselves in dynamic environments, explore unknown optimal policies, and transfer knowledge to practical networks.


\subsection{Scope of This Paper}
This paper provides a review of domain knowledge (i.e., analytical tools and optimization frameworks) and deep learning algorithms (i.e., supervised, unsupervised, and reinforcement learning). Then, we illustrate how to combine them to optimize URLLC in a cross-layer manner. The scope of this paper is summarized as follows,
\begin{enumerate}
\item In Section II, we introduce the background of URLLC including research challenges, methodologies, and a road-map toward URLLC.
\item In Section III, we review promising network architectures and deep learning frameworks for URLLC and summarize the design principles of deep learning frameworks in 6G networks.
\item We revisit analytical tools in information theory, queuing theory, and communication theory in Section IV, and show how to apply them in cross-layer optimization for URLLC in Section V.
\item Then, we examine the potential of applying supervised/unsupervised deep learning and Deep Reinforcement Learning (DRL) in URLLC in Sections VI, VII, and VIII, respectively. Related open problems are summarized in each section.
\item In Section IX, we provide some simulation and experimental results to illustrate how to integrate domain knowledge into deep learning algorithms for URLLC.
\item Finally, we highlight some future directions in Section X and conclude this paper in Section XI.
\end{enumerate}

\section{Background of URLLC}

\begin{figure}[tbp]
        \centering
        \begin{minipage}[t]{0.45\textwidth}
        \includegraphics[width=1\textwidth]{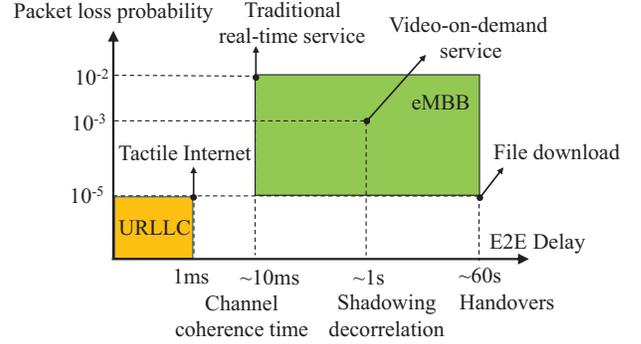}
        \end{minipage}
        \caption{Delay and reliability requirements of URLLC and eMBB \cite{Mahyar2018Short,dymarski2011qos,WirelessCom}.}
        \label{fig:tradeoff}
        \vspace{-0.2cm}
\end{figure}

The requirements on E2E delay and overall packet loss probability of URLLC and enhanced Mobile BroadBand (eMBB) services are illustrated in Fig. \ref{fig:tradeoff}. For eMBB services, communication systems can trade reliability with delay by retransmissions. For URLLC services, the requirements on the E2E delay and reliability are much more stringent than eMBB services. In addition, the wireless channel fading, shadowing, and handovers will have significant impacts on the reliability of URLLC services (See some existing results in \cite{ResourceShe,she2019ultra,David2016Availability}.).  Therefore, ensuring the QoS requirements of URLLC is more challenging than eMBB services.

\begin{table*}[htbp]
\vspace{-0.0cm}\small
\renewcommand{\arraystretch}{1.3}
\caption{KPIs and Research Challenges of Different URLLC Applications \cite{3gpp.22.104,Mahyar2018Short,Philipp2017Latency,holfeld2016wireless,sun2019communications, mavromatis2018multi,aijaz2019tactile,thuemmler2018requirements,cosovic20175g,lema2017business,she2019ultra,liu2019taming}}
\begin{center}\vspace{-0.0cm}\label{T:Applications}
\begin{tabular}{|c|c|c|}
  \hline
  \multicolumn{3}{|c|}{Indoor large-scale scenarios}\\\hline
  {\bf{Applications}} & {\bf{KPIs (except E2E delay \& reliability)}} & {\bf{Research Challenges}}\\\hline
  Factory automation &   SE, EE \& AoI & Scalability \& network congestions \\\hline
  VR/AR applications &  SE \& throughput & Processing/transmission 3D videos \\\hline
  \multicolumn{3}{|c|}{Indoor wide-area scenarios}\\\hline
  {\bf{Applications}} & {\bf{KPIs (except E2E delay \& reliability)}} & {\bf{Research Challenges}}\\\hline
  Tele-surgery &  {Round-trip delay, throughput \& jitter } & {Propagation delay \& high data rate}\\\hline
  eHealth monitoring &   EE \& network availability  & Propagation delay \& localization \\\hline
  \multicolumn{3}{|c|}{Outdoor large-scale scenarios}\\\hline
  {\bf{Applications}} & {\bf{KPIs (except E2E delay \& reliability)}} & {\bf{Research Challenges}}\\\hline
  Vehicle safety&  AoI, SE, security \& network availability & High mobility, scalability \& network congestions\\\hline
  \multicolumn{3}{|c|}{Outdoor wide-area scenarios}\\\hline
  {\bf{Applications}} & {\bf{KPIs (except E2E delay \& reliability)}} & {\bf{Research Challenges}}\\\hline
  Smart grid &  SE & Propagation delay \& scalability \\\hline
  Tele-robotic control &  SE, security, network availability \& jitter & Propagation delay \& high data rate \\\hline
  UAV control & EE, security, network availability \& AoI & Propagation delay \& high mobility \\\hline
\end{tabular}
\end{center}
\vspace{-0.2cm}
\end{table*}

\subsection{Other Key Performance Indicators (KPIs) and Challenges}\label{Sec:KPIs}
In the existing literature, the application scenarios, KPIs, and research challenges of URLLC have been extensively investigated. In Table \ref{T:Applications}, we classify these applications into four categories according to the communication scenarios, including indoor large-scale networks, indoor wide-area networks, outdoor large-scale networks, and outdoor wide-area networks.

\subsubsection{Indoor Large-scale Networks} Typical applications in indoor large-scale networks include factory automation and VR/AR applications such as immersed VR games \cite{holfeld2016wireless,German,Mischa2018Towards}. As the density of devices increases, improving the \emph{spectrum efficiency}, say the number of services that can be supported with a given total bandwidth, becomes an urgent task. Since the battery capacities of mobile devices are limited, \emph{energy efficiency} is another KPI in this scenario. In factory automation, a large number of sensors keep updating their status to the controller and actuator \cite{liu2019taming}, where the freshness of information is one of the KPIs measured by Age of Information (AoI) \cite{kam2018age}.

In wireless communications, the numbers of constraints and optimization variables increase with the number of devices. Since most of the optimization algorithms in \cite{boyd} only work well for small and medium scale problems, as the density of devices grows, \emph{scalability} becomes the most challenging issue in indoor large-scale networks. In addition, stochastic packet arrival processes from a large number of users will lead to strong interference in the air interface. The burstiness and correlation of packet arrivals will result in severe \emph{network congestions} in queuing systems and computing systems. How to alleviate network congestions in large-scale networks remains an open problem.

\subsubsection{Indoor Wide-area Networks} For long-distance applications in Tactile Internet, like telesurgery and remote training, the system stability and teleoperation quality are very sensitive to \emph{round-trip delay} \cite{lawrence1993stability,aijaz2019tactile}. Thus, the round-trip delay is a KPI in bilateral teleoperation systems \cite{Mischa2018Towards}. On the other hand, to maintain stability and transparency, the signals from a haptic sensor are sampled, packetized, and transmitted at $1000$~Hz or even higher \cite{Mischa2018Towards}. As a result, the required \emph{throughput} is very high. Meanwhile, the \emph{jitter} of delay is crucial for the stability of control systems as stated in \cite{3gpp.22.104}. If the jitter is larger than the inter-arrival time between two packets (less than $1$~ms in telesurgery), the order of the control commands arriving at the receiver can be different from the order of commands generated by the controller. In this case, the control system is unstable.

Due to long communication distance, the propagation delay in core networks might be much longer than the required E2E delay, and thus will become the bottleneck of URLLC. To handle this issue, prediction and communication co-design is a promising solution, which, however, will introduce extra prediction errors \cite{feng2019towards}. To guarantee the round-trip delay, overall reliability, and throughput, we need to analyze the fundamental tradeoffs among them and design practical solutions that can approach the performance limit \cite{Hou2019Prediction}.

\subsubsection{Outdoor Large-scale Networks} One of the most important applications in outdoor large-scale networks is the vehicle safety application \cite{mavromatis2018multi}. With the current technologies for autonomous vehicles, no information is shared among vehicles. Thus, vehicles are not aware of potential threats from the blind areas of their image-based sensors and distance sensing devices \cite{luettel2012autonomous}. URLLC can enhance road safety by sharing street maps and safety messages among vehicles. Like indoor large-scale networks, the \emph{spectrum efficiency} is one of the major KPIs in outdoor large-scale networks with high user density. To maintain current state information from nearby vehicles, minimizing AoI is helpful for improving road safety in vehicular networks \cite{kaul2011minimizing}.  In addition, outdoor wireless networks are more vulnerable than indoor networks. Potential eavesdroppers can receive signals and crack the information with high probabilities. Therefore, \emph{security} should be considered for URLLC in outdoor wireless networks \cite{chen2019physical}.

Current cellular networks can achieve $95$\% coverage probability, which is satisfactory for most of the existing services. In URLLC, the required \emph{network availability} can be up to $99.999$\% \cite{Popovski2014METIS}. As a result, we need to improve the network availability by several orders of magnitude. Due to high mobility, service interruptions caused by frequent handovers and network congestions are the bottlenecks for achieving high availability. Besides, existing tools for analyzing availability are only applicable in small-scale networks \cite{David2018VTC,She2018availability}. How to analyze and improve network availability in large-scale networks remains unclear.

\subsubsection{Outdoor Wide-area Networks} Unmanned Aerial Vehicle (UAV) control, telerobotic control, and smart grids are typical applications in outdoor wide-area networks \cite{Philipp2017Latency,she2019ultra,cosovic20175g}. Establishing secure, reliable, and real-time control and non-payload communication (CNPC) links between UAVs and ground control stations/devices or satellites in a wide-area has been considered as one of the major goals in future space-air-ground integrated networks \cite{liu2018space}. Thus, the KPIs in outdoor wide-area networks include \emph{security}, \emph{network availability}, \emph{round-trip delay}, \emph{AoI} and \emph{jitter}.

In UAV control, there are two ways to maintain CNPC links: ground-to-air communications and satellite communications \cite{R2015control,gupta2016survey}. Nevertheless, ground-to-air links may not be available in rural areas, where the density of ground control stations is low. If packets are sent via satellites, they suffer from long propagation delays and coding delays. In smart grids, although the E2E delay and overall reliability are less stringent than factory automation, the long communication distance leads to long propagation delay \cite{Philipp2017Latency}. As a result, achieving the target KPIs in outdoor wide-area networks is still very challenging with current communication technologies.

\subsection{Tools and Methodologies for URLLC}\label{Sec:tools}
To achieve the target KPIs, we need to revisit analytical tools and design methodologies in wireless networks.

\subsubsection{Analytical Tools} To analyze the performance of a wireless network, a variety of theoretical tools have been developed in the existing literature \cite{WirelessCom,Mor2013Queue}. For example, to reduce transmission delay, several kinds of channel coding schemes with short blocklength have been developed in the existing literature \cite{Mahyar2018Short}. To obtain the achievable rate in the short blocklength regime, a new approximation was derived by Y. Polyanskiy \emph{et al.} \cite{Yury2010Channel}. It indicates that the decoding error probability does not vanish for arbitrarily finite Signal-to-Noise Ratio (SNR) when the blocklength is short. Such an approximation can characterize the relationship between data rate and decoding error probability, and is fundamentally different from Shannon's capacity that was widely used to design wireless networks.

If we can derive the closed-form expressions of KPIs in Table \ref{T:Applications}, then it is possible to predict the performance of a solution or decision. The disadvantage is that theoretical analytical tools are based on some assumptions and simplified models that may not be accurate enough for URLLC applications. Moreover, to analyze the E2E performance in URLLC, the models may be very complicated, and closed-form results may not be available.

Another approach is to evaluate the performance of the solution with simulation or experiment. The advantage is that they do not rely on any theoretical assumptions. On the negative side, the evaluation procedure is time-consuming, especially for URLLC. For example, to validate the packet loss probability (around $10^{-5}\sim 10^{-7}$ in URLLC) of a user, the system needs to send $10^{8} \sim 10^{10}$ packets.

\subsubsection{Cross-layer Optimization}
The E2E delay and the overall packet loss probability depend on the solutions in different layers \cite{jiang2018low}. To optimize the performance of the whole system, we need cross-layer optimization frameworks, from which it is possible to reveal some new technical issues by analyzing the interactions among different layers. For example, as shown in \cite{Tang2007QoS}, when the required queuing delay is shorter than the channel coherence time, the power allocation policy is a type of channel inverse policy, which can be unbounded over typical wireless channels, such as the Nakagami-$m$ fading channel. One way to analyze the reliability of a wireless link is to analyze the outage probability \cite{caire1999optimum}. However, such a performance metric cannot characterize the queuing delay violation probability in the link layer. To keep the decoding error probability and the queuing delay violation probability below the required thresholds, some packets should be dropped proactively when the channel is in deep fading \cite{Cross2018she}. By cross-layer design, one can take the delay components and packet losses in different layers into account, and hence it is possible to achieve the target E2E delay and overall packet loss probability in URLLC.

Although cross-layer optimization has the potential to achieve better E2E performances than dividing communication systems into separated layers, to implement cross-layer optimization algorithms in practical systems, one should address the following issues,
\begin{itemize}
\item \emph{High computing overheads:} Wireless networks are highly dynamic due to channel fading and traffic load fluctuations. As a result, the system needs to adjust resource allocation frequently according to these time-varying factors. Without the closed-form expression of the optimal policy, the system needs to solve the optimization problem in each TTI. This will bring very high computing overheads. Even if the problem is convex and can be solved by existing searching algorithms, such as the interior-point method \cite{boyd}, the algorithms are still too complicated to be implemented in real-time.
\item \emph{Intractable problems: } Owning to complicated models from different layers, cross-layer optimization problems are usually non-convex or Non-deterministic Polynomial-time (NP)-hard. Solving NP-hard optimization problems usually takes a long time, and the resulting optimization algorithms can hardly be implemented in real-time. Low-complexity numerical algorithms that can be executed in practical systems for URLLC are still missing.
\item \emph{Model mismatch:} Since the available models in different layers are not exactly the same as practical systems, the model mismatch may lead to severe QoS violations in real-world networks. How to implement the solutions and policies obtained from the analysis and optimization in practical systems remains an open problem.
\end{itemize}

\subsubsection{Deep Learning for URLLC} Unlike optimization algorithms, deep learning approaches can be model-based or model-free \cite{he2019model,balatsoukas2019deep,luong2019applications}, and have the potential to be implemented in real-world communication systems \cite{dorner2017deep}. The advantages of deep learning algorithms can be summarized as follows.
\begin{itemize}
\item \emph{Real-time implementation:} After off-line training, a DNN serves as a mapping from the observed state to a near-optimal action in communication systems, where the forward propagation algorithm is used to compute the output of a given input. According to the analysis in \cite{Dora2020Deep}, the computational complexity of the forward propagation algorithm is much lower than searching algorithms for cross-layer optimizations. Therefore, by using DNN in 5G NR, it is possible to obtain a near-optimal action within each TTI.
\item \emph{Searching optimal policies numerically:} When optimal solutions are not available, there is no labeled sample for supervised learning. To handle this issue, the authors of \cite{Fei2019Power,eisen2019learning,sun2019PIMRC} proposed to use unsupervised deep learning algorithms to search the optimal policy. Unlike optimization algorithms that find an optimal solution for a given state of the system, unsupervised deep learning algorithms find the form of the optimal policy numerically. The basic idea is to approximate the optimal policy with a DNN and optimize the parameters of the DNN towards a loss function reflecting the design goal.
\item \emph{Exploring optimal policies in real-world networks:} When there is no labeled sample or theoretical model to formulate an optimization problem, we can use DRL to explore optimal policies from the real-world network \cite{luong2019applications}. For example, an actor-critic algorithm uses two DNNs to approximate the optimal policy and value function, respectively. From the feedback of the network, the two DNNs are updated until convergence \cite{lillicrap2015continuous}.
\end{itemize}

Although deep learning algorithms have the above advantages, to apply them in URLLC, the following issues remain unclear.
\begin{itemize}
\item \emph{QoS guarantee:} When designing wireless networks for URLLC, the stringent QoS requirements should be satisfied. When using a DNN to approximate the optimal policy, the approximation should be accurate enough to guarantee the QoS constraints. Although the universal approximation theorem of DNNs indicates that a DNN can be arbitrarily accurate when approximating a continuous function \cite{rumelhart1986learning,hornik1989multilayer}, how to design structures and hyper-parameters of DNNs to achieve a satisfactory accuracy remains unclear.
\item \emph{Learning in non-stationary networks:} When applying supervised deep learning in URLLC, the DNN is trained off-line with a large number of training samples \cite{Dora2019Deep}. When using unsupervised deep learning to find the optimal policy, we need to formulate an optimization problem by assuming the system is stationary \cite{sun2019learning}. Both approaches can perform very well in stationary networks \cite{Dora2019Deep,sun2019learning}. However, when the environment changes, the DNN trained off-line can no longer guarantee the QoS constraints of URLLC. To handle this issue, the system needs to adjust the DNN in non-stationary environments with few training samples \cite{Dora2020Deep}.
\item \emph{Exploration safety of DRL:} To improve the policy in the unknown environment, a DRL algorithm will try some actions randomly to estimate the reward of different actions \cite{sutton2018reinforcement}. During explorations, the DRL algorithm may try some bad actions, which will deteriorate the QoS significantly and may lead to unexpected accidents in URLLC systems. Thus, the exploration safety will become a bottleneck for applying DRL in URLLC.
\end{itemize}

\subsection{Road-map Toward URLLC in 6G Networks}
\begin{figure*}[tbp]
        \centering
        \begin{minipage}[t]{0.6\textwidth}
        \includegraphics[width=1\textwidth]{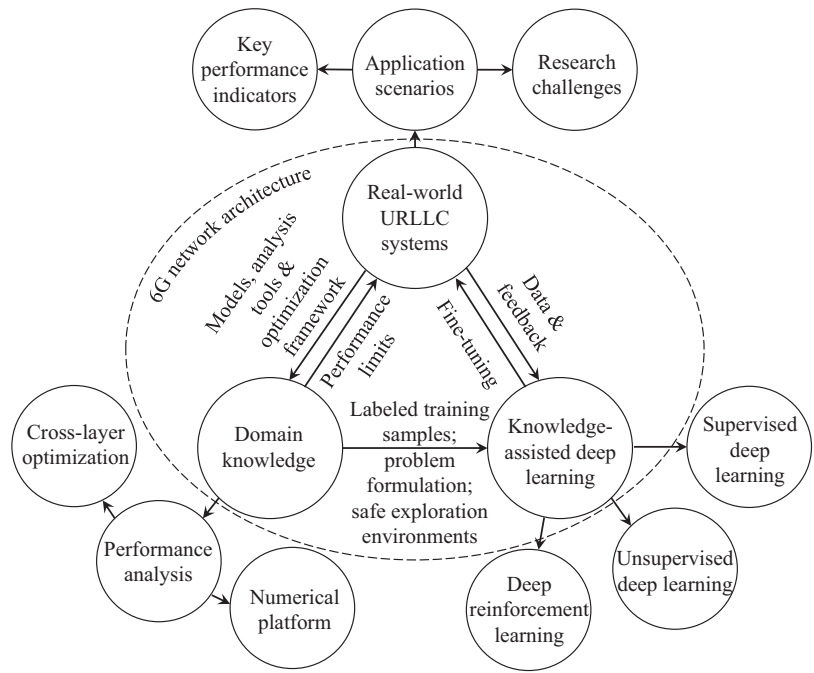}
        \end{minipage}
        \caption{Road map toward URLLC in 6G networks \cite{Dora2019Deep,Changyang2020Deep}.}
        \label{fig:roadmap}
        \vspace{-0.2cm}
\end{figure*}

Based on the KPIs and challenges of different applications in Section \ref{Sec:KPIs} and the tools and methodologies in Section \ref{Sec:tools}, the road-map toward URLLC in 6G networks is summarized in Fig. \ref{fig:roadmap}. As indicated in \cite{Changyang2020Deep}, by integrating domain knowledge into deep learning algorithms, it is possible to provide in-depth understandings and practical solutions for URLLC.

\emph{Step 1 (Knowledge-based analysis and optimization}): To satisfy the QoS requirements of URLLC in real-world systems, one should first take the advantage of the knowledge in information theory, queuing theory, and communication theory \cite{she2017radio}. From theoretical analytical tools and cross-layer optimization frameworks, we can obtain the performance limits on the tradeoffs among different KPIs \cite{Yury2010Channel,Yury2014Quasi,Cross2018she}.

\emph{Step 2 (Knowledge-assisted training of deep learning):} Based on analytical tools and optimization frameworks in communications and networking, one can build a simulation platform to train deep learning algorithms \cite{he2019model,balatsoukas2019deep}. In supervised deep learning,  we can find labeled training samples from the simulation platform  \cite{Dora2019Deep}. In unsupervised deep learning, the knowledge of the communication systems can be used to formulate the loss function \cite{sun2019learning}. In DRL, by initializing algorithms off-line, the exploration safety and convergence time of DRL algorithms can be improved significantly \cite{Zhouyou2020Knowledge}.

\emph{Step 3 (Fine-tuning deep learning in real-world networks)} From the previous step, DNNs are pre-trained with the help of models, analytical tools, and optimization frameworks that capture some key features of real-world systems. Since the models are not exactly the same as the real-world systems, we need to fine-tune DNNs in non-stationary networks. As shown in \cite{luong2019applications}, data-driven deep learning updates the initialized DNNs by exploiting real-world data and feedback. After fine-tuning, we can obtain practical solutions from the outputs of the DNNs in real-time.

\subsection{Related Works}
\subsubsection{Survey and Tutorial of URLLC} URLLC was included in the specification of 3GPP in 2016. Since then, great efforts have been devoted to this area \cite{Giuseppe2016Toward,antonakoglou2018towards,nasrallah2018ultra,jiang2018low,kim2018ultrareliable,parvez2018survey,Mehdi2018Ultra}. The review from the information theory aspect was published in 2016 \cite{Giuseppe2016Toward}, where the analyses of the impacts of short blocklength channel codes on the latency and reliability were discussed. With a focus on standard activities, A. Nasrallah \emph{et al.} \cite{nasrallah2018ultra} provided a comprehensive overview of the IEEE $802.1$ Time-Sensitive Networking (TSN) standard and Internet Engineering Task Force Deterministic Networking standards as the key solutions in the link and network layers, respectively. X. Jiang \emph{et al.} \cite{jiang2018low} proposed a holistic analytical and classification of the main design principles and technologies that will enable low-latency wireless networks, where the technologies from different layers and cross-layer design were summarized. With the focus on the physical layer, K. S. Kim   introduced solutions for spectrally efficient URLLC techniques \cite{kim2018ultrareliable}. G. Sutton \emph{et al.} reviewed the technologies for URLLC in the physical layer and the medium access control layer technologies \cite{sutton2019enabling}. K. Antonakoglou \emph{et al.} \cite{antonakoglou2018towards} introduced haptic communication as the third media stream that will complement audio and vision over the Internet. The evaluation methodologies and necessary infrastructures for haptic communications were discussed in \cite{antonakoglou2018towards}. I. Parvez \emph{et al.} \cite{parvez2018survey} presented a general view on how to meet latency and other 5G requirements with Software-Defined Network (SDN), network function virtualization, caching, and Mobile Edge Computing (MEC). M. Bennis \emph{et al.} noticed that the existing design approaches mainly focus on average performance metrics and hence are not applicable in URLLC (e.g., average throughput, average delay, and average response time)  \cite{Mehdi2018Ultra}. They summarized the tools and methodologies to characterize the tail distribution of the delay, service quality, and network scalability.

\subsubsection{Survey and Tutorial of Deep Learning in Wireless Networks} Deep learning has been considered as one of the key enabling technologies in the intelligent 5G \cite{li2017intelligent} and the beyond 5G cellular networks \cite{saad2019vision,letaief2019roadmap}. To combine deep learning with wireless networks, several comprehensive surveys and reviews have been carried out \cite{Wang2019thirty,mao2018deep,zappone2019wireless,zhang2019deep,luong2019applications,liang2019veh,sun2019application,park2019wireless}. J. Wang \emph{et al.} presented a thorough survey of machine learning algorithms and their use in the next-generation wireless networks in \cite{Wang2019thirty}. Q. Mao \emph{et al.} \cite{mao2018deep} reviewed how to use deep learning in different layers of the OSI model and cross-layer design. With the focus on the smart radio environment, A. Zappone \emph{et al.} proposed to integrate deep learning with traditional model-based technologies in future wireless networks \cite{zappone2019wireless}. A comprehensive survey of the crossovers between deep learning and wireless networks was presented by C. Zhang \emph{et al.} in \cite{zhang2019deep}, where the authors illustrated how to tailor deep learning to mobile environments. N. C. Luong \emph{et al.} presented a tutorial on DRL and reviewed the applications of DRL in wireless communications and networking \cite{luong2019applications}. A tutorial on how to use deep/reinforcement learning for wireless resource allocation in vehicular networks was presented by L. Liang \emph{et al.} in \cite{liang2019veh}. Y. Sun \emph{et al.} summarized some key technologies and open issues of applying machine learning in wireless networks in \cite{sun2019application}. J. Park \emph{et al.} provided a tutorial on how to enable wireless network intelligence at the edge \cite{park2019wireless}.

Although the above two branches of existing works either investigated model-based approaches for URLLC or reviewed deep learning in wireless networks. None of them discussed how to optimize wireless networks for URLLC by integrating domain knowledge of communications and networking into deep learning, which is the central goal of our work. Different from existing papers, we revisit the theoretical models, analytical tools, and optimization frameworks, review different deep learning algorithms for URLLC. Furthermore, we illustrate how to combine them together to achieve URLLC.

\section{Network Architectures and Deep Learning Frameworks in 6G}
The application of machine learning in cellular networks has gained tremendous attention in both academia and industry. Recently, how to operate cellular networks with machine learning algorithms had been considered as one of the 3GPP work items in \cite{RP2020New}. Since computing and storage resources will be deployed at both edge and central servers of 6G networks, 6G will have the capability to train deep learning algorithms \cite{park2019wireless}. In this section, we review some network architectures and illustrate how to develop deep learning frameworks.

\begin{figure*}[tbp]
        \centering
        \begin{minipage}[t]{0.6\textwidth}
        \includegraphics[width=1\textwidth]{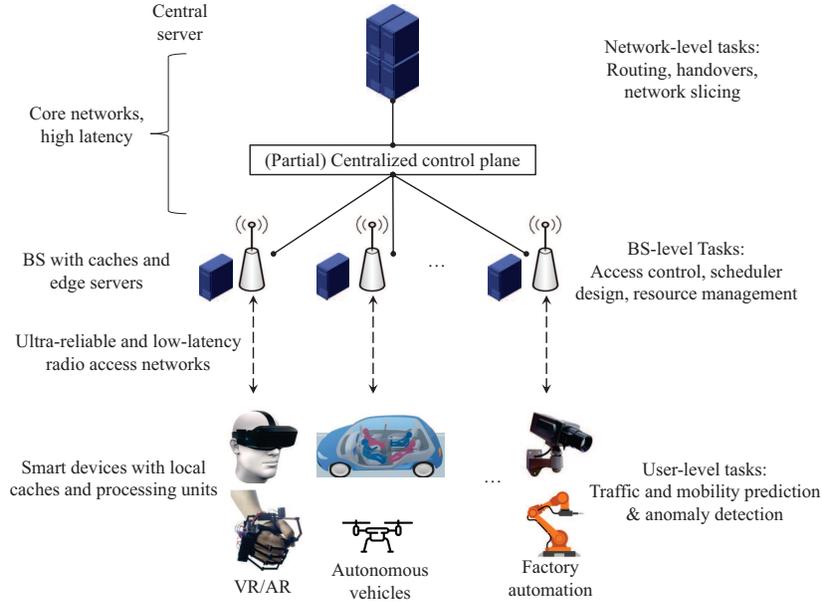}
        \end{minipage}
        \caption{A general architecture in 6G networks \cite{Changyang2020Deep,song2020artificial,tang2019future}.}
        \label{fig:architecture}
        \vspace{-0.2cm}
\end{figure*}

\subsection{Network Architectures}
A general network architecture for various application scenarios (such as IoT networks \cite{song2020artificial} and vehicular networks \cite{tang2019future}) in 6G networks is illustrated in Fig. \ref{fig:architecture}. As shown in \cite{Changyang2020Deep}, the communication, computing, and caching resources are integrated into a multi-level system to support tasks at the user level, the Base Station (BS) level, and  the network level. Such an architecture can be considered as a ramification of the following architectures.

\subsubsection{MEC} As a new paradigm of computing, MEC is believed to be a promising architecture for URLLC \cite{Yuyi2017MEC}. Unlike centralized mobile cloud computing with routing delay and propagation delay in backhauls and core networks, the E2E delay in MEC systems consists of Uplink (UL) and Downlink (DL) transmission delays, queuing delays in the buffers of users and BSs, and the processing delay in the MEC \cite{Changyang2019IoT}. The results in \cite{Changyang2019IoT} indicate that it is possible to achieve $1$~ms E2E delay in MEC systems by optimizing task offloading and resource allocation in RANs.


To analyze the E2E delay and overall packet loss probability of URLLC, Y. Hu \emph{et al.} considered both decoding error probability in the short blocklength regime and the queuing delay violation probability  \cite{hu2018delay}. C.-F. Liu \emph{et al.} minimized the average power consumption of users subject to the constraint on delay bound violation probability \cite{liu2017latency}. In multi-cell MEC networks, J. Liu \emph{et al.} investigated how to improve the tradeoff between latency and reliability \cite{liu2018offloading}, where a typical user is considered. To further exploit the computing resources from the central server, a multi-level MEC system was studied in heterogeneous networks, where computing-intensive tasks can be offloaded to the central server \cite{wang2019hetmec}.

To implement the existing solutions in real-world MEC systems, some problems remain open:
\begin{itemize}
\item \emph{Optimization complexity:} Optimization problems in MEC systems are general non-convex. To find the optimal solution, the computing complexity for executing searching algorithms is too high to be implemented in real-time.

\item \emph{Mobility of users:} In vehicle networks, users are moving with high speed. The key issue here is adjusting task offloading during frequent handovers. How to optimize task offloading for high mobility users remains unclear.

\item \emph{High overhead for exchanging information:} Current task offloading and resource allocation algorithms are executed at the central controller. This will lead to high overhead for exchanging information among each BS and the central controller. To avoid this, we need distributed algorithms for URLLC, which are still unavailable in the existing literature.

\item \emph{Serving hybrid services:} VR/AR applications are expected to provide immersion experience to users \cite{elbamby2018toward}. To achieve this goal, the system needs to send $360^{\rm o}$ video as well as tactile feedback to each user. How to optimize task offloading for both video and tactile services in VR/AR applications needs further research.
\end{itemize}


\subsubsection{Multi-Connectivity and Anticipatory Networking}
To achieve a high network availability, an effective approach is to serve each user with multiple links, e.g., multi-connectivity \cite{Tom2017Applying}. Such an approach will bring two new research challenges: 1) improving the fundamental tradeoff between spectrum efficiency and network availability \cite{Jimmy2018URLLC} and 2) providing seamless service for moving users \cite{Kwang2019ULL}.

As illustrated in \cite{David2018VTC}, one way to improve network availability without sacrificing spectrum efficiency is to serve each user with multiple BSs over the same subchannel (or subcarriers). Such an approach is referred to as intra-frequency multi-connectivity. The disadvantage of this intra-frequency multi-connectivity is that the failures of different links are highly correlated. For example, if there is a strong interference on a subchannel, then the signal to interference plus noise ratios of all the links are low. To alleviate cross-correlation among different links, different nodes can connect to one user with different subchannels or even with different communication interfaces \cite{Jimmy2018URLLC}. To achieve a better tradeoff between spectrum efficiency and availability, network coding is a viable solution \cite{walsh2009optimal}. Most of the existing network coding schemes are too complicated to be implemented in URLLC. To reduce the complexity of signal processing, configurable templates for network coding were developed in \cite{Xiaoli2019Minimum}.

To provide seamless service for moving users, anticipatory networking is a promising solution \cite{bui2017survey}. The basic idea of anticipatory networking is to predict the mobility of users according to their mobility pattern and reserve resources before handovers proactively. N. Bui \emph{et al.} validated that anticipatory networking allows network operators to save half of the resources in Long Term Evolution systems \cite{bui2018data}. More recently, K. Chen \emph{et al.} proposed a proactive network association scheme, where each user can proactively associate with multiple BSs \cite{Kwang2019ULL}. As proved in \cite{hung2018delay}, the requirements on queuing delay and queuing delay violation probability can be satisfied with this scheme.


\subsubsection{SDN and Network Slicing} SDN architectures have been adopted by 5G Infrastructure Public Private Partnership (5G-PPP) architecture working group \cite{View20175GPPP}. By splitting the control plane and the user plane, the controller can manage radio resources and data flows in fully centralized, partially centralized, and fully decentralized manners. Essentially, there is a tradeoff between control-plane overheads and user-plane QoS. To achieve a better tradeoff, 5G-PPP structured network functions into the following three parts according to their position in the protocol stack \cite{View20175GPPP}: high layer, medium layer, and low layer. The high layer control plane manages resource coordination and long-term load balancing for QoS guarantee and network slicing. The medium layer control plane handles mobility and admission control. The low layer control plane deals with short-term scheduling and physical-layer resource management, which requires low control-plane latency and high control-plane reliability. To address these issues, short frame structure and grant-free scheduling have been developed in 5G NR to reduce latency; Polar codes will be applied in control signaling with short blocklength \cite{3GPP2017Agree}. In addition, IEEE TSN standard in wired communications and other standards on synchronization, cyclic queuing and forwarding, frame preemption, stream reservation have specified the control signaling to achieve low latency, low jitter, and low packet loss probability \cite{won20203gpp}.

Since the control plane is deployed at different layers, there is no need to collect all the network state information at the central controller. Network resources (e.g., computing resources \cite{Antonio2019Computational}, storage resources \cite{Nguyen2019Real}, and physical resource blocks \cite{Behnam2019RAN}) are allocated to different slices according to the slow varying traffic loads and QoS requirements. To guarantee the QoS of different services, the system needs to map the QoS requirements to network resources \cite{tello2018sdn}. However, the E2E delay and overall packet loss probability depend on physical-layer resource allocation, link-layer transmission protocols and schedulers, as well as network architectures. Thus, how to quantify the network resources required by URLLC remains unclear. Most of the existing papers on network slicing either assumed that the required resources to guarantee the QoS of each service are known at the central controller \cite{Nguyen2019Real} or considered a simplified model to formulate QoS constraints \cite{popovski20185g}. To fully address this issue, a cross-layer model for E2E optimization is needed.

\subsection{Deep Learning Frameworks}
\begin{figure*}[tbp]
        \centering
        \begin{minipage}[t]{0.75\textwidth}
        \includegraphics[width=1\textwidth]{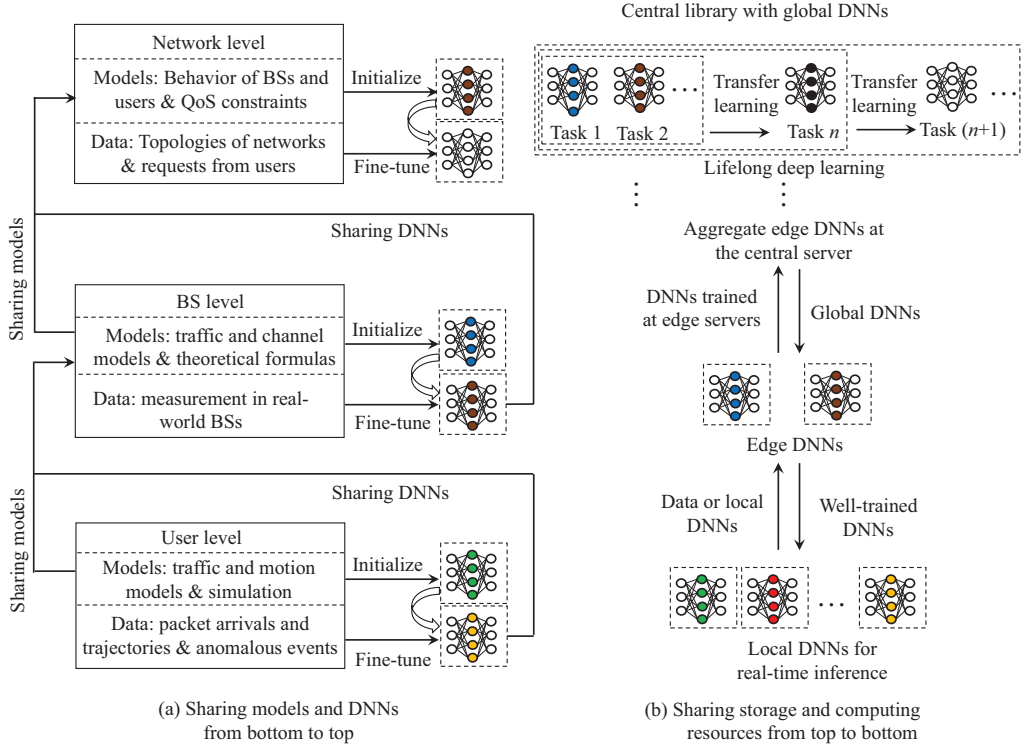}
        \end{minipage}
        \caption{A general deep learning framework \cite{Changyang2020Deep,liu2019lifelong}.}
        \label{fig:framework}
        \vspace{-0.2cm}
\end{figure*}

To cope with the tasks at different levels in Fig. \ref{fig:architecture}, a deep learning framework should exploit computing and storage resources at different entities \cite{Changyang2020Deep}. A general deep learning framework is illustrated in Fig. \ref{fig:framework}, which is built upon the following subframeworks.

\subsubsection{Federated Learning} In the multi-level computing system in Fig. \ref{fig:architecture}, we can hardly obtain well-trained DNNs with local resources and data since the computing resources and the number of training samples at a user or a BS are not enough. Federated learning is capable to exploit local DNNs trained with local data to learn global DNNs \cite{mcmahan2017communication}. To obtain a global DNN, the basic idea is to compute the weighted sum of the parameters of local DNNs. In this way, each user or BS only needs to upload its local or edge DNN to the central server, and does not need to share its data. Such a framework can avoid high communication overheads and the security issue caused by collecting private training data from all the users \cite{Howard2019FL}.

Federated learning was used to evaluate the trail probability of delay in URLLC \cite{samarakoon2018federated}, where the basic idea is to collect the estimated parameters from all the users and perform a global average. Such an approach highly relies on the assumption that the local training data at different users are Independent and Identically Distributed (IID). To implement federated learning with non-IID data sets, solutions have been proposed in two of the most recent papers  \cite{zhao2018federated,liu2019edge}. However, both of these works were not focused on URLLC. How to guarantee the QoS requirements of users with non-IID data deserves further research.

\subsubsection{Deep Transfer Learning} Transfer learning aims to apply knowledge learned in previous tasks to novel tasks \cite{pan2009survey}. Recently, transfer learning was combined with DNNs, and the combined approach is referred to as deep transfer learning \cite{Chuanqi2018transfer}. Different from training a new DNN from scratch, deep transfer learning only needs a few training samples to update the DNN trained for the previous task. Such a feature of deep transfer learning brings several advantages in wireless networks:

\begin{itemize}
\item \emph{Handle model mismatch problem:} In wireless networks, the models for performance analysis and optimization are not exactly the same as real-world systems. To implement a DNN that is trained from theoretical models, we can use deep transfer learning to fine-tune the DNN \cite{Dora2019Deep}.

\item \emph{Deep learning in non-stationary networks:} In non-stationary wireless networks, the pre-trained DNN needs to be updated according to the non-stationary environments. Since it is difficult to obtain a large number of new training samples within a short time in the new environment, deep transfer learning should be applied.

\item \emph{Applying global DNNs at local devices:} Considering that the data sets at different devices or BSs follow different distributions \cite{zhao2018federated,liu2019edge}, if a global DNN is directly used by a device or a BS for inference, it can neither guarantee the QoS requirements nor achieve optimal resource utilization efficiency. To avoid this issue, deep transfer learning should be used to fine-tune a global DNN with local training samples.
\end{itemize}

\subsubsection{Lifelong Deep Learning} As shown in \cite{chen2018lifelong}, lifelong learning is a kind of cumulative learning algorithm that consists of multiple steps. In each step, transfer learning is applied to improve the learning efficiency of the new task. As illustrated in Fig. \ref{fig:framework}, the knowledge learned from the previous $(n-1)$ tasks will be reused to initialize the $n$-th new task. After the training stage, the $n$-th task becomes a source task and will be reused to train new tasks in the future.
For example, by combining the federated learning framework with lifelong learning, a lifelong federated reinforcement learning framework for cloud robotic systems was proposed by B. Liu \emph{et al.} \cite{liu2019lifelong}. Such a framework can be adopted in the network architecture in Fig. \ref{fig:architecture}, where the DNNs for different tasks are saved in a central library at the central server. For each device or BS, it fetches the global DNN of the relevant task and fine-tunes it according to the environment. Finally, a local DNN can be obtained for real-time inference. 

\subsubsection{Distributed Deep Learning} To avoid high overheads for exchanging information, the control plane of a cellular network should not be fully centralized \cite{View20175GPPP}. In other words, each control plane makes its' own decisions according to the available information. To guarantee the QoS requirements of URLLC with a partially centralized or a fully decentralized control plane, a distributed algorithm is needed \cite{lee2019deep}.

In a system, where users and BSs make decisions in a distributed way, the problem can be formulated as a fully cooperative game to improve the global network performance \cite{liang2019spectrum}. To find stationary solutions of the game, multi-agent (deep) reinforcement learning turns out to be a promising approach as shown in  \cite{sharma2019distributed,doan2019finite}.

\subsection{Summary of Design Principles}
The principles for designing deep learning frameworks in 6G networks are summarized as follows.

\subsubsection{Sharing Models and DNNs from Bottom to Top} As illustrated in Fig. \ref{fig:framework}(a), the tasks in the network level and the BS level depend on the behaviors of BSs and users, respectively. When optimizing policies in the higher levels, the system needs to acquire the models at lower levels \cite{Dora2019Deep}. Alternatively, the system can share DNNs that mimic the behaviors of BSs and users with the centralized control plane. For example, when the theoretical models are inaccurate and the number of real-world data samples is limited, Generative Adversarial Networks (GANs) can be used to approximate the distributions of packet arrival processes or wireless channels \cite{kasgari2019experienced}. When the generator networks are available at the centralized control plane, there is no need to share traffic and channel models.

\subsubsection{Sharing Storage and Computing Resources from Top to Bottom}
To guarantee the data-plane latency, the control plane needs to make decisions according to dynamic traffic states and channel states in real-time \cite{View20175GPPP}. With well-trained DNNs, it is possible for BSs and mobile devices to obtain inference results within a short time. However, the training phase of a deep learning algorithm usually needs huge storage and computing resources. As a result, a BS or a device can hardly obtain a well-trained DNN with local training samples and computing resources. As shown in Fig. \ref{fig:framework}(b), to enable deep learning at lower levels, the central server needs to share storage and computing resources to BSs and devices by federated learning.

\subsubsection{Tradeoffs among Communications, Computing, and Caching} Sharing information among different levels helps saving computing and caching resources at edge servers and mobile devices at the cost of introducing extra communication overheads. Meanwhile, communication delay and reliability affect the performance of learning algorithms \cite{ang2019robust,zhu2019broadband}. Thus, when developing a deep learning framework in a wireless network, the tradeoffs among communications, computing, and caching should be examined carefully. Novel approaches that can improve these tradeoffs are in urgent need \cite{Howard2019FL,zhu2019broadband}.

\section{Tools for Performance Analysis in URLLC}\label{S:Tools}
To illustrate how to improve deep learning with theoretical tools and models, we revisit tools in information theory, queuing theory, and communication theory for performance analysis in URLLC.

\subsection{Information Theory Tools}
The fundamental relationship between the achievable rate and radio resources is crucial for formulating optimization problems in communication systems. In 1948, the maximal error-free data rate that can be achieved in a communication system was derived in \cite{shannon1948mathematical}, which is known as Shannon's capacity. For example, over an Additive White Gaussian Noise (AWGN) channel, Shannon's capacity can be expressed as follows,
\begin{align}
R = W\log_2\left(1+\gamma\right)\;\text{(bits/s)},\label{eq:Shannon}
\end{align}
where $W$ is the bandwidth of the channel and $\gamma$ is the SNR. In the past decades, researchers tried to develop a variety of coding schemes to approach the performance limit. Meanwhile, Shannon's capacity was widely applied in communication system design.

It is well known that Shannon's capacity can be approached when the blocklength of channel codes goes to infinity. However, to avoid a long transmission delay in URLLC, the blocklength should be short. As a result, Shannon's capacity cannot be achieved, and the decoding error probability is non-zero for arbitrary high SNRs.

There are two branches of research in the finite blocklength regime. The first one is analyzing the delay (i.e., coding blocklength) and reliability tradeoffs that can be achieved with different coding schemes \cite{Jeong2004Lower,Guosen2007Analysis,Dina2014Improved,Seyed2014Finite,Kai2014Polar,Pablo2015A,Dan2017On,Vahid2018Finite}.
For example, a lower bound on the bit error rate of finite-length turbo codes was derived in \cite{Jeong2004Lower}. The performance of finite-length Low-Density Parity-Check codes was analyzed in  \cite{Guosen2007Analysis}. The authors of \cite{Seyed2014Finite} investigated how to adjust the blocklength of polar codes to keep the BLock Error Rate (BLER) constant.

The second branch of research aims to derive the performance limit on the achievable rate in the short blocklength regime \cite{Yury2010Channel,Yury2011Scalar,Yury2014Quasi,Austin2014Orthogonal,Giuseppe2016Toward,Giuseppe2016Short}. The milestones of the achievable rate in the short blocklength regime are summarized in Table~\ref{T:Rate}.

\begin{table}[htbp]
\vspace{-0.0cm}\small
\renewcommand{\arraystretch}{1.3}
\caption{Achievable Rate in the Short Blocklength Regime}
\begin{center}\vspace{-0.0cm}\label{T:Rate}
\begin{tabular}{|p{0.5cm}|p{1.2cm}|p{5cm}|}
  \hline
  {\bf{Year}} & {\bf{Reference}} & {\bf{Channel Model}} \\\hline
  2010 &  \cite{Yury2010Channel} & AWGN channel\\\hline
  2011 &  \cite{polyanskiy2011dispersion} & Gilbert-Elliott Channel \\\hline
  2011 &  \cite{Yury2011Scalar} & Scalar coherent fading channel\\\hline
  2014 &  \cite{Yury2014Quasi} & Quasi-static Multiple-Input-Multiple-Output (MIMO) channel\\\hline
  2014 &  \cite{Austin2014Orthogonal} & Coherent Multiple-Input-Single-Output (MISO) channel\\\hline
  2016 &  \cite{Giuseppe2016Short}& Multi-antenna Rayleigh fading channel \\\hline
  2019 &  \cite{collins2019coherent} & Coherent multiple-antenna block-fading channels \\\hline
  2019 &  \cite{lancho2019saddlepoint} & Saddlepoint Approximations, Rayleigh block-fading channels \\\hline
\end{tabular}
\end{center}
\vspace{-0.2cm}
\end{table}

To show the difference between the achievable rate in the short blocklength regime and Shannon's capacity, we rephrase some results in \cite{Yury2010Channel} here. The achievable rate in short blocklength regime over an AWGN channel can be accurately approximated by the normal approximation, i.e.,
\begin{align}
R \approx \frac{W}{\ln2}\left[\ln\left(1+\gamma\right) - \sqrt{\frac{V}{L_{\rm B}}}f^{-1}_{\rm Q}(\varepsilon_{\rm c})\right]\;\text{(bits/s)},\label{eq:Polyanskiy}
\end{align}
where $L_{\rm B}$ is the blocklength, $\varepsilon_{\rm c}$ is the decoding error probability, $f^{-1}_{\rm Q}(\cdot)$ is the inverse of Q-function, and $V$ is the channel dispersion, which can be expressed as $V=1-\frac{1}{(1+\gamma)^2}$ over the AWGN channel. The blocklength $L_{\rm B}$ is the number of symbols in each block. To transmit $L_{\rm B}$ symbols, the amount of time and frequency resources can be obtained from $D_{\rm t}W = L_{\rm B}$, where $D_{\rm t}$ is the transmission duration of the $L_{\rm B}$ symbols.

The difference between \eqref{eq:Shannon} and \eqref{eq:Polyanskiy} lies in the second term in \eqref{eq:Polyanskiy}. When the blocklength $L_{\rm B}$ goes to infinite, then the achievable rate in \eqref{eq:Polyanskiy} approaches Shannon's capacity. To transmit $b$~bits in one block with duration $D_{\rm t}$, the decoding error probability can be derived as follows,
\begin{align}
\varepsilon_{\rm c} \approx {f_{\rm{Q}}}\left( {\sqrt {\frac{{{D_{\rm{t}}}W}}{V}} \left[ {\ln \left( {1 + \gamma } \right) - \frac{{b\ln 2}}{{{D_{\rm{t}}}W}}} \right]} \right), \label{eq:decode}
\end{align}
which is obtained by substituting \eqref{eq:Polyanskiy} into $D_{\rm t}R = b$. From \eqref{eq:decode} we can see that to keep the decoding error probability constant, the required bandwidth decreases with the transmission duration. Essentially, there are tradeoffs among the transmission delay, the spectrum efficiency, and the decoding error probability.

Considering that the result in \eqref{eq:decode} is an approximation that cannot guarantee the required decoding error probability, the lower and upper bounds of the achievable rate were provided in \cite{lancho2019saddlepoint}. The authors further illustrated that the Saddlepoint Approximation derived in this paper lies between the lower and upper bounds, and is more accurate than the normal approximation over Rayleigh fading channel. However, Saddlepoint Approximation has no closed-form expression in general, and hence we cannot obtain closed-form expressions of QoS constraints by using Saddlepoint Approximation.  As shown in \cite{Mahyar2018Short}, when the BLER is $10^{-7}$, the gap between the normal approximation and practical channel coding schemes, such as the extended Bose, Chaudhuri, and Hocquenghem (eBCH) code, is less than $0.1$~dB. Thus, \eqref{eq:Polyanskiy} and \eqref{eq:decode} are still good choices in practical system design.


%
%

\subsection{Queuing Theory Tools}\label{subSec:toolqueue}
Depending on the delay metrics, the existing queuing theory research papers can be classified into three categories, i.e., average delay \cite{Berry2002Gallager,Neely2007Optimal,Yuzhou2014Energy,Niu2015Characterizing}, hard deadline \cite{Elif2002Energy,MITTRANS,Hou2013Timely,Gongzheng2016Delay,Shengfeng2016Convexity}, and statistical QoS \cite{EB,EC,Wenchi2013Joint,Leila2015Effective,Cross2018she,ren2018low,Gross2015Delay,Qiuming2017Per}. 

\subsubsection{Average Queuing Delay}
Let us consider a queuing system with average packet arrival process $\bar{a}(t)$, where $a(t)$ is the data arrival rate at time $t$. Then, the average queuing delay $\bar{D}(t)$ and the average queue length $\bar{Q}(t)$ satisfy a simple and exact relation \cite{Mor2013Queue}, i.e.,
\begin{align}
\bar{Q}(t) = \bar{D}(t)\bar{a}(t), \nonumber
\end{align}
which is the famous Little's Law. It is a very general conclusion that does not depend on the service order of the queuing system, the length of the buffer, or the distributions of the inter-arrival time between packets and the service time of each packet. The only assumption is that the arrival processes and service processes are stationary.

The average delay metric is suitable for services with no strict delay requirement, such as file download. However, it is not applicable to URLLC. This is because the average delay metric cannot reflect jitter or the packet loss probability caused by delay violations. Even if the average delay is shorter than $1$~ms, the delay experienced by a certain packet could be much longer than $1$~ms.

\subsubsection{Hard Deadline} If a service requires a hard deadline $T_{\rm d}$, it means that the packets should be transmitted to the receiver before the deadline with probability one. To illustrate how to guarantee a hard deadline, we consider a fluid model in a First-Come-First-Serve (FCFS) system as that in \cite{MITTRANS}. The total amount of data arriving at the buffer during the time interval $[0,t]$ is defined as $A(t)=\int_0^t {a( \tau )d\tau }$. The total amount of data leaving the buffer during time interval $[0,t]$ is denoted by $S(t)=\int_0^t {s( \tau )d\tau }$, where $s( t )$ is the service rate at time $t$. To satisfy the hard deadline $T_{\rm d}$, the following constraints should be satisfied,
\begin{align}
&S(t) \leq A(t), \forall t \in [0,T] \label{eq:UB},\\
&S(t) \geq A(t-T_{\rm d}), \forall t \in [T_{\rm d},T+T_{\rm d}] \label{eq:LB},
\end{align}
where $T$ is the total service time. Constraint \eqref{eq:UB} ensures that the amount of data transmitted by the system does not exceed the amount of data that has arrived at the buffer. Constraint \eqref{eq:LB} ensures the hard deadline requirement.

The hard deadline is widely applied in wireline communications, where the channel is deterministic \cite{MITTRANS,MITTRANS2}. However, constraint \eqref{eq:LB} cannot be satisfied in most wireless communication systems \cite{Shengfeng2016Convexity}.

\subsubsection{Statistical QoS}
For most of the real-time services and URLLC, statistical QoS is the best choice among the three kinds of metrics. Statistical QoS is characterized by a delay bound and a threshold of the maximal tolerable delay bound violation probability, $(D_{\rm q}^{\max},\varepsilon_{\rm q})$, respectively. For Voice over Internet Protocol services, the requirement in radio access networks is $(50~\text{ms}, 2\%)$ \cite{3GPPQoS}. For URLLC, the queuing delay bound should be less than the E2E delay and the queuing delay violation probability should be less than the overall packet loss probability, e.g., $(1~\text{ms},10^{-7})$ \cite{3GPP2017Scenarios}.

The existing publications on the statistical QoS or the distribution of queuing delay mainly focus on specific arrival models or service models. Some useful results with FCFS servers and Processor-Sharing (PS) servers are summarized in Table~\ref{T:Statistic}, where Kendall's Notation indicates arrival types, service types, number of servers, and scheduling principles. Here ``M" and ``D" represent memoryless processes and deterministic processes, respectively. ``G" means that the inter-arrival time between packets or the service time of each packet may follow any general distribution. The abbreviation ``$\text{Geo}^{[X]}$" in \cite{devassy2019reliable} means that the packets arrived in a time slot constitutes a bulk, and the bulk arrival process is stationary and memoryless with Binomial marginal distribution. Considering that PS servers are widely deployed in computing systems \cite{Mor2013Queue}, A. P. Zwart \emph{et al.} derived the distribution of the delay in the large-delay regime, which is referred to as a tail probability \cite{Sojourn2000Zwart}. Since the delay experienced by short packets in the short delay regime is more relevant to URLLC, C. She \emph{et al.} derived an approximation of the delay experienced by short packets in a PS server \cite{Changyang2019IoT}.

\begin{table}[htbp]
\vspace{-0.0cm}\small
\renewcommand{\arraystretch}{1.3}
\caption{Statistical QoS of Specific Queuing Models}
\begin{center}\vspace{-0.0cm}\label{T:Statistic}
\begin{tabular}{|p{1.2cm}|p{2.6cm}|p{3.5cm}|}
  \hline
  {\bf{Reference}} & {\bf{Kendall's Notation}} & {\bf{Result}}  \\\hline
  \cite{Gross1985MD1} & M/D/1/FCFS & Distribution of delay \\\hline
  \cite{stewart2009probability} & M/M/1/FCFS & Distribution of delay \\\hline
  \cite{devassy2019reliable} & $\text{Geo}^{[X]}$/G/1/FCFS & Upper bound of delay violation probability \\\hline
  \cite{Sojourn2000Zwart} & M/G/1/PS & Tail probability of delay \\\hline
  \cite{Changyang2019IoT} & M/G/1/PS & Delay experienced by short packets \\\hline
\end{tabular}
\end{center}
\vspace{-0.2cm}
\end{table}

There are two principal tools for analyzing statistical QoS in more general queuing systems, i.e., network calculus and effective bandwidth/capacity \cite{le2001network,EB,EC}.

The basic idea of network calculus is to convert the accumulated arrival and service rates from the bit domain to the SNR domain, and then derive the upper bounds of the Mellin transforms of the arrival and service processes \cite{al2016network}. From the upper bounds of the Mellin transforms, the upper bound of the delay violation probability can be obtained \cite{Gross2015Delay}.

Network calculus provides the upper bound of the Complementary Cumulative Distribution Function (CCDF) of the steady-state queuing delay. However, network calculus is not convenient in wireless network design since it is very challenging to derive the relation between $(D_{\rm q}^{\max},\varepsilon_{\rm q})$ and the required radio resources in a closed form. In URLLC, we are interested in the tail probability of the queuing delay, i.e., the case that $\varepsilon_{\rm q}$ is very small. Thus, there is no need to derive the upper bound of CCDF for all values of $\varepsilon_{\rm q}$.

To analyze the asymptotic case when $\varepsilon_{\rm q}$ is very small, effective bandwidth and effective capacity can be used \cite{EB,EC}. Based on Mellin transforms in network calculus, we can derive the effective bandwidth of arrival processes and effective bandwidth of service processes with $\text{G}\ddot{\text{a}}\text{rtner-}\text{Ellis}$ theory \cite{EB,EC,al2016network}.

Effective bandwidth is defined as the minimal constant service rate that is required to guarantee $(D_{\rm q}^{\max},\varepsilon_{\rm q})$ for a random arrival process $a(t)$ \cite{EB}. Inspired by the concept of effective bandwidth, D. Wu \emph{et al.} introduced the concept of effective capacity over wireless channels \cite{EC}. Effective capacity is defined as the maximal constant arrival rate that can be served by a random service process $s(t)$ subject to the requirement $(D_{\rm q}^{\max},\varepsilon_{\rm q})$. We denote the effective bandwidth of $a(t)$ and the effective capacity of $s(t)$ as $E_B$ and $E_C$, respectively. When both the arrival process and the service process are random, $(D_{\rm q}^{\max},\varepsilon_{\rm q})$ can be satisfied if \cite{liu2007resource}
\begin{align}
E_C \geq E_B. \label{eq:ECEB}
\end{align}

The formal definitions of effective bandwidth and effective capacity were summarized in \cite{she2015energy}, i.e.,
\begin{align}
{E_B}& = \mathop {\lim }\limits_{t \to \infty }
\frac{1}{{\theta t}}\ln \mathbb{E}\left[ {{e^{\theta \int_0^t {a\left( \tau  \right)d\tau }}}} \right],\nonumber\\
{E_C}& = -\mathop {\lim }\limits_{t
\to \infty } \frac{1}{{\theta t}}\ln \mathbb{E}\left[
{{e^{-\theta \int_0^t {s\left( \tau  \right)d\tau } }}} \right],\nonumber
\end{align}
where $\theta$ is a QoS exponent. The value of $\theta$ decreases with $D_{\rm q}^{\max}$ and $\varepsilon_{\rm q}$ according to the following expression \cite{she2016energy},
\begin{align}
\Pr\left(D_{\rm q}^{\infty} \geq D_{\rm q}^{\max}\right) \approx \exp\left(-{\theta} E_B D_{\rm q}^{\max}\right) = \varepsilon_{\rm q},\label{eq:theta}
\end{align}
where $D_{\rm q}^{\infty}$ is the steady-state queuing delay. The approximation in \eqref{eq:theta} is accurate for the tail probability. For URLLC, the accuracy of \eqref{eq:theta} has been validated in \cite{Cross2018she}, where a Poisson process, an interrupted Poisson process, or a switched Poisson process is served by a constant service rate (since the delay requirement is shorter than the channel coherence time in most cases in URLLC, the channel is quasi-static and the service rate is constant).



Effective bandwidth and effective capacity have been extensively investigated in the existing literature with different queuing models. Some useful results have been summarized in Table \ref{T:Queue}. A more comprehensive survey on effective capacity in wireless networks is presented by Amjad \emph{et al.} \cite{amjad2019effective}.


\begin{table}[htbp]
\vspace{-0.0cm}\small
\renewcommand{\arraystretch}{1.3}
\caption{Useful results for analyzing statistical QoS}
\begin{center}\vspace{-0.0cm}\label{T:Queue}
\begin{tabular}{|p{0.5cm}|p{1.2cm}|p{5.5cm}|}
  \hline
  \multicolumn{3}{|c|}{Effective Bandwidth} \\\hline
  {\bf{Year}} & {\bf{Reference}} & {\bf{Sources}} \\\hline
  1995 &  \cite{EB} & General definition \\\hline
  1996 &  \cite{Frank1996Notes} & Periodic sources, Gaussian sources, ON-OFF sources, and compound Poisson sources \\\hline
  2016 &  \cite{ozmen2015wireless} & Interrupted Poisson sources \\\hline
  2018 &  \cite{Cross2018she} & Poisson sources in closed form \\\hline
  \multicolumn{3}{|c|}{Effective Capacity} \\\hline
  {\bf{Year}} & {\bf{Reference}} & {\bf{Channels}} \\\hline
  2003 &  \cite{EC} & General definition\\\hline
  2007 &  \cite{liu2007resource} & ON-OFF channel \\\hline
  2007 &  \cite{tang2007quality} & Nakagami-$m$ fading channel \\\hline
  2010 &  \cite{soret2010capacity}& Correlated Rayleigh channel \\\hline
  2013 & \cite{gursoy2013throughput}& AWGN, finite blocklength regime \\\hline
  2016 & \cite{Yulin2016Blocklength}& Rayleigh channel, finite blocklength regime \\\hline
\end{tabular}
\end{center}
\vspace{-0.2cm}
\end{table}


\subsection{Communication Theory Tools}
\subsubsection{Characterizing Correlations of Multiple Links} To improve the reliability of URLLC in wireless communications, we can exploit different types of diversities, such as frequency diversity \cite{David2014Achieving}, spatial diversity \cite{Martin2015GCworkshop}, and interface diversity \cite{Jimmy2017URLLC}. The basic idea is to send signals through multiple links in parallel. If one or some of these links experience a low channel quality or congestions, the receiver can still recover the packet in time with the signals from the other links.



For example, $N_{\rm P}$ copies of one packet are transmitted over $N_{\rm P}$ paths. The packet loss probability of each path is denoted by $P_{\rm L}(n), n = 1,...,N_{\rm P}$. If the packet losses of different paths are uncorrelated, then the overall packet loss probability can be expressed as $P^{\rm tot}_{\rm L} = \prod\limits_{n = 1}^{{N_{\rm{P}}}} {{P_{\rm{L}}}(n)}$.

\begin{figure}[htbp]
        \centering
        \begin{minipage}[t]{0.25\textwidth}
        \includegraphics[width=1\textwidth]{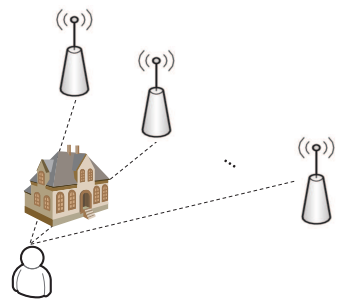}
        \end{minipage}
        \caption{Cross-correlation of packet losses \cite{she2019ultra,WirelessCom}.}
        \label{fig:DisAntenna}
        \vspace{-0.0cm}
\end{figure}

However, the packet losses of different paths may be highly correlated. To illustrate the impact of cross-correlation of packet losses on the reliability, we consider a multi-antenna system with $N_{\rm P}$ distributed antennas in Fig. \ref{fig:DisAntenna}. Let ${\bf{1}}_{\rm LoS}(n_{\rm P})$ be the indicator that the there is a Line-of-Sight (LoS) path between the user and the $n_{\rm P}$-th antenna. If there is a LoS, ${\bf{1}}_{\rm LoS}(n_{\rm P})=1$, otherwise, ${\bf{1}}_{\rm LoS}(n_{\rm P})=0$. Then, the probability that there is a LoS path between the user and each of the distributed antennas can be defined as $P^{\rm one}_{\rm LoS} \triangleq {\mathbb{E}}\{{\bf{1}}_{\rm LoS}(n_{\rm P})\}$. When the distances among the $N_{\rm P}$ antennas are comparable to the typical scale of obstacles (e.g., the widths and lengths of buildings) in the environment, the values of ${\bf{1}}_{\rm LoS}(n_{\rm P}), n_{\rm P}=1,...,N_{\rm P},$ are correlated. The correlation coefficient of two adjacent antennas is denoted by $\rho_{\rm LoS}$.

For URLLC, we are interested in the probability that at least one of the antennas can receive the packet from the user. Let's consider a toy example: if there are one or more LoS paths, the antenna can receive the packet, otherwise, the packet is lost. Then, the reliability of the system is the probability that at least one of the antennas has LoS path to the user, i.e. \cite{minkova2014new},
\begin{align}
P^{\rm all}_{\rm LoS} = 1-\left(1-P^{\rm one}_{\rm LoS}\right)\left[1-P^{\rm one}_{\rm LoS}(1-\rho_{\rm LoS})\right]^{N_{\rm P}-1}.\nonumber
\end{align}
The above expression indicates that $P^{\rm all}_{\rm LoS}$ decreases with $\rho_{\rm LoS}$. If $\rho_{\rm LoS} = 0$, then $P^{\rm all}_{\rm LoS} = 1-\left(1-P^{\rm one}_{\rm LoS}\right)^{N_{\rm P}}$. If $\rho_{\rm LoS} = 1$, then $P^{\rm all}_{\rm LoS} = P^{\rm one}_{\rm LoS}$.

For a particular antenna, the value of $P^{\rm one}_{\rm LoS}$ depends on the communication environment and the locations of both the antenna and the user. According to 3GPP LoS model, $P^{\rm one}_{\rm LoS}$ in a terrestrial cellular network can be expressed as follows \cite{3GPPQoS},
\begin{align}
P^{\rm one}_{\rm LoS} = \min\left(\frac{18}{r_{\rm u}},1\right)\left[1-\exp(-\frac{r_{\rm u}}{36})\right] + \exp{-\frac{r_{\rm u}}{36}},\nonumber
\end{align}
where $r_{\rm u}$ is the distance between the user and the antenna. For ground-to-air channels, the value of $P^{\rm one}_{\rm LoS}$ depends on the elevation angle according to the following expression \cite{Akram2014Modeling,altitude2014CL},
\begin{align}
P^{\rm one}_{\rm LoS}  = \frac{1}{{1 + \phi_{\rm e} \exp \left[ { - \psi_{\rm e} \left( {{\theta_{\rm u}} - \phi_{\rm e} } \right)} \right]}},\nonumber
\end{align}
where ${\theta_{\rm u}}$ is the elevation angle of the user, and $\phi_{\rm e}$ and $\psi_{\rm e}$ are two constants, which depend on the communication environment, such as suburban, urban, dense urban, and highrise urban. The typical values of $\phi_{\rm e}$ and $\psi_{\rm e}$ can be found in \cite{she2019ultra}.

In practice, the packet loss probability depends not only on whether there is a LoS path but also on the shadowing of a wireless channel. To characterize shadowing correlation, S. Szyszkowicz \emph{et al.} established useful shadowing correlation models in \cite{szyszkowicz2010feasibility}. Even with these models, deriving the packet loss probability in URLLC is very challenging. To overcome this difficulty, a numerical method for evaluating packet loss probability was proposed by D.~$\ddot{\text{O}}$hmann \emph{et al.} \cite{David2016WCL}. More recently, C. She \emph{et al.} analyzed the impact of shadowing correlation on the network availability, where a packet is transmitted via a cellular link and a Device-to-Device (D2D) link \cite{She2018improving}.

Furthermore, the correlation of multiple subchannels in MIMO channels and frequency-selective channels also affects micro-diversity gains. The achievable rate of massive MIMO systems with spatial correlation was investigated in \cite{jiang2015achievable}, where the authors maximized the data rate over correlated channels. The reliability over frequency-selective Rayleigh fading channels was considered in \cite{David2014Achieving}, where an approximation of the outage probability over correlated channels was derived. Considering that there are very few results on the reliability over correlated channels, how to guarantee QoS requirements of URLLC over correlated channels deserves further research.

\subsubsection{Stochastic Geometry for Delay Analysis in Large-Scale Networks} Most of the existing publications on delay analysis consider the systems with small scale and deterministic network topologies. However, in practical systems, the locations of a large number of users are stochastic. Users can either communicate with each other via D2D links or communicate with BSs via cellular links. Caused by dynamic traffic loads, network congestions lead to a significant delay in large-scale wireless networks with high user density. Yet, how to analyze delay, especially in the low-latency regime, remains an open problem.

In 2012, M. Haenggi applied stochastic geometry to analyze delays in large-scale networks \cite{haenggi2012local}, where the average access delay was derived. One year later, P. Kong studied the tradeoff between the power consumption and the average delay experienced by packets \cite{Peng2013Power}, where existing results on M/G/1 queue were fitted within the stochastic geometry framework. More recently, Y. Zhong \emph{et al.} proposed a notation of delay outage to evaluate the delay performance of different scheduling policies, defined as the probability that the average delay experienced by a typical user is longer than a threshold conditioned on the spatial locations of BSs, i.e. \cite{Yi2017JSAC},
\begin{align}
\Pr\{\bar{D} > T_{\rm th}| \Phi_{\rm b}\},\nonumber
\end{align}
where $\bar{D}$ is the mean delay experienced by the user, $T_{\rm th}$ is the required threshold, and $\Phi_{\rm b}$ is a realization of spatial locations of BSs.

As discussed in Section \ref{subSec:toolqueue}, the requirement of URLLC should be characterized by a constraint on the statistical QoS. However, most of the existing studies analyzed the average delay in large-scale networks. To the best knowledge of the authors, there is no available tool for analyzing statistical QoS in large-scale networks.

\subsection{{Summary of Analytical Tools}}
The analytical tools enable us to evaluate the performance of wireless networks without extensive simulations, and serve as the key building blocks of optimization frameworks. Yet, there are some open issues:
\begin{itemize}
\item Each of these tools is used to analyze the performance of one part of a wireless network. A whole picture of the system is needed for E2E optimization, which is analytically intractable in most of the cases.
\item To apply these tools, we need simplified models and ideal assumptions. The impacts of model mismatch on the performance of URLLC remain unclear. For example, if Rayleigh fading (without LoS path between transmitters and receivers) is adopted in analytical, one can obtain an upper bound of the latency or the packet loss probability (The LoS component exists in practical communication scenarios). However, such a model will lead to conservative resource allocation.
\item The tools in Tables \ref{T:Rate}, \ref{T:Statistic}, and \ref{T:Queue} rely on simplified channel and queuing models. For more complicated and practical models, we can hardly derive the tail probability of delay and the packet loss probability.
\end{itemize}
To address the above issues, we will illustrate how to optimize the whole network with cross-layer design in Section \ref{S:cross}, introduce deep transfer learning to handle model mismatch in Section \ref{Sec:SDLNet}, and discuss model-free approaches in Sections \ref{S:unsupervised} and \ref{Sec:DRL}.

\section{Cross-Layer Design}\label{S:cross}
By decomposing communication systems into seven layers \cite{jiang2018low}, we can develop practical, but suboptimal solutions for URLLC in each layer. To reflect the interactions among different layers, and to optimize the delay and reliability of the whole system, we turn to cross-layer optimization. Although cross-layer design in wireless networks has been studied in the existing literature, the solutions for URLLC are limited. The most recent works on cross-layer design for URLLC are summarized in Table \ref{T:Cross}.

\begin{table*}[htbp]
\vspace{-0.0cm}\footnotesize
\renewcommand{\arraystretch}{1.3}
\caption{Cross-layer optimization for URLLC}
\begin{center}\vspace{-0.0cm}\label{T:Cross}
\begin{tabular}{|p{2.3cm}|p{3.6cm}|p{4.0cm}|p{3.0cm}|p{3.3cm}|}
  \hline
 {\bf{Design Objective}} & {\bf{Physical-layer issues}} & {\bf{Link-layer issues}}  & \bf{Network-layer issues} & \bf{Challenges/Results}  \\\hline
 Power control optimization \cite{gursoy2013throughput}  & Normal approximation over AWGN channel& Ensuring the statistical QoS with effective capacity & N/A & No closed-form expression in the short blocklength regime\\\hline
 Throughput analysis \cite{ozcan2013throughput} & Normal approximation in cognitive radio systems & Ensuring the statistical QoS with effective capacity & N/A & No closed-form result in general \\\hline
 Delay analysis \cite{Gross2015Delay} & Normal approximation over quasi-static fading channel & Analyzing the statistical QoS with network calculus & N/A & No closed-form result in general \\\hline
 Packet scheduling \cite{Shengfeng2016Convexity} & Normal approximation over quasi-static fading channel & DL packet scheduling before a hard deadline& N/A & An online algorithm with small performance losses \\\hline
 Throughput analysis \cite{Yulin2016Blocklength} & Throughput of relay systems the in short blocklength regime& Ensuring the statistical QoS with effective capacity& N/A & No closed-form result in general \\\hline
 Power minimization \cite{Cross2018she} & Normal approximation over quasi-static fading channel & Ensuring the statistical QoS with effective bandwidth & N/A & Near-optimal solutions of the non-convex problem\\\hline
 Improving the reliability-SE trade-off \cite{Jimmy2017URLLC} & N/A & Improving the tradeoff between spectrum efficiency and reliability with network coding & Packet cloning and splitting on multiple communication interfaces & Intractable in large-scale network\\\hline
 Improving reliability \cite{tuninetti2018scheduling} & Decoding errors in short blocklength regime & Scheduler design under a hard deadline constraint& N/A & Reducing outage probability with combination strategies\\\hline
 Bandwidth minimization \cite{Joint2018She} & Achievable rate in short blocklength regime& E2E optimization (i.e., UL/DL transmission delay and queuing delay)& N/A & Near-optimal solutions of the non-convex problem\\\hline
 Power allocation \cite{Hu2018TWC} & Maximizing the sum throughput of multiple users in short blocklength regime & Ensuring statistical QoS for different kinds of packet arrivals & N/A  & A sub-optimal algorithm for solving the non-convex problem\\\hline
 Bandwidth minimization \cite{Hou2018Burstiness} & Normal approximation over quasi-static fading channel & Burstiness aware resource reservation under statistical QoS constraints& N/A & Reducing $40$\% bandwidth by traffic state prediction\\\hline
 Availability analysis \cite{She2018improving} & Normal approximation over quasi-static fading channel & Different transmission protocols via cellular links and D2D links& Improving network availability with multi-connectivity & An accurate approximation of network availability \\\hline
 Coding-queuing analyses \cite{devassy2019reliable} & Variable-length-stop-feedback codes & Retransmission under latency and peak-age violation guarantees& N/A  & Accurate approximations of target tail probabilities\\\hline
 Coded retransmission design \cite{malak2019tiny} & Tiny codes with just $2$ packets & Automatic Repeat Request (ARQ) protocols with delayed and unreliable feedback& N/A & Improving $40$\% throughput with latency and reliability guarantees \\\hline
 Optimizing resource allocation \cite{tang2019service} & Ensuring transmission delay in the short blocklength regime & N/A& Network slicing for eMBB and URLLC in cloud-RAN  & Searching near-optimal solutions of mixed-integer programming\\\hline
 Delay analysis \cite{schiessl2019delay} & Multi-user MISO with imperfect CSI and finite blocklength coding & Analyzing statistical QoS with network calculus & N/A  & No closed-form solution in general\\\hline
 Delay, reliability, throughput analysis \cite{sahin2019delay} & Modulation and coding scheme & Incremental redundancy-hybrid ARQ over correlated Rayleigh fading channel& N/A  & No closed-form solution in general\\\hline
 Resource allocation \cite{elayoubi2019radio} & Modulation and coding scheme& Individual resource reservation or contention-based resource sharing& N/A  & Analytical expressions of reliability\\\hline
 Spectrum and power allocation \cite{guo2019resource} &Interference management for vehicle-to-vehicle and vehicle-to-infrastructure links& Ensuring statistical QoS with effective capacity & N/A  & Solving subproblems on spectrum/power allocation in polynomial time \\\hline
 Distortion minimization \cite{zhou2019lossy} & Joint source and channel codes& N/A & Improving reliability over parallel AWGN channels  & Tight approximations and bounds on distortion level, but not in closed form \\\hline
 Network architecture design \cite{khoshnevisan20195g} & CoMP communications  & E2E design of 5G networks for industrial factory automation & A novel architecture for TSN  & Around $2$~ms round-trip delay in their prototype\\\hline
 Offloading optimization \cite{Changyang2019IoT} &Normal approximation over quasi-static fading channel & Delay analysis in processor-sharing servers with both short and long packets & User association \& task offloading of mission-critical IoT in MEC  & Closed-form approximations of delay and low-complexity offloading algorithms\\\hline
\end{tabular}
\end{center}
\vspace{-0.2cm}
\end{table*}

\begin{figure*}[ht]
        \centering
        \begin{minipage}[t]{0.7\textwidth}
        \includegraphics[width=1\textwidth]{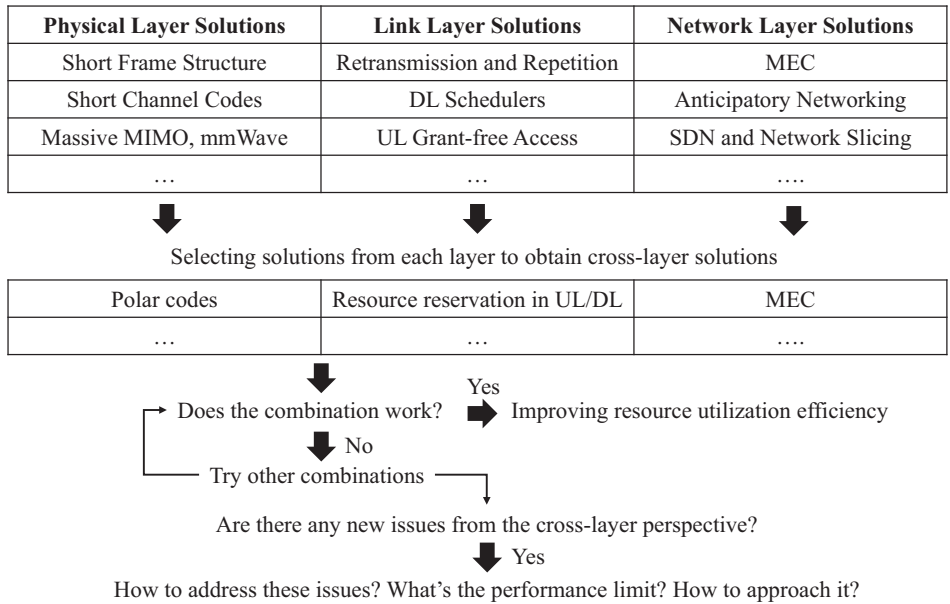}
        \end{minipage}
        \caption{From single-layer design to cross-layer design.}
        \label{fig:Crosslayer}
        \vspace{-0.2cm}
\end{figure*}

\subsection{Design Principles and Fundamental Results}
\subsubsection{Design Principles} The design principles in the cross-layer design are shown in Fig. \ref{fig:Crosslayer}. A straightforward approach is selecting technologies from each layer, and see whether this combination works. If the combination can guarantee the QoS requirement of URLLC, then we can further optimize resource allocation to improve the resource utilization efficiency. Otherwise, we need to try some other combinations. Meanwhile, we may identify new issues from a cross-layer perspective. By cross-layer optimization, we aim to find out the performance limit and show how to approach the performance limit.

\subsubsection{Fundamental Results} R. Berry \emph{et al.} investigated how to guarantee average delay constraints over fading channels \cite{Berry2002Gallager}. They obtained the Pareto optimal power-delay tradeoff, i.e., the average transmit power decreases with the average queuing delay according to $1/{\bar{D}^2}$ in the large delay regime. To achieve the optimal tradeoff, a power control policy should take both Channel State Information (CSI) and Queue State Information (QSI) into account. If we apply any power control policy that does not depend on QSI, such as the water-filling policy \cite{WirelessCom}, then the average transmit power decreases with the delay according to $1/{\bar{D}}$. In other words, directly combining the water-filling policy (i.e., the optimal policy that maximizes the average throughput in the physical layer) with a queuing system cannot achieve the optimal tradeoff between the average transmit power and the average queuing delay.

Following this fundamental result, M. J. Neely extended the power-delay tradeoff into multi-user scenarios \cite{Neely2007Optimal} and proposed a packet dropping mechanism that can exceed the Pareto optimal power-delay tradeoff \cite{neely2009intelligent}. Considering that the average delay metric is not suitable for real-time services, such as video and audio, an optimal power control policy that maximizes the throughput subject to the statistical QoS requirement was derived by J. Tang \emph{et al.} in \cite{tang2007quality}. The result in \cite{tang2007quality} shows that when the required delay bound goes to infinite, the optimal power control policy converges to the water-filling policy in \cite{WirelessCom}. When the required delay bound approaches the channel coherence time, the optimal power control policy converges to the channel inverse.


\begin{figure}[ht]
        \centering
        \begin{minipage}[t]{0.45\textwidth}
        \includegraphics[width=1\textwidth]{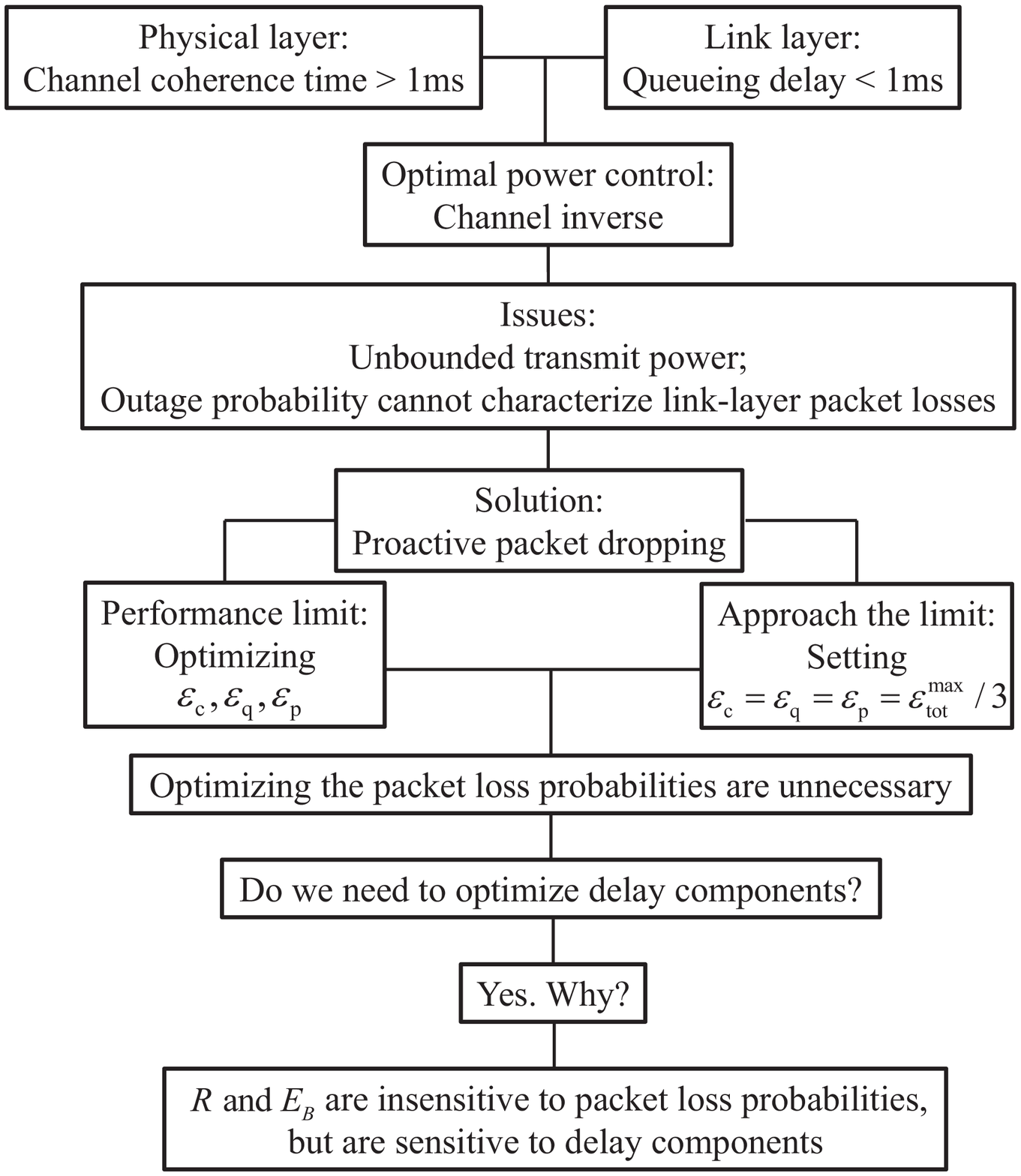}
        \end{minipage}
        \caption{Diagram of physical- and link-layer optimization \cite{she2017radio,Cross2018she}.}
        \label{fig:PHYLINK}
        \vspace{-0.2cm}
\end{figure}

\subsection{Physical- and Link-Layer Optimization}
A diagram of cross-layer optimization for physical and link layers is shown in Fig. \ref{fig:PHYLINK}. Since the E2E delay requirement of URLLC is around $1$~ms, the required delay bound is shorter than the channel coherence time \cite{David2005Fundamentals}. To guarantee a queuing delay that is shorter than the channel coherence time, the power control policy should be the channel inverse \cite{tang2007quality}. For typical wireless channels, such as Rayleigh fading or Nakagami-$m$ fading, the required transmit power of the channel inverse is unbounded.

The issue observed from cross-layer design is similar to the outage probability defined in \cite{caire1999optimum}. Different from the physical-layer analysis, outage probability does not equal to the packet loss probability in cross-layer design. As shown in \cite{Cross2018she}, by dropping some packets proactively, it is possible to reduce the overall packet loss probability. Such an idea was inspired by the result in \cite{neely2009intelligent}. Denote the proactive packet dropping probability by $\varepsilon_{\rm p}$. By optimizing decoding error probability, queuing delay violation probability, and proactive packet dropping probability, we can minimize the required maximal transmit power at the cost of high computational complexity. To avoid complicated optimization, a straightforward solution is setting $\varepsilon_{\rm c} = \varepsilon_{\rm q} = \varepsilon_{\rm p} = \varepsilon_{\rm tot}^{\max}/3$. The results in \cite{Cross2018she} show that the performance loss (i.e., required transmit power) without optimization is less than $5$~\% when the number of antennas is larger than eight. In other words, setting the packet loss probabilities as equal is near-optimal, and there is no need to optimize the values of $\varepsilon_{\rm c}$, $\varepsilon_{\rm q}$, and $\varepsilon_{\rm p}$.

%
%
%
%
%
%

%

From the above-mentioned conclusion, it is natural to raise the following question: Is it necessary to optimize the delay components subject to the E2E delay? The results in \cite{Joint2018She} show that by optimizing UL/DL delays and queuing delay subject to the E2E delay constraint, a large amount of bandwidth can be saved. This is because the achievable rate in the short blocklength regime is insensitive to the decoding error probability, but very sensitive to transmission delay. Therefore, optimizing the delay components subject to the E2E delay requirement is necessary for maximizing resource utilization efficiencies of URLLC.

%


Apart from the above solutions, some other cross-layer solutions for physical and link layers have been developed in URLLC recently. By considering the hard deadline metric, an energy-efficient scheduling policy for short packet transmission was proposed in \cite{Shengfeng2016Convexity}, and delay bound violation probabilities of different schedulers were evaluated in \cite{tuninetti2018scheduling}. How to guarantee statistical QoS in the short blocklength regime has been studied in UL Tactile Internet \cite{Hou2018Burstiness}, DL multi-user scenarios with Markovian sources \cite{Hu2018TWC}, point-to-point communications with imperfect CSI \cite{schiessl2019delay}, and vehicle networks with strong interferences between vehicle-to-vehicle and vehicle-to-infrastructure links \cite{guo2019resource}. Coded ARQ and incremental redundancy-hybrid ARQ schemes were developed in \cite{malak2019tiny} and \cite{sahin2019delay}, respectively. Considering that the density of devices in smart factory can be very large, the individual resource reservation scheme and the contention-based resource sharing scheme were considered in \cite{elayoubi2019radio}.

\subsection{Physical-, Link-, and Network-Layer Optimization} The cross-layer models will be very complicated if more than two layers are considered. Thus, joint optimization of physical, link, and network layers is very challenging. Although a few papers optimize network management from a cross-layer approach, some simplified models on the physical and link layers are used in these works \cite{Jimmy2017URLLC,khoshnevisan20195g,tang2019service}.

When optimizing packet splitting on multiple communication interfaces, the authors of \cite{Jimmy2017URLLC} assumed that the distribution of latency (i.e., the latency-reliability function) of each interface is available. With the simplified link-layer model, coded packets are distributed through multiple interfaces. As in \cite{Jimmy2017URLLC}, joint source and channel codes over parallel channels were developed in \cite{khoshnevisan20195g}, where simplified AWGN channels with independent fading are considered in the analysis.  To optimize network slicing for eMBB and URLLC in cloud-RAN, the resources that are required to satisfy the data rate constraint of eMBB and the transmission delay constraint of URLLC were derived in \cite{tang2019service}. The latency in the link-layer was not considered in the framework in \cite{tang2019service}. Otherwise, the problem will become intractable. Resource allocation in vehicle networks was optimized to maximize the throughput of URLLC in \cite{guo2019TWC}, where Shannon's capacity is used to simplify the complicated achievable rate over interference channel and the average packet loss probability and average latency are analyzed in the queuing system for analytical tractability.

To validate the E2E design in a time-sensitive network architecture with CoMP for URLLC, issues in the three lower layers were considered in \cite{khoshnevisan20195g}, where most of the conclusions are obtained via simulation. Since the models are too complicated, one can hardly obtain any closed-form result. To gain some useful insights from theoretical analysis, the authors of \cite{Changyang2019IoT} analyzed the E2E delay in an MEC system, where the UL and DL transmission delays and the processing delay at the servers are considered. In addition, the packet loss probabilities caused by decoding errors and queuing delay violations were also derived. Based on these physical- and link-layer models, user association and computing offloading of URLLC service were optimized in a multi-cell MEC system, where eMBB services are considered as background services (i.e., not optimized).

\subsection{From Lower Layers to Higher Layers} The E2E performance of URLLC depends on all the seven layers in the OSI model \cite{Irfan2016A}. By considering more layers in communication system design, better performance can be achieved. However, most of the working groups in standardization mainly focus on one or two layers \cite{nasrallah2018ultra}. For example, the Institute of Electronics and Electrical Engineers (IEEE) and the Internet Engineering Task Force (IETF) have formed some working groups on the standardization of TSN \cite{Farkas2018TSN}. The IEEE 802.1 working group mainly focuses on the physical and link layers, while the deterministic network working group of IETF focuses on the network layer and higher layers \cite{nasrallah2018ultra}. By including more protocol layers in one design framework, the problem will become more complicated, but the obtained policy has the potential to achieve a better performance. Essentially, there is a tradeoff between complexity and performance. To move from the three lower layers to higher layers, novel technologies that can capture complicated features and can be implemented in real-time are in urgent need. For example, driven by the requirement from the application layer, the freshness of the knowledge becomes one of the KPIs in remote monitoring/control systems \cite{kaul2011minimizing}. To achieve a better tradeoff between latency and AoI, the authors of \cite{devassy2019reliable} developed variable-length-stop-feedback codes that outperform the simple automatic repetition request in terms of the delay violation probability. In addition, their results indicated that by changing the service order of a queuing system, it is possible a smaller peak-age violation probability.

\subsection{Summary of Cross-layer Optimization}
From a cross-layer optimization framework, one can discover new research challenges in URLLC. By solving cross-layer optimization problems, the performance limits on latency, reliability, and resource utilization efficiency can be obtained. In addition, the optimal policies provide some useful insights on how to find near-optimal solutions and how to approach the performance limits. Some open issues of cross-layer optimization are summarized as follows:

\begin{itemize}
\item The problem formulation relies on the analytical tools in Section \ref{S:Tools}. Thus, some open issues of the analytical tools, such as model mismatch, also exist in the optimization frameworks.
\item The optimization problems are NP-hard in most cases. The optimal policies or fast optimization algorithms are developed on a case by case basis. A general methodology for designing fast algorithms is missing.
\item The results obtained in the existing literature (See Table \ref{T:Cross}) only considered two or three bottom layers. Thus, these frameworks cannot address the technical problems in the upper layers. If the issues in the upper layers are taken into account, the optimization problems could be intractable.
\end{itemize}
We will discuss how to take advantage of existing domain knowledge and how to transfer the knowledge to practical scenarios in the rest part of this paper.







\section{Supervised Deep Learning for URLLC}\label{Sec:SDLNet}
Deep learning is a family of machine learning techniques based on DNNs. Depending on the training methods, it can be categorized as supervised, unsupervised, and reinforcement learning. Supervised deep learning has been applied in different fields in computer science, including computer vision \cite{voulodimos2018deep}, speech recognition \cite{amodei2016deep}, natural language processing \cite{collobert2008unified}, and so on. Recently, supervised deep learning algorithms were also adopted in wireless networks \cite{qin2019deep}. The existing publications that used supervised deep learning for channel esimation/prediction, traffic prediction, mobility/trajectory prediction, QoS/Quality-of-Experience (QoE) prediction, and policy approximation are summarized in Table \ref{T:Supervised}. Although the algorithms in this table are not designed for URLLC, they can be applied in the prediction and communication co-design framework for URLLC in \cite{Hou2018Burstiness}.

\begin{table*}[tbp]
\vspace{-0.0cm}\footnotesize
\renewcommand{\arraystretch}{1.3}
\caption{Supervised deep learning in wireless networks}
\begin{center}\vspace{-0.0cm}\label{T:Supervised}
\begin{tabular}{|p{1.1cm}|p{6.8cm}|p{2.8cm}|p{2.5cm}|p{2.5cm}|}
  \hline
  \multicolumn{5}{|c|}{\bf{Channel Estimation or Prediction}}\\\hline
  {\bf{Reference}} & {\bf{Summary}} & {\bf{Accuracy}}& {\bf{Complexity}} & {\bf{System Performance}}\\\hline
  \cite{guo2019convolutional} & Proposing a multiple-rate compressive sensing deep learning framework for CSI feedback in massive MIMO systems & Relative error: below $0.01$ & Linear in the size of the channel matrix & Not available \\\hline
  \cite{dong2019ICASSP} & Employing CNN for wideband channel estimation in mmWave massive MIMO systems & Relative error: below $0.01$ &  Lower than Minimum Mean Square Error (MMSE) & Not available \\\hline
  \cite{Yang2019Deep} & Applying Long Short Term Memory (LSTM) to make channel predictions in body area networks & Relative error: below $0.05$ & Not available & Outage probability: $1\sim10$\% \\\hline
  \cite{Herath2019Deep} & Predicting future signal strength with LSTM and gated recurrent unit algorithms & Better than linear regression & Not available & Not available \\\hline
  \multicolumn{5}{|c|}{\bf{Traffic Prediction}}\\\hline
  {\bf{Reference}} & {\bf{Summary}} & \bf{Accuracy}& \bf{Complexity} & \bf{System Performance} \\\hline
  \cite{wang2017spatiotemporal} & Proposing a hybrid deep learning model for spatiotemporal traffic prediction in cellular networks & Reduce $20\sim40$\% prediction error& Not available & Not available \\\hline
  \cite{qiu2018spatio} & Applying RNN in spatiotemporal traffic prediction in cellular networks & The n-to-n RNN outperforms other schemes& Not available & Not available \\\hline
  \cite{azari2019user} & Comparing the performance of statistic learning and deep learning methods in user traffic prediction & LSTM is better than other schemes& Not available & $45$\% decrease in delay \\\hline
  \cite{Zhang2019TrafficPred} & Developing deep learning architecture for traffic prediction and applying transfer learning among different scenarios & $5\sim20$\% improvements in different datasets & Not available & Not available \\\hline
  \multicolumn{5}{|c|}{\bf{Mobility or Trajectory Prediction}}\\\hline
  {\bf{Reference}} & {\bf{Summary}} & {\bf{Accuracy}}& {\bf{Complexity}} & {\bf{System Performance}} \\\hline
  \cite{zhang2018trajectory} & Comparing LSTM, echo state network, and an optimal linear predictor, and Kalman filter, in trajectory prediction & LSTM is better than other schemes& Not available & Not available \\\hline
  \cite{Gebrie2019What} & Comparing FNN, Extreme Gradient Boosting Trees (XGBoost), and two other benchmarks in location prediction & XGBoost is better than other schemes (error probability: $< 10$\%) & Semi-Markov-based model has the lowest complexity & XGBoost performs best ($80.68$\% energy saving) \\\hline
  \cite{Ke2019Trajectory} & Applying RNN in unmanned aerial vehicle trajectory prediction & Almost $98$\% prediction accuracy & Not available & Not available \\\hline
  \cite{Xueshi2019GC} & Applying LSTM in head and body motion prediction to reduce latency in VR applications & Prediction error: $< 1^{\text{o}}$ and $<1$~mm & Not available & Mismatched pixels: $1\sim3$\% \\\hline
  \multicolumn{5}{|c|}{\bf{QoS or QoE Prediction}}\\\hline
  {\bf{Reference}} & {\bf{Summary}}& {\bf{Accuracy}}& {\bf{Complexity}} & {\bf{System Performance}} \\\hline
  \cite{lopez2018deep} & Proposing a combination of CNN, RNN, and Gaussian process classifier for a binary QoE prediction & From $60$\% to $90$\% depending on datasets& Not available & Not available \\\hline
  \cite{Gary2018Forecasting} & Applying LSTM in QoS prediction for IoT applications & LSTM outperforms other schemes& Not available & Not available \\\hline
  \cite{Hatem2019Processing} & Predicting the processing time of radio and baseband network functions with mathematical and deep learning models & DNN outperforms other schemes & Not available & Not available \\\hline
  \multicolumn{5}{|c|}{\bf{Approximating Optimal Policies}}\\\hline
  {\bf{Reference}} & {\bf{Summary}} & {\bf{Accuracy}}& {\bf{Complexity}} & {\bf{System Performance}} \\\hline
  \cite{Haoran2018learning} & Approximating Weighted Minimum Mean Square Error (WMMSE) power control scheme with FNN & Up to $98$\% & CPU time: $10$\% of WMMSE & Near-optimal in small-scale networks \\\hline
  \cite{lee2018cnn} & Using SE/EE as the loss function to train CNNs for power control& Not available& Lower than FNN and WMMSE & Higher SE/EE than WMMSE \\\hline
  \cite{Jia2018VTC} & Employing an FNN to approximate the optimal predictive resource allocation policy for delay-tolerant services  & $70 \sim 90$\% depending on the density of users& Lower than Interior-Point algorithm & $97$\% of users with satisfactory QoS \\\hline
\end{tabular}
\end{center}
\vspace{-0.2cm}
\end{table*}

\subsection{Basics of DNNs} \label{subSec:FNN}
In practice, a complicated DNN is usually made up of some basic structures, such as  feed-forward neural networks (FNNs), recurrent neural networks (RNNs), and convolutional neural networks (CNNs) in Fig. \cite{lecun2015deep}. How to build a DNN for a specific problem depends on the structure of data. In this section, we take FNNs as an example to introduce some basics of DNNs. The motivation is to provide a quick review of DNN. In general, FNNs work well with small-scale problems. For an in-depth introduction of different kinds of DNNs, please refer to \cite{zhang2019deep}.

\begin{figure}[htbp]
        \centering
        \begin{minipage}[t]{0.36\textwidth}
        \includegraphics[width=1\textwidth]{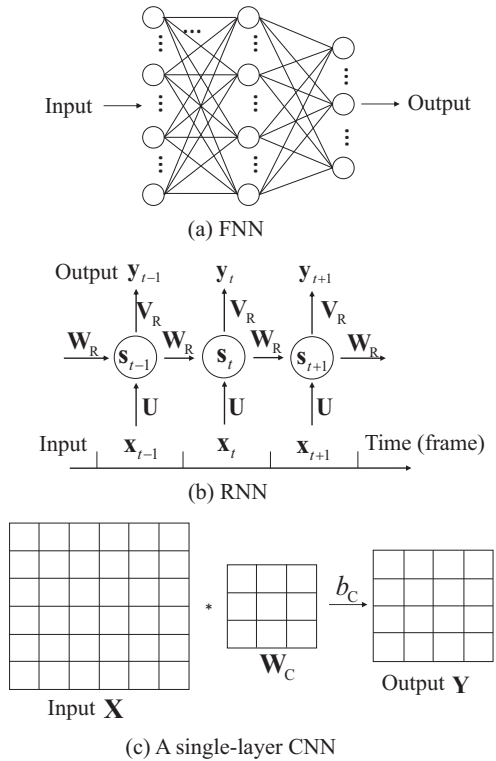}
        \end{minipage}
        \caption{Different kinds of neural networks \cite{lecun2015deep}.}
        \label{fig:DNN}
        \vspace{-0.2cm}
\end{figure}

%

\subsubsection{Basics of FNN} An FNN includes an input layer, an output layer, and $N_L$ hidden layers \cite{lecun2015deep}. The output of the $n$-th layer is a column vector ${\bf{x}}^{[n]}\in {\mathbb{R}}^{M^{[n]} \times 1}$, where $n=0,1,...,N_L+1$ and $M^{[n]}$ is the number of neurons in the $n$-th layer. For the input layer, $n=0$. For the output layer, $n=N_L+1$. The relation between ${\bf{x}}^{[n]}$ and ${\bf{x}}^{[n-1]}$ is determined by the following expression,
\begin{align}
{{\bf{x}}^{[n]}} = f_{\rm{A}}^{[n]}\left( {{{\bf{W}}^{[n]}}{{\bf{x}}^{[n - 1]}} + {{\bf{b}}^{[n]}}} \right), \nonumber
\end{align}
where ${{\bf{W}}^{[n]}} \in {\mathbb{R}}^{M^{[n]}\times M^{[n-1]}}$ and ${{\bf{b}}^{[n]}} \in {\mathbb{R}}^{M^{[n]} \times 1}$ are the weights and bias of the $n$-th layer, and $f_{\rm{A}}^{[n]}(\cdot)$ is the activation function of the $n$-th layer, and it is an element-wise operation. The widely used activation functions include the sigmoid function,
\begin{align}\label{eq:sigmoid}
f_{\rm{A}}({\bf{x}}) = \frac{1}{1+\exp({\bf{x}})},
\end{align}
$\tanh({\bf{x}})$, the Rectified Linear Unit (ReLU) function,
\begin{align}\label{eq:ReLU}
f_{\rm{A}}({\bf{x}}) = \max \left({\bf{x}},0\right),
\end{align}
as well as its generalizations.

Given an FNN, we can compute the output from any given input according to $ {\bf{\hat{y}}} = \Phi\left({\bf{x}}^{[0]};\Omega\right)$, where $\Omega \triangleq \left\{{\bf{W}}^{[n]},{{\bf{b}}^{[n]}}, n = 0,..., N_L+1\right\}$ is the set of parameters of the FNN. Let us denote a labeled training sample as $({\bf{x}}^{[0]},{\bf{y}})$. In the training phase, a supervised deep learning algorithm optimizes the parameters of the FNN to minimize the expectation of the difference between outputs of the FNN and labels, i.e.,
\begin{align}\label{eq:Loss}
\min {{\mathbb{E}}_{{\bf{x}}^{[0]}}}{\left( {{\bf{\hat y}} - {{\bf{y}}}} \right)^2}
\end{align}
where the expectation is taken over input data, ${\bf{x}}^{[0]}$.

\subsubsection{Universal Approximation Theorem} In wireless networks, an optimal policy maps the state of the system ${\bf{x}}^{[0]}$ into the optimal action: ${\bf{u}}^*({\bf{x}}^{[0]})$. If we know some optimal state-action pairs, but do not have the expression of ${\bf{u}}^*(\cdot)$, then we can use an FNN, $\Phi\left(\cdot;\Omega\right)$, to approximate the policy. The following theorem shows how accurate the FNN can be \cite{hornik1989multilayer,Haoran2018learning}.

\begin{theorem}\label{Th:Uni}
Universal Approximation Theorem: \emph{Let ${\bf{u}}^*(\cdot)$ be a deterministic continuous function defined over a compact set $\mathcal{D}$, e.g., ${\bf{x}}^{[0]} \in {\mathbb{R}}^{K}$. Then, for any $\varepsilon_{\rm NN} > 0$, there exists an FNN $\Phi\left(\cdot;\Omega\right)$ constructed by sigmoid activations in \eqref{eq:sigmoid}, such that}
\begin{align}
\mathop {\sup }\limits_{{{\bf{x}}^{[0]}} \in \mathcal{D}} \left\| {\Phi\left({\bf{x}}^{[0]};\Omega\right) - {\bf{u}}^*({\bf{x}}^{[0]})} \right\| \le \varepsilon_{\rm NN}.\nonumber
\end{align}
\end{theorem}
The above theorem has been extended to other kinds of activation functions, including the ReLU function in \eqref{eq:ReLU} and its variants \cite{leshno1993multilayer}. More recently, the results in \cite{Haoran2018learning} showed that to achieve a target accuracy, the size of network scales with the approximation accuracy $\varepsilon_{\rm NN}$ according to ${\mathcal{O}}\left(\ln(1/\varepsilon_{\rm NN})\right)$.

\subsection{{Predictions in URLLC}}\label{Sec:prediction}
As shown in Table \ref{T:Supervised}, supervised deep learning has been employed for channel prediction, mobility prediction, traffic prediction, and QoS or QoE prediction. The reason why supervised deep learning algorithms are very powerful in prediction is that it is very convenient to collect a large number of labeled training samples from the history data, such as trajectories \cite{Hou2019Prediction,Xueshi2019GC}, channel and traffic variations \cite{Herath2019Deep,Hou2018Burstiness}, and QoS experienced by users \cite{lopez2018deep}.

\subsubsection{Channel Prediction} Unreliable wireless channels lead to a high packet loss probability in URLLC \cite{Cross2018she}. If the transmitter can predict which subchannel will be in deep fading, then packet losses over deep fading channels can be avoided.

By exploiting the temporal correlation of wireless channels, different RNN variants were developed in \cite{Herath2019Deep} to predict future signal strength variations. Wideband channels are correlated in the frequency domain. Based on this feature, CNN was used to predict channels for mmWave massive MIMO systems \cite{dong2019ICASSP}. Although these algorithms are data-driven, the understanding of wireless channel models plays an important role. For example, the structures of neural networks depend on the features of wireless channels (e.g., correlations in temporal and frequency domains). To validate whether the learning algorithms work or not, the model-based methods serve as benchmarks for performance comparison.

\subsubsection{Traffic Prediction} As shown in \cite{wang2017spatiotemporal,qiu2018spatio,Zhang2019TrafficPred}, predicting the traffic load in a wireless network is helpful for network deployment. These works mainly focus on traffic prediction in a large area. However, to improve the latency, reliability, and resource utilization efficiency of URLLC, the system needs to predict the traffic state of each user. Per-user traffic prediction was studied in \cite{azari2019user}, the results indicate that LSTM outperforms statistical learning for both short and long-term future predictions. To improve the bandwidth utilization efficiency subject to the latency and reliability requirements, both model-based and data-driven schemes were proposed in \cite{Hou2018Burstiness}. Since URLLC has a stringent requirement on reliability, prediction errors cannot be ignored. With the model-based scheme, the authors of \cite{Hou2018Burstiness} analyzed the prediction error probability in the traffic state prediction. With the data-driven scheme, a better bandwidth utilization efficiency can be achieved, but the prediction error probability can only be evaluated via experiments \cite{Hou2018Burstiness}.

\subsubsection{Mobility Prediction} To reveal the potential gain of mobility prediction in URLLC, a model-based method is applied in the prediction and communication co-design framework in \cite{Hou2019Prediction}. By predicting the motion of the head and body with LSTM, a predictive pre-rendering approach was proposed in \cite{Xueshi2019GC}. The results in \cite{Hou2019Prediction} and \cite{Xueshi2019GC} indicate that mobility prediction can help reduce user-experienced latency in remote control and VR applications.

It is worth noting that prediction is not error-free. The accuracies of different algorithms in trajectory and location predictions were evaluated in \cite{zhang2018trajectory} and \cite{Gebrie2019What}. Compared with model-based methods, data-driven methods can achieve better prediction accuracies. On the other hand, deriving the performance achieved by data-driven methods is very difficult. To obtain some insights on fundamental trade-offs, such as the latency and reliability trade-off, we still need the help of models \cite{Hou2019Prediction}.

\subsubsection{QoS Prediction} If a control plane can predict the data plane QoS, it can update resource allocation to avoid QoS violation. For example, by predicting the delay caused by baseband signal processing, a better offloading scheme in could-RAN or MEC systems was developed in \cite{Hatem2019Processing}. According to the evaluation in \cite{lopez2018deep,Gary2018Forecasting}, supervised deep learning outperforms other baselines in predicting QoE for video services and QoS of IoT applications. Nevertheless, how to predict the QoS or QoE of URLLC services and how to optimize resource allocation for URLLC according to the predicted information remain unclear.

\subsection{Approximating Optimal Policies}
With the model-based cross-layer optimization in Section~\ref{S:cross}, we can obtain optimal policies in wireless networks but with high computing complexity. As a result, these policies can hardly be implemented in real time. To handle this issue, FNNs were used to approximate optimal policies in the existing literature, such as \cite{Haoran2018learning,Dora2020Deep}.

\subsubsection{Function Approximator} To apply supervised deep learning in URLLC, the analytical tools and optimization algorithms discussed in Sections \ref{S:Tools} and \ref{S:cross} can be used to find labeled samples. Recently, the results in \cite{Haoran2018learning} showed that an FNN can be used to approximate an iterative algorithm, such as WMMSE for power control. Based on the Universal Approximation Theorem, it is possible to achieve high approximating accuracy. Inspired by these theoretical results, the authors of \cite{lee2018cnn} further showed that by using a CNN model, the required number of weights to achieve the same performance is much smaller than that in an FNN model. DNNs can be also applied to approximate other optimal policies such as bandwidth allocation and user association \cite{Jia2018VTC,Dora2019Deep}.

\subsubsection{Deep Transfer Learning}\label{subSec:transfer}
To implement supervised deep learning in dynamic wireless networks, the DNN needs to adapt to non-stationary environments and diverse QoS requirements with a few training samples \cite{Changyang2020Deep}. To achieve this, network-based deep transfer learning is an optional approach \cite{azari2019user}. To illustrate the concept of deep transfer learning, we consider two FNNs, $\Phi_0\left(\cdot;\Omega_0\right)$ and $\Phi_1\left(\cdot;\Omega_1\right)$. The first FNN is well-trained on a data set with a large number of labeled samples, $\mathcal{S}_0 = \{({\bf{x}}^{\rm{[0]}}_{n_t},{\bf{y}}_{n_t}), {n_t} = 1,2,...,N_{\rm tr}\}$. Now we train a new FNN, $\Phi_1\left(\cdot;\Omega_1\right)$, from another data set $\mathcal{S}_1 = \{(\tilde{\bf{x}}^{\rm{[0]}}_{n_f},{\tilde{\bf{y}}}_{n_f}), {n_t} = 1,2,...,N_{\rm tf}\}$, where the number of training samples in $\mathcal{S}_1$ is much smaller than that in $\mathcal{S}_0$, i.e., $N_{\rm tf} \ll N_{\rm tr}$. Since the number of labeled samples in $\mathcal{S}_1$ is small, we can hardly train $\Phi_1\left(\cdot;\Omega_1\right)$ from randomly initialized parameters. If the data sets $\mathcal{S}_0$ and $\mathcal{S}_1$ are related to each other, then we can initialize the new FNN with $\Omega_0$ and update it by using new labeled samples in $\mathcal{S}_1$.

\begin{figure}[ht]
        \centering
        \begin{minipage}[t]{0.28\textwidth}
        \includegraphics[width=1\textwidth]{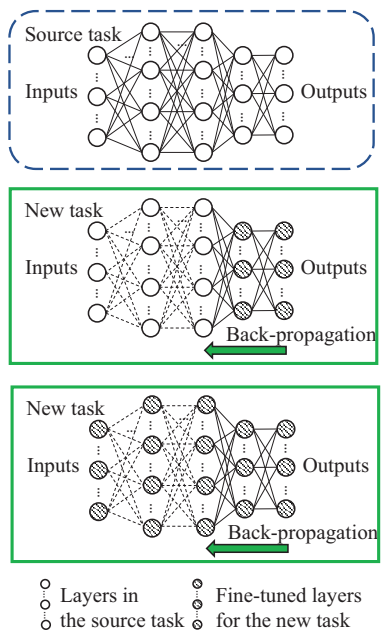}
        \end{minipage}
        \caption{Illustration of network-based deep transfer learning \cite{Changyang2020Deep,Dora2020Deep}.}
        \label{fig:DeepTransfer}
        \vspace{-0.2cm}
\end{figure}

As illustrated in Fig. \ref{fig:DeepTransfer}, $\Omega_1$ in the new FNN is initialized with $\Omega_0$. In the training phase, the first three layers of $\Omega_1$ remain constant, and the back-propagation stops in the fourth layer. Then, fine-tuning is used to refine the weights and bias of the last two layers on the new training data set with small learning rates \cite{shen2018transfer}. Alternatively, we can fine-tune all the parameters of the FNN. The results in \cite{Dora2020Deep} show that the latter approach requires more training samples and longer training time, but does not bring significant performance gain. To determine how many layers should be updated, we need to try different options and find the best one by trial-and-error \cite{Dora2020Deep}.

\subsection{Summary of Supervised Deep Learning} \label{subSec:SumSDL}
With existing optimization algorithms, a large number of labeled training samples can be obtained off-line for supervised deep learning. After the off-line training phase, the system further fine-tunes DNNs in non-stationary networks. In this way, domain knowledge help to initialize DNNs before online implementation. Deep transfer learning handles the model-mismatch problem with the help of few training samples in new environments.

Although supervised deep learning shows great potential in prediction and function approximation, there are some open problems when using it in URLLC:
\begin{itemize}
\item The prediction or approximation accuracy of DNNs will have impacts on the QoS of URLLC. The relative errors achieved by most of the existing supervised learning algorithms in Table \ref{T:Supervised} is around $0.01$, which is good enough for traditional services, but far from the requirement of URLLC.
\item In some NP-hard problems, labeled training samples are unavailable. For these problems, supervised deep learning is not applicable.
\item The number of users and the types of services in a wireless network are dynamic. When features and the dimension of the input data change, deep transfer learning cannot be applied directly.
\end{itemize}
In the coming two sections, we will review unsupervised deep learning and DRL that do not need labeled training samples.

\section{Unsupervised Deep Learning for URLLC}\label{S:unsupervised}
Unlike supervised deep learning, unsupervised deep learning can find the optimal policy without labeled training samples. In this subsection, we review two branches of existing works that applied unsupervised deep learning in wireless communications, 1) autoencoder for physical-layer communications and 2) FNN trained without labels for resource allocation. The existing works are summarized in Table \ref{T:unsupervised}, where the basic idea is to formulate a functional optimization problem and then approximate the optimal function with a DNN trained toward a loss function that represents the goal of the problem .

\begin{table*}[tbp]
\vspace{-0.0cm}\footnotesize
\renewcommand{\arraystretch}{1.3}
\caption{Unsupervised deep learning in wireless networks}
\begin{center}\vspace{-0.0cm}\label{T:unsupervised}
\begin{tabular}{|p{0.8cm}|p{1.2cm}|p{7.5cm}|p{6cm}|}
  \hline
  \multicolumn{4}{|c|}{\bf{Part I: Autoencoder for Physical-layer Communications (Unconstrained Functional Optimization)}}\\\hline
  {\bf{Year}} &{\bf{Reference}} & {\bf{Summary}} & {\bf{Performance}}\\\hline
  2018 & \cite{dorner2017deep} & An autoencoder was used to learn a better constellation & The BLER gap between fine-tuned Autoencoder and an existing constellation is $1$~dB \\\hline
  2018 & \cite{felix2018ofdm} & An autoencoder was applied for E2E learning in OFDM systems for reliable communications over multipath channels & Autoencoder achieves better BLER than Quadrature Phase Shift Keying (QPSK) \\\hline
  2018 &  \cite{farsad2018deep} & A joint source and channel coding scheme was proposed to minimize the E2E distortion & The learning-based method outperforms an existing separate source and channel coding scheme \\\hline
  2019 & \cite{xue2019unsupervised} & An autoencoder was used for multiuser waveform design at the transmitter side & The learning-based method outperforms linear MMSE in terms of lower BLER \\\hline
  \multicolumn{4}{|c|}{\bf{Part II: Primal-dual Method for Radio Resource Allocation (Constrained Functional Optimization)}}\\\hline
  {\bf{Year}} &{\bf{Reference}} & {\bf{Summary}} & {\bf{Performance}} \\\hline
  2019 & \cite{eisen2019learning} & The primal-dual method is used to train an FNN to find the optimal resource allocation subject to statistic constraints & The sum-capacity of the learning-based method is close to WMMSE \\\hline
  2019 & \cite{lee2019deep} & Applying the primal-dual method to train DNNs for distributed wireless resource management problems & Distributed DNNs can achieve near-optimal sum-capacity \\\hline
  2019 & \cite{sun2019PIMRC} & Applying the primal-dual method to train a DNN for resource allocation subject to the QoS constraints of URLLC & The QoS of URLLC can be satisfied by reserving $1$\% of extra bandwidth \\\hline
  2019 & \cite{kalogerias2019model} & Developing a model-free primal-dual algorithm for learning optimal ergodic resource allocations & The sum-capacity achieved by the model-free approach is close to WMMSE \\\hline
  2020 & \cite{Dong2020NetMag} & Developing a reinforced-unsupervised learning framework for the model-free optimization problems & Unsupervised learning outperforms supervised learning when the number of training samples is large \\\hline
  2020 & \cite{eisen2020optimal} & Training a Graph Neural Network (GNN) to find the optimal resource allocation in large-scale wireless networks & GNN outperforms WMMSE in large-scale networks \\\hline
\end{tabular}
\end{center}
\vspace{-0.2cm}
\end{table*}

\subsection{Introduction of Functional Optimization}
A lot of problems in communications can be described by a mapping from an input message (or system state), ${\bf{s}}$, to an output signal (or action), ${\bf{a}}^{\rm s}$, i.e.,
\begin{align}
{\bf{a}}^{\rm s} = {\bf{u}}({\bf{s}}). \nonumber
\end{align}
For example, a modulation and coding scheme maps the input bits stream to coding blocks. A scheduler maps QSI and CSI to resource blocks allocated to different users.

\subsubsection{Variable Optimization Problems} A regular variable optimization problem can be formulated in a general form as follows,
\begin{align}
\mathop {\min }\limits_{{\bf{a}}^{\rm s}} \; &  {F\left( {{\bf{s}},{{\bf{a}}^{\rm s}}} \right)} \label{eq:Utility1}\\
\text{s.t.}\; &G_j \left( {{\bf{s}},{{\bf{a}}^{\rm s}}} \right) \le 0, j = 1,...,J,\label{eq:instant1}\tag{\theequation a}
\end{align}
where $F\left( {{\bf{s}},{{\bf{a}}^{\rm s}}} \right)$ is the objective function, \eqref{eq:instant1} are the constraints on available resources or QoS.

By solving problem \eqref{eq:Utility1}, we can obtain the optimal solution ${{\bf{a}}^{\rm s}}^*$ for a given input message or system state ${\bf{s}}$. When the input message or the state varies, the system needs to update ${{\bf{a}}^{\rm s}}^*$ by solving problem \eqref{eq:Utility1} numerically, unless the optimal solution can be derived in a closed-form expression with respect to ${\bf{s}}$. However, in most of the cases, we can hardly derive the closed-form expression, and optimization algorithms for solving problem \eqref{eq:Utility1} are too complicated to be implemented in real-time.

\subsubsection{Functional Optimization Problems} As proved in \cite{sun2019PIMRC}, variable optimization problem \eqref{eq:Utility1} is equivalent to the following functional optimization problem,
\begin{align}
\mathop {\min }\limits_{{\bf{u}}(\cdot)} \; & {{\mathbb{E}}_{\bf{s}}}\left[ {F\left( {{\bf{s}},{\bf{u}}({\bf{s}})} \right)} \right]\label{eq:Utility2}\\
\text{s.t.}\; & G_j \left( {{\bf{s}},{\bf{u}}({\bf{s}})} \right) \le 0, \forall {\bf{s}} \in \mathcal{D}_{\rm s}, j = 1,...,J.\label{eq:instant2} \tag{\theequation a}  \nonumber
\end{align}
In problem \eqref{eq:Utility2}, the elements to be optimized are resource allocation or scheduler functions with respect to the states of users. By solving the above problem, we can obtain the expression of the optimal policy. However, solving a functional optimization problem is more challenging than solving variable optimization problems \cite{osmolovskii1998calculus,boyd}. This is because the optimization variables are functions, each of which is equivalent to an infinite-dimensional vector (e.g., Fourier series with infinite terms). Moreover, each constraint in \eqref{eq:instant2} is equivalent to an infinite number of constraints in variable optimization problems (constraints for any given value of $\bf{s}$). 


\subsubsection{Methods for Solving Functional Optimization Problems} To derive the closed-form expression of a functional optimization problem, we can find the solutions that satisfy the first-order necessary conditions in \cite{gregory2018constrained}. For example, optimizing the instantaneous power control policy to maximize the ergodic capacity under the maximal transmit power constraint belongs to functional optimization problems \cite{WirelessCom}. By solving the first-order necessary conditions, we can derive that the optimal power control policy is the water-filling policy. Compared with this physical-layer optimization problem, cross-layer optimization problems are much more complicated and the closed-form expressions of optimal policies are usually unavailable. For example, C. She \emph{et al.} investigated how to adjust transmit power and subcarrier allocation according to QSI in MIMO-Orthogonal Frequency Division Multiplexing (OFDM) systems. The problem was formulated as a functional optimization problem, and a closed-form solution can only be obtained for a specific packet arrival process in an asymptotic case: the number of antennas at the BS goes to infinite \cite{she2015energy}.

When there is no closed-form solution, we need numerical methods to find near-optimal solutions. There are several numerical methods for solving functional optimization problems, such as the finite element method (FEM) \cite{Zienkiewicz1977FEM}. The basic idea of the FEM is to parameterize ${\bf{u}}({\bf{s}})$ with a finite sampled points, and approximate the optimal solution by optimizing the values on the sampled points. However, the number of the sampled points increases exponentially with the dimension of ${\bf{s}}$. In addition, these numerical methods are usually used to solve problems formulated in the three-dimensional space, and heavily rely on the theorems in dynamics \cite{wendt2008computational}. Whether they can work well with high-dimensional communication problems remains unclear.

Inspired by the Universal Approximation Theorem in \cite{hornik1989multilayer, leshno1993multilayer}, we can use a DNN to approximate the optimal policy ${\bf{u}}^*(\cdot)$ that maps the state ${\bf{s}}$ to the optimal solution ${{\bf{a}}^{\rm s}}^*$ and find the parameters of the DNN with unsupervised deep learning methods.


\subsection{Autoencoder for E2E Design in Physical-Layer Communications} Autoencoder is a useful tool for data compression and feature extraction \cite{baldi2012autoencoders}. The basic idea is to train DNNs with the input data. An autoencoder consists of three components: an encoder, a decoder, and a distortion function that compares the input and the output of the autoencoder.

\begin{figure}[ht]
        \centering
        \begin{minipage}[t]{0.45\textwidth}
        \includegraphics[width=1\textwidth]{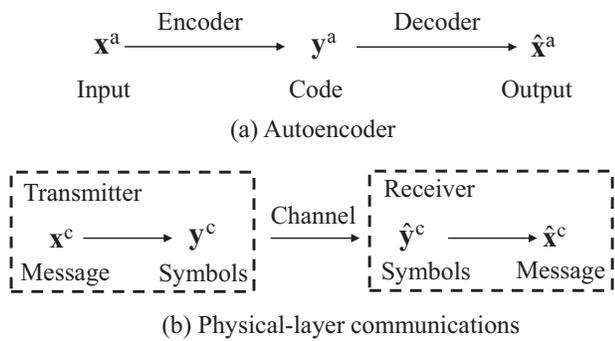}
        \end{minipage}
        \caption{Autoencoder for physical-layer communications \cite{baldi2012autoencoders,dorner2017deep}.}
        \label{fig:autoencoder}
        \vspace{-0.2cm}
\end{figure}

As shown in Fig. \ref{fig:autoencoder}(a), the input and the output of the autoencoder are ${\bf{x}}^{\rm a} \in {\mathbb{R}}^{N^{\rm a}_{\rm x}\times 1}$ and $\hat{\bf{x}}^{\rm a} \in {\mathbb{R}}^{N^{\rm a}_{\rm x}\times 1}$, respectively. The compressed code is ${\bf{y}}^{\rm a} \in {\mathbb{R}}^{N^{\rm a}_{\rm y} \times 1}$, where $N^{\rm a}_{\rm y} < N^{\rm a}_{\rm x}$. Denote the encoder and the decoder by $f_{\rm e}(\cdot)$ and $f_{\rm d}(\cdot)$, respectively. The autoencoder problem is to find $f_{\rm e}(\cdot)$ and $f_{\rm d}(\cdot)$ that minimize the average of the distortion function \cite{baldi2012autoencoders},
\begin{align}
\mathop {\min }\limits_{{f_{\rm e}}\left( \cdot \right),{f_{\rm d}}\left( \cdot \right)} {\mathbb{E}}{\left( {{{{\bf{\hat x}}}^{\rm{a}}} - {{\bf{x}}^{\rm{a}}}} \right)^2}\label{eq:autoencoder}
\end{align}
where the average is taken over the input, and $\left( {{{{\bf{\hat x}}}^{\rm{a}}} - {{\bf{x}}^{\rm{a}}}} \right)^2$ is the distortion function. Problem \eqref{eq:autoencoder} is an unconstrained functional optimization problem \cite{osmolovskii1998calculus}. To solve the functional optimization problem, a widely applied approach is approximating the encoder and the decoder with two DNNs \cite{droniou2013gated}.

The components of physical-layer communications are shown in Fig. \ref{fig:autoencoder}(b), which are very similar to that of an autoencoder. The transmitter sends message ${\bf{x}}^{\rm{c}}$ to the receiver over a certain channel. First, the transmitter encodes the message into a coding block that consists of multiple symbols, ${\bf{y}}^{\rm{c}}$, such as $M$-QAM symbols. Then, the block is sent over a channel. Owning to the noise and interference, the received symbols, denoted by $\hat{\bf{y}}^{\rm{c}}$, are different from ${\bf{y}}^{\rm{c}}$. Based on the received symbols, the receiver tries to decode the block and recover the message $\hat{\bf{x}}^{\rm{c}}$. A communication problem is to find the optimal encoder and decoder that minimize the BLER, i.e.,
\begin{align}
\mathop {\min }\limits_{{f_{\rm e}}\left( \cdot \right),{f_{\rm d}}\left( \cdot \right)} {\mathbb{E}}\{{\bf{1}}\left( {{{{\bf{\hat x}}}^{\rm{a}}} \ne {{\bf{x}}^{\rm{a}}}} \right)\},\nonumber
\end{align}
where the average is taken over input messages and stochastic channels, and ${\bf{1}}\left( {{{{\bf{\hat x}}}^{\rm{a}}} \ne {{\bf{x}}^{\rm{a}}}} \right)$ is an indicator of event ${{{{\bf{\hat x}}}^{\rm{a}}} \ne {{\bf{x}}^{\rm{a}}}}$. If ${{{{\bf{\hat x}}}^{\rm{a}}} \ne {{\bf{x}}^{\rm{a}}}}$ happens, then the indicator equals to $1$. Otherwise, it equals to $0$.

The encoder and the decoder of the communication system can be approximated by two DNNs \cite{dorner2017deep,farsad2018deep}. To allow for back-propagation, a channel should be approximated by a DNN \cite{farsad2018deep}. Considering that the real-world channel is not available, the autoencoder is first trained with a stochastic channel model and then implemented in real hardware for over-the-air transmissions \cite{dorner2017deep}. To compensate for the mismatch between the model and the real-world channel, the system can fine-tune the receiver part of the autoencoder.

As summarized in Part I of Table \ref{T:unsupervised}, the above method has been implemented in several communication systems. Since the goal is to minimize the difference between the input data and output data, the autoencoder is suitable for URLLC. For example, the results in \cite{xue2019unsupervised} show that the scheme obtained with the autoencoder can achieve much lower BLER than the linear minimum mean-square error MIMO receiver. In addition, autoencoders inherently learn how to deal with hardware impairments \cite{dorner2017deep,felix2018ofdm}. In other words, they do not rely on the accuracy of models in communication systems. Furthermore, after the training phase, the outputs of the encoder and the decoder, represented by DNNs, can be computed from the forward propagation algorithm, which has much lower complexity than existing modulation and coding schemes.

%


\subsection{Optimizing Resource Allocation}

\subsubsection{Model-based learning} In the case that the objective function and constraints can be derived with the model-based method in Section \ref{S:Tools}, a resource allocation problem can be formulated in a general form as follows,
\begin{align}
\mathop {\min }\limits_{{\bf{u}}(\cdot)} \; & {{\mathbb{E}}_{\bf{s}}}\left[ {F\left( {{\bf{s}},{\bf{u}}({\bf{s}})} \right)} \right]\label{eq:Utility}\\
\text{s.t.}\; &{{\mathbb{E}}_{\bf{s}}}\left[ G_i^{{\rm{ave}}}\left( {\bf{s}},{\bf{u}}({\bf{s}}) \right) \right] \le 0,i = 1,...,I,\label{eq:longterm}\tag{\theequation a}\\
&G_j^{{\rm{ins}}}\left( {\bf{s}},{\bf{u}}({\bf{s}}) \right) \le 0,\forall {\bf{s}} \in \mathcal{D}_{\rm s},j = 1,...,J,\label{eq:instant}\tag{\theequation b}\\
&{\bf{u}}({\bf{s}}) \in \mathcal{D}_{\rm a}.\nonumber
\end{align}
In addition to the instantaneous constraints in problem \eqref{eq:Utility1}, long-term statistic constraints in \eqref{eq:longterm} are considered. The statistic constraints can be the average transmit power constraint \cite{WirelessCom} or the statistical QoS constraint \cite{tang2007quality}. The instantaneous constraints ensure instantaneous requirements, such as the maximal transmit power constraint. Here, the objective function and the constraints are differentiable but not necessary to be convex.

\begin{figure*}[tbh]
        \centering
        \begin{minipage}[t]{0.65\textwidth}
        \includegraphics[width=1\textwidth]{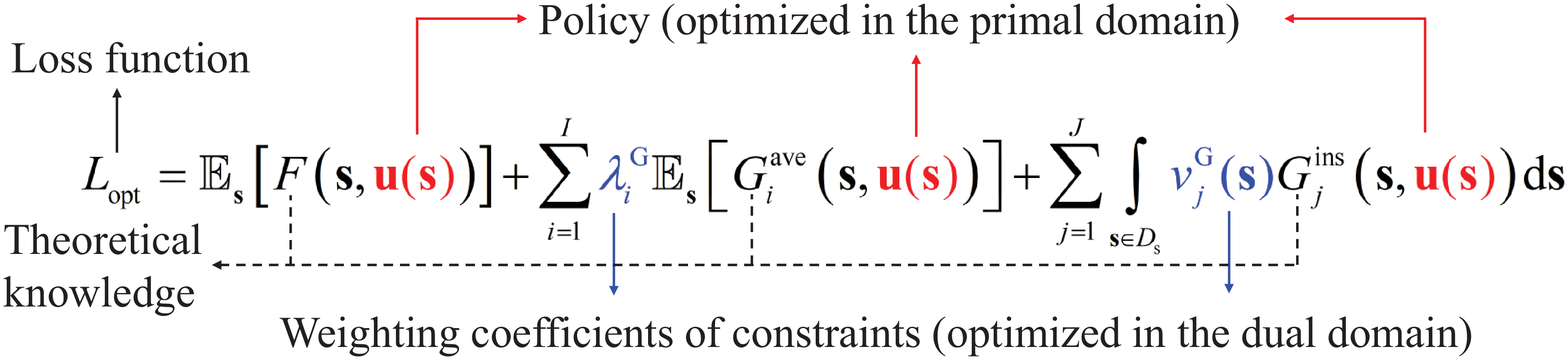}
        \end{minipage}
        \caption{Definition of Lagrangian \cite{sun2019learning}.}
        \label{fig:Lag}
        \vspace{-0.2cm}
\end{figure*}

The primal-dual method for solving functional optimization problems has been extensively used in optimal control theory \cite{Prussing2017Control}. To apply the method, we first define the Lagrangian of problem \eqref{eq:Utility} in Fig. \ref{fig:Lag}, where the Lagrange multipliers, $\lambda_i^{\rm G} \geq 0$ and $v_j^{\rm{G}}({\bf{s}}) \geq 0, \forall {\bf{s}} \in {D_{\rm{s}}}$, are the weighing coefficients of the constraints. To solve problem \eqref{eq:Utility}, we turn to solve the following problem:
\begin{align}
\mathop {\mathop {\max }\limits_{{\lambda^{\rm G}_i},i = 1,...,I} }\limits_{{v_j^{\rm{G}}}\left( {\bf{s}} \right),j = 1,...,J} \mathop {\min }\limits_{{\bf{u}}\left( {\bf{s}} \right)}  \;& {L_{{\rm{opt}}}},\label{eq:dualobj}\\
\text{s.t.}\;& \lambda_i^{\rm G} \geq 0, v_j^{\rm{G}}({\bf{s}}) \geq 0, \forall {\bf{s}} \in {D_{\rm{s}}}.\label{eq:multi}\tag{\theequation a}
\end{align}
where the policy ${\bf{u}}\left( {\bf{s}} \right)$ and the Lagrangian multipliers are optimized in the primal and dual domains, iteratively.

To obtain ${\bf{u}}({\bf{s}})$ and Lagrange multipliers numerically, two neural networks, $\Phi^{\rm u}({\bf{s}};\Omega^{\rm u})$ and $\Phi^{\rm v}({\bf{s}};\Omega^{\rm v})$, are used to approximate ${\bf{u}}({\bf{s}})$ and ${\bf{v}}^{\rm{G}}({\bf{s}})$, respectively, where ${\bf{v}}^{\rm{G}}({\bf{s}}) \triangleq [v_1^{\rm{G}}({\bf{s}}),...,v_J^{\rm{G}}({\bf{s}})]^T$. Then, by using the Lagrangian in Fig. \ref{fig:Lag} as the loss function, we can update between primal variables $\Omega^{\rm u}$, and dual variables ${{\lambda}}_i^{\rm G}$ and $\Omega^{\rm v}$ iteratively with the stochastic gradient descent algorithm. As will be shown in the case studies, by integrating the theoretical formulations of objective function and constraints into the loss function, unsupervised deep learning outperforms supervised deep learning in terms of approximation accuracy.

For the applications in Table \ref{T:Applications}, communication systems need to satisfy multiple conflicting KPIs, which are either considered in the constraints or formulated as the objective function. A straightforward approach is to minimize/maximize a weighted sum of different KPIs by setting the weighting coefficients manually. From the definition of Lagrangian in Fig. \ref{fig:Lag}, we can see that with different weighting coefficients, the optimal policies that minimize the Lagrangian are different. The primal-dual method is able to find the optimal weighting coefficients achieving the target balance between the objective function and constraints. As such, the objective function is minimized and the constraints are satisfied \cite{lee2019deep}.

%

\subsubsection{Model-free learning} In the case that there is no theoretical models of the considered system, the expressions of $F$, $G_i^{{\rm{ave}}}$, and $G_j^{{\rm{ins}}}$ and the gradients or derivatives of these functions are unavailable. For an optimization problem as in \eqref{eq:Utility}, one can train neural networks to approximate these functions according to the observed values of these functions with the supervised deep learning method. Then, the unknown gradients or derivatives can be estimated by the neural networks through backward propagation. The details on the model-free unsupervised deep learning method can be found in \cite{eisen2019learning,Chengjian2019model1,kalogerias2019model,Dong2020NetMag}.

\subsection{{Summary of Unsupervised Deep Learning}}
Theoretical models can be used to initialize autoencoders and formulate optimization problems. Since autoencoders are trained from the input data, they do not require analytical models of communication systems and can handle hardware impairment. For analytically intractable optimization problems, an approximation of the optimal solution can be obtained by combining the formulas into the loss function of unsupervised deep learning, i.e., Lagrangian.

Although unsupervised deep learning shows above advantages over supervised deep learning, it also suffers from some critical issues.

\begin{itemize}
\item When applying autoencoders in physical-layer design, the function that maps the input to the output of real-world channels is not available, and hence we cannot obtain the gradient required to fine-tune the transmitter.
\item When applying model-based unsupervised deep learning in functional optimization, the objective function and constraints should be differentiable. Thus, it cannot be applied to integer or mixed-integer programming directly.
\item Like supervised deep learning, how to update the neural network according to dynamic features, such as the dimension of the input data, needs further investigation.
\end{itemize}
In the next section, we will illustrate DRL that can optimize both continuous and discrete variables.

\section{Deep Reinforcement Learning for URLLC}\label{Sec:DRL}
DRL was widely applied in optimal control of stochastic processes. The most important feature of DRL is that it learns from the feedback that evaluates the actions taken rather than learns from correct actions. In this section, we first introduce the basics in DRL and then summarize the existing studies that applied DRL in wireless networks. To guarantee the QoS requirements in URLLC, we review a promising candidate: constrained DRL. Finally, we illustrate how to combine model-based tools with DRL in URLLC.

\subsection{Introduction of DRL}\label{Sec:IntroDRL}

\subsubsection{Elements of Reinforcement Learning} In reinforcement learning, there is an agent (or multiple agents) that observes the \emph{reward signal} from the environment, takes actions according to a \emph{policy}, and tries to maximize a \emph{value function}. If a \emph{model} of the environment is available, it can help improve learning efficiency.

{\emph{Policy:}} Let ${\bf{s}}(t) \in \mathcal{D}_{\rm s}$ and ${\bf{a}}^{\rm s}(t) \in \mathcal{D}_{\rm a}$ denote the state and the action in the $t$-th time slot. A policy is a mapping from states of the environment to actions an agent may take when in those states. We denote a stochastic policy by $\pi$. When the numbers of states and actions are finite,  $\pi$ consists of the probabilities to perform different actions in every state.

{\emph{Reward signal:}} In each time slot, a real number, the reward signal, is sent to the agent from the environment. It defines the goal of the problem, and the agent aims to maximize the long term reward. The reward signal depends on the states and actions taken by the agent, but the agent cannot change the function that generates the reward signal. We denote the reward signal in the $t$-th slot by $r({\bf{s}}(t),{\bf{a}}^{\rm s}(t))$, which can be simplified as $r(t)$.

{\emph{Value functions:}}  In general, a reinforcement learning algorithm aims to maximize the long-term return, such as the accumulated discounted reward,
\begin{align}
G(t) = \sum_{i=t}^{\infty} \gamma_{\rm d}^{i-t}r({\bf{s}}(i),{\bf{a}}^{\rm s}(i)), \label{eq:return}
\end{align}
where $\gamma _{\rm{d}} \in [0,1)$ is the discounting factor. $G(t)$ reflects the long-term return that the agent will receive after the $t$-th slot in an infinite horizon. Similarly, we can also formulate the discounted reward in a finite horizon if the task ends after finite steps, $G(t) = \sum_{i=t}^{t_{\max}} \gamma_{\rm d}^{i-t}r({\bf{s}}(i),{\bf{a}}^{\rm s}(i))$. This definition is also referred to as partial return in \cite{sutton2018reinforcement}. The value of taking action ${{\bf{a}}^{\rm s}}\left( t \right)$ in state ${{\bf{s}}\left( t \right)}$ under a policy $\pi$, denoted by $Q^{\pi}\left( {{\bf{s}}\left( t \right),{{\bf{a}}^{\rm s}}\left( t \right)} \right)$, is the expected return starting from state ${{\bf{s}}\left( t \right)}$, taking the action ${{\bf{a}}^{\rm s}}\left( t \right)$, and thereafter following policy $\pi$. $Q^{\pi}\left( {{\bf{s}}\left( t \right),{{\bf{a}}^{\rm s}}\left( t \right)} \right)$ is also known as the state-action value function. Similarly, the expected return of visiting state ${{\bf{s}}\left( t \right)}$ is defined as $V^{\pi}\left( {{\bf{s}}\left( t \right)} \right)$, which is referred to as the state value function.

\subsubsection{Exploitation and Exploration} The trade-off between exploitation and exploration is one of the challenges that raise in reinforcement learning. By exploiting the experience, the agent can obtain a good reward that maximizes the estimated value function. To explore better actions, the agent needs to try some actions that are not optimal according to the previous experience. There are many sophisticated methods to balance exploitation and exploration, such as $\varepsilon$-greedy methods \cite{rodrigues2009dynamic}, using optimistic initial values \cite{sutton2018reinforcement}, and the Bayesian sampling approach \cite{asmuth2009bayesian}. However, when using reinforcement learning in URLLC, we need to satisfy the QoS constraints during explorations. Thus, random explorations are not applicable. We will discuss how to re-design exploration schemes for URLLC in Section \ref{subSec:DTforDRL}.

\subsubsection{Markov Decision Process (MDP)}\label{subsubSec:MDP} To apply reinforcement learning in optimal control of a dynamic system, the uncertainty of the system should be formulated in the framework of MDP \cite{sutton2018reinforcement}. An MDP is a controlled stochastic process defined by a state space, $ \mathcal{D}_{\rm s}$, an action space, $ \mathcal{D}_{\rm a}$, a reward function, $\mathcal{D}_{\rm s} \times \mathcal{D}_{\rm a} \times \mathcal{D}_{\rm s} \to \mathbb{R}$, that generates rewards signal according to the states and the action, and transition probabilities, i.e.,
\begin{align}
&\Pr[{\bf{s}}(t+1),r(t+1)|{\bf{s}}(t),{\bf{a}}(t)]\nonumber\\
&= \Pr[{\bf{s}}(t+1),r(t+1)|{\bf{s}}(0),{\bf{a}}(0),r(0),..., {\bf{s}}(t),{\bf{a}}(t),r(t)].\label{eq:transition}
\end{align}
The transition probabilities in \eqref{eq:transition} indicate that we can predict the state and reward in the $(t+1)$-th slot based solely on the state and action in the $t$-th time slot just as well as we know the states and actions in all the previous slots. Such a property is referred to as Markov property or ``memorylessness".

In reinforcement learning, the transition probabilities of the MDP are referred to as the ``model" of the problem. This ``model" is definitely different from the models in communications and networking. Depending on whether the transition probabilities are available or not, reinforcement learning algorithms are classified into two categories: model-based and model-free \cite{sutton2018reinforcement,gu2016continuous}. In this work, we only review the model-free reinforcement learning that does not need the transition probabilities, and use the term knowledge-assisted reinforcement learning to represent another category of algorithms that exploit domain knowledge to improve the learning efficiency.

\subsubsection{Bellman Equation}
In the framework of MDP, $Q^{\pi}\left( {{\bf{s}}\left( t \right),{{\bf{a}}^{\rm s}}\left( t \right)} \right)$ can be formally defined as follows,
\begin{align}
Q^{\pi}\left( {{\bf{s}}\left( t \right),{{\bf{a}}^{\rm s}}\left( t \right)} \right) = {{\mathbb{E}}}\left[ {G(t) \bigg|{\bf{s}}\left( t \right),{{\bf{a}}^{\rm s}}\left( t \right)} \right]. \label{eq:valueQ}
\end{align}
and the expectation in \eqref{eq:valueQ} is taken over $\{r\left( {\bf{s}}\left( i \right),{{\bf{a}}^{\rm s}}\left( i \right)\right), {\bf{s}}\left( i+1 \right), {{\bf{a}}^{\rm s}}\left( i+1 \right), i \geq t\}$, following policy $\pi$. We call the function in \eqref{eq:valueQ} the action-value function for policy $\pi$.

By applying $G(t) = r({{\bf{s}}\left( t \right),{{\bf{a}}^{\rm s}}\left( t \right)}) + G(t+1)$, we can express the value function in recursive form, known as the Bellman equation \cite{sutton2018reinforcement}. To avoid complicated expectations, we illustrate the Bellman equation with a deterministic policy, which can be described as ${\bf{a}}^{\rm s}(t) = {\bf{u}}({\bf{s}}(t))$. In this case, the value function in \eqref{eq:valueQ} can be simplified as follows,
\begin{align}
&{Q^{\bf{u}}}\left( {{\bf{s}}\left( t \right),{{\bf{a}}^{\rm s}}\left( t \right)} \right) \nonumber\\
& = {{\mathbb{E}}}\left[ {r\left( {{\bf{s}}\left( t \right),{{\bf{a}}^{\rm s}}\left( t \right)} \right) + {\gamma _{\rm{d}}}{Q^{\bf{u}}}\left( {{\bf{s}}\left( {t + 1} \right),{\bf{u}}\left( {{\bf{s}}\left( {t + 1} \right)} \right)} \right)} \right], \label{eq:bellmanQ}
\end{align}
where the expectation is taken over $r\left( {{\bf{s}}\left( t \right),{{\bf{a}}^{\rm s}}\left( t \right)} \right)$ and ${\bf{s}}\left( {t + 1} \right)$.

\subsubsection{Actor-Critic Algorithm}
For a reinforcement learning problem, we aim to find the optimal policy that maximizes the value function in \eqref{eq:bellmanQ}. It is well known that when the transition probabilities are known at the agent, then the optimal policy can be obtained by dynamic programming \cite{bellman1966dynamic}. However, due to the curse of dimensionality, dynamic programming can hardly be implemented when the dimensions of observation/action spaces are high. To handle this issue, the actor-critic algorithm was proposed in \cite{konda2000actor}. The basic idea of the actor-critic algorithm is combining DNN with reinforcement learning by approximating the policy and the value function with two DNNs \cite{mnih2015human}. For example, if the policy is a deterministic and continuous function, the deep deterministic policy gradient algorithm can be applied \cite{lillicrap2015continuous}.

\begin{figure}[ht]
        \centering
        \begin{minipage}[t]{0.45\textwidth}
        \includegraphics[width=1\textwidth]{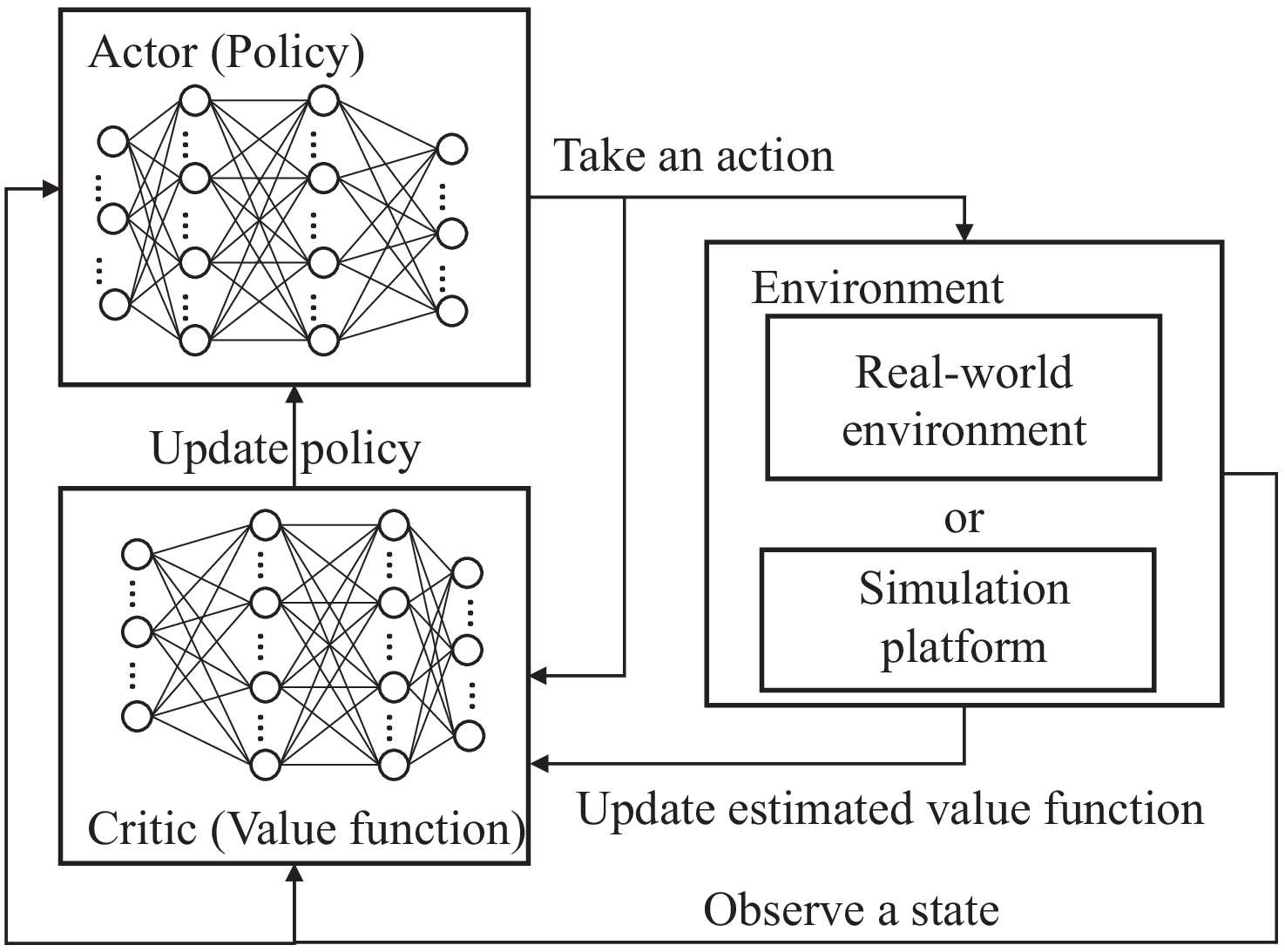}
        \end{minipage}
        \caption{Illustration of actor-critic algorithm with FNNs \cite{lillicrap2015continuous,sutton2018reinforcement}.}
        \label{fig:ActorCritic}
        \vspace{-0.2cm}
\end{figure}

An actor-critic algorithm with FNNs is illustrated in Fig.~\ref{fig:ActorCritic}. Given the state in the current time slot, ${\bf{s}}(t)$, the actor takes action according to the output of the FNN. Then, the reward from the environment is sent to the actor and the critic for updating the policy and the approximation of the value function, respectively. For a deterministic policy, an actor-critic, model-free algorithm, was proposed in \cite{lillicrap2015continuous}, where the algorithm directly learn from the real-world environment. To avoid QoS violations during the training phase, the authors of \cite{Dora2019Deep} train the policy in a simulation platform that is built upon theoretical models and practical configurations.

\begin{table*}[htbp]
\vspace{-0.0cm}\footnotesize
\renewcommand{\arraystretch}{1.3}
\caption{DRL in Wireless Networks}
\begin{center}\vspace{-0.0cm}\label{T:DRL}
\begin{tabular}{|p{1.2cm}|p{7cm}|p{7cm}|}
  \hline
  \multicolumn{3}{|c|}{\bf{Surveys or Tutorials}}\\\hline
  {\bf{Reference}} & \multicolumn{2}{|l|}{{\bf{Contributions}}} \\\hline
  \cite{luong2019applications} & \multicolumn{2}{|l|}{{A comprehensive survey on applications of DRL in communications and networking}}  \\\hline
  \cite{xiong2019deep} & \multicolumn{2}{|l|}{{A tutorial on the implementation of DRL in beyond 5G systems}}  \\\hline
  \cite{Deze2019Resource} & \multicolumn{2}{|l|}{{A tutorial on the applying DRL in resource allocation at the network edge}}  \\\hline
 \multicolumn{3}{|c|}{\bf{Mobile Networking}}\\\hline
 {\bf{Reference}} & {\bf{Contributions}} & {\bf{Performance}} \\\hline
  \cite{wang2018handover} & Optimizing handover controller at each user via a model-free asynchronous advantage actor-critic (A3C) framework & A3C outperforms the upper confidence bandit algorithm and the 3GPP protocol in terms of throughput and handover rate \\\hline
  \cite{jay2019deep} & Mitigating internet congestion by adjusting the data rate of each sender via DRL  & The proposed framework can achieve a better latency-throughput trade-off than existing approaches \\\hline
  \cite{de2019deep} & A fast-learning DRL model that dynamically optimizing network slicing in WiFi networks & The proposed model-free approach with pre-training is promising in dynamic environments  \\\hline
 \multicolumn{3}{|c|}{\bf{Scheduler Design}}\\\hline
 {\bf{Reference}} & {\bf{Contributions}} & {\bf{Performance}} \\\hline
 \cite{atallah2017deep} & Energy-efficient scheduling policy for vehicle safety and QoS concerns & The DRL approach outperforms other methods in terms of incomplete requests, battery lifetime, and latency \\\hline
 \cite{wei2018user} & Energy-efficient user scheduling and resource allocation in heterogeneous networks & Actor-critic learning can achieve higher reward than Q-learning \\\hline
 \cite{chinchali2018cellular} & Actor-critic DRL for traffic scheduling in cellular networks with measured data & DRL method can achieve higher reward (network utilization, throughput, congestion) than some benchmarks \\\hline
 \cite{comcsa2018towards} & Scheduler selection for minimizing packet delays and packet drop rates  & The proposed policy outperforms existing scheduling policies in terms of both delay and packet drop rate \\\hline
 \cite{meng2019delay} & Task scheduling and offloading for minimizing average slowdown and timeout period in MEC  & DRL method outperforms existing schedulers in terms of average slowdown and average timeout period \\\hline
 \multicolumn{3}{|c|}{\bf{Resource Allocation}}\\\hline
 {\bf{Reference}} & {\bf{Contributions}} & {\bf{Performance}} \\\hline
 \cite{xu2017deep} & DRL for switching remote radio heads ON and OFF in could radio access networks & DRL outperforms single BS association and full coordinated association in terms of power consumption \\\hline
 \cite{li2018wcnc} & Minimizing execution delay and energy consumption in MEC by optimizing offloading and resource allocation& DRL outperforms benchmarks in terms of execution delay and energy consumption \\\hline
 \cite{ye2018deep} & Decentralized resource allocation for vehicle-to-vehicle communications & DRL achieves higher data rates in vehicle-to-vehicle and vehicle-to-infrastructure links than existing methods \\\hline
 \cite{rongpeng2018deep} & Maximizing weighted sum of spectrum efficiency and quality-of-experience in network slicing & DRL can enhance the effectiveness and agility for network slicing by matching the allocated resource to users' demand \\\hline
 \cite{sun2018deepFog} & Minimizing long-term system power consumption in Fog-RAN & The energy consumption achieved by DRL is lower than benchmarks and can be further reduced by transfer learning \\\hline
 \cite{chen2019iraf} & Minimizing average energy consumption or average latency in MEC IoT networks by using Monte Carlo tree search and multi-task learning& The proposed DRL framework outperforms other DRL algorithms and benchmarks in terms of average energy consumption and average service latency \\\hline
 \cite{qi2019deep} & Applying the model-based DRL in \cite{gu2016continuous} for resource management in network slicing & By introducing discrete normalized advantage functions into DRL, better SE and QoE can be achieved \\\hline
 \cite{shafin2019self} & Applying DRL in broadcast beam optimization for both periodic and Markov mobility patterns& DRL converges to the optimal solutions and can adjust actions according to user distributions \\\hline
 \multicolumn{3}{|c|}{{\bf{Routing}}}\\\hline
 {\bf{Reference}} & {\bf{Contributions}} & {\bf{Performance}} \\\hline
 \cite{rusek2020routenet,almasan2020deep} & Developing a GNN-based framework, RouteNet, to predict mean delay, jitter and packet loss \cite{rusek2020routenet} and combining GNN in DRL for routing optimization \cite{almasan2020deep} & RouteNet outperforms model-based analysis in terms of prediction accuracy and GNN+DRL outperforms existing DRL, especially when the topology is changing \\\hline
\end{tabular}
\end{center}
\vspace{-0.2cm}
\end{table*}

\subsection{DRL in Wireless Networks} \label{subSec:DRLWN}
As summarized in Table \ref{T:DRL}, DRL algorithms have been applied in wireless networks for mobile networking \cite{wang2018handover,jay2019deep,de2019deep}, scheduler design \cite{atallah2017deep,wei2018user,chinchali2018cellular,comcsa2018towards,meng2019delay}, resource allocation \cite{xu2017deep,li2018wcnc,rongpeng2018deep,chen2019iraf,qi2019deep,shafin2019self}, and routing \cite{rusek2020routenet,almasan2020deep}.

\emph{Key Challenges:} According to \cite{chinchali2018cellular}, research challenges when applying DRL in wireless networks lie in the following aspects,
\begin{itemize}
\item \emph{Non-Markovian Processes:} In practical systems, optimal control problems may not be Markovian. For example, the network dynamics in the real-world scenarios in \cite{chinchali2018cellular} are not Markovian. However, to implement DRL, we need to formulate the problem in the framework of MDP. Otherwise, the Bellman equation does not hold.

\item \emph{Safe Exploration:} Exploration in the unknown environment may lead to bad feedback, especially for mission-critical IoT in factory automation or vehicle networks. If an action cannot guarantee ultra-high reliability, then it will result in unexpected accidents. Thus, we need to avoid bad actions during explorations to improve the safety of a system.

\item \emph{Long Convergence Time:} With high-dimensional observation/action spaces, the convergence time of model-free DRL will be long. In addition, estimating the reliability of URLLC in real-world networks is time-consuming. For example, to evaluate the reward of an action in URLLC, the system needs to send a large number of packets for estimating the reliability of an action \cite{Dora2019Deep}. As a result, it takes a very long time for approximating the value function.
\end{itemize}

In the following subsections, we review some methods to address the above challenges.

\subsection{Techniques for Improving DRL in URLLC}\label{Sec:enhanceDRL}
When using DRL, there is no strict proof of the convergence owning to the approximations of the policy and the value function \cite{sutton2018reinforcement}. Empirical evaluations in some existing publications showed that DRL converges slowly when the state and action spaces are large \cite{lillicrap2015continuous,mnih2015human}. Thus, how to reduce the convergence time of DRL is one of the most important topics in this area \cite{gu2016continuous}. On the other hand, the approximation errors of the policy and value function may have impacts on the reliability of URLLC. How to achieve high reliability when using DRL to optimize communication systems has not been investigated in the existing literature. In this subsection, we review some promising techniques for reducing the convergence time and improving the approximation accuracy of DRL.

\subsubsection{Reward Shaping} As discussed in \cite{marom2018belief}, delayed reward remains one of the key challenges in many reinforcement learning problems. It means that an agent receives a delayed reward after taking several steps. For example, when playing chess, the agent only receives a positive or negative reward by the end of a game.

\begin{figure}[ht]
        \centering
        \begin{minipage}[t]{0.38\textwidth}
        \includegraphics[width=1\textwidth]{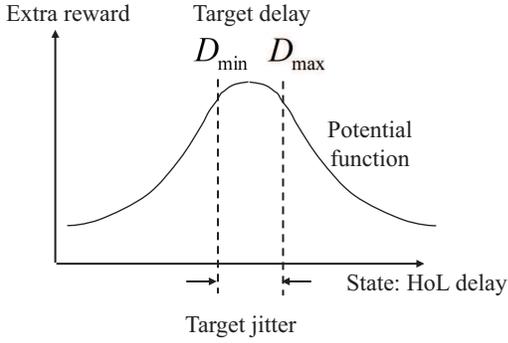}
        \end{minipage}
        \caption{An example of potential function \cite{ng1999policy,Zhouyou2020Knowledge}.}
        \label{fig:RewardShaping}
        \vspace{-0.2cm}
\end{figure}

Reward shaping is a technique to handle this issue by modifying the reward function according to the design goal of a system \cite{ng1999policy}. With reward shaping, the instantaneous reward in the $t$-th slot is redefined as
\begin{align}
\tilde{r}\left( {{\bf{s}}\left( t \right),{{\bf{a}}^{\rm s}}\left( t \right)} \right) = r\left( {{\bf{s}}\left( t \right),{{\bf{a}}^{\rm s}}\left( t \right)} \right) - \Psi\left({\bf{s}}\left( t \right)\right) + \gamma_{\rm d} \Psi\left({\bf{s}}\left( t+1 \right)\right),\label{eq:shap}
\end{align}
where $\Psi\left({\bf{s}}\left( t \right)\right)$ is referred to as the potential function in \cite{ng1999policy}. It is a function that depends on the state and does not rely on the action. Such a function reflects the potential benefit of visiting a state. For example, to meet the latency and the jitter requirements of URLLC, the Head-of-Line (HoL) delays of packets should lie in a small range, $[D_{\min},D_{\max}]$. The agent will receive a positive reward only when a packet is transmitted within this range. Otherwise, ${r}\left( {{\bf{s}}\left( t \right),{{\bf{a}}^{\rm s}}\left( t \right)} \right) = 0$. To avoid zero rewards, we can design a potential function in Fig. \ref{fig:RewardShaping}. If the HoL delay of a packet is lower than $D_{\min}$, the potential function increases with the HoL delay. In this region, the packet should not be transmitted in order to get a positive $\tilde{r}\left( {{\bf{s}}\left( t \right),{{\bf{a}}^{\rm s}}\left( t \right)} \right)$. If the HoL delay is higher than $D_{\min}$, the potential function decreases with the HoL delay. In this region, the packet should be transmitted as soon as possible, otherwise $\tilde{r}\left( {{\bf{s}}\left( t \right),{{\bf{a}}^{\rm s}}\left( t \right)} \right)$ is negative.

As proved in \cite{ng1999policy}, for arbitrary potential function, reward shaping does not change the optimal policy of a reinforcement learning algorithm. With this conclusion, we have the flexibility to design any potential function according to our understanding of a problem. How to design potential functions for different problems has been widely studied in the existing literature, such as \cite{marom2018belief,zou2019reward}. We will evaluate the performance of using reward shaping in URLLC by providing a concrete example in Section \ref{Sec:cases}.

\subsubsection{Multi-head Critic} 
\begin{figure}[ht]
        \centering
        \begin{minipage}[t]{0.4\textwidth}
        \includegraphics[width=1\textwidth]{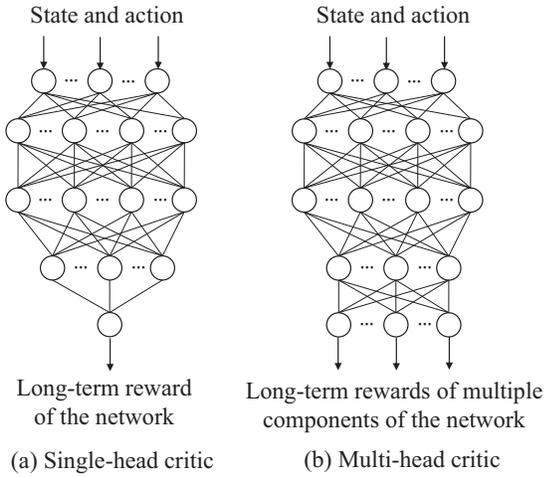}
        \end{minipage}
        \caption{Difference between single-head and multi-head critics \cite{van2017hybrid,Zhouyou2020Knowledge}.}
        \label{fig:multihead}
        \vspace{-0.2cm}
\end{figure}

The DRL introduced in Section \ref{Sec:IntroDRL} evaluates the long-term reward of the system with a single-head critic, where the output of the critic is a scalar that approximates the value of system when an action is taken at a state. However, a wireless network consists of multiple components, such as multiple BSs and multiple users. As shown in \cite{van2017hybrid}, the single-head critic is not aware of the rewards of different components of a system, and hence the agent learns slowly and the algorithm can be unstable. To handle this issue, a multi-head critic was proposed in \cite{van2017hybrid} to learn a separate value function for each component of the system. The difference between single-head and multi-head critics is illustrated in Fig. \ref{fig:multihead}. In Section \ref{Sec:cases}, we will apply the multi-head critic in a multi-user URLLC system to improve the learning efficiency of DRL and the reliability of each user.

\subsubsection{Importance Sampling} In actor-critic DRL algorithms,  the state-value function is approximated by a DNN, i.e., the critic. Due to approximation errors, the action that maximizes the output of the critic may be different from the action that maximizes the true state-value function. As a result, the wireless communication system may take a wrong action leading to packet losses in URLLC. To improve the reliability of URLLC, we need to reduce the approximation errors of DRL.

In the training phase, DRL selects a batch of transitions from the replay memory to train the critic by using the Bellman equation. Let us denote the critic by $\hat{Q}({\bf{s}}(t),{\bf{a}}(t)|\Omega_Q)$, where $\Omega_Q$ represents the parameters of the DNN. The approximation error at the state-action pair $({\bf{s}}(t),{\bf{a}}(t))$ can be characterized by
\begin{align}
\left[y(t) - \hat{Q}({\bf{s}}(t),{\bf{a}}^{\rm s}(t)|\Omega_Q)\right]^2,
\end{align}
where $y(t) \triangleq r(t) + \hat{Q}({\bf{s}}(t+1),{\bf{a}}^{\rm s}(t+1)|\Omega_Q)$ is the approximation of the right-hand side of the Bellman equation in \eqref{eq:bellmanQ}.

In most of the existing publications in Table \ref{T:DRL}, the transitions are selected from the replay memory with the same probability. For the state-action pairs that are rarely visited, the transitions are rarely selected in the training phase. Thus, the approximation errors could be high. To improve the approximation accuracy, importance sampling proposed in \cite{schaul2015prioritized} is a promising approach. The basic idea is to define a weight of each transition according to the approximation error $w(t) = \left[y(t) - \hat{Q}({\bf{s}}(t),{\bf{a}}^{\rm s}(t)|\Omega_Q)\right]^2$. Then, the probability that a transition will be selected is proportional to its weight. In this way, the transitions with higher approximation errors are more likely to be selected than the transitions with lower approximation errors. In Section \ref{Sec:cases}, we will illustrate how does importance sampling help improve the reliability of URLLC.

%

\subsection{{Constrained DRL for URLLC}}
When applying DRL for the URLLC applications in Table~\ref{T:Applications}, the long term reward is usually defined as a weighted sum of different KPIs. With different weight coefficients, the final achieved KPIs are different. To guarantee the requirements on latency and reliability, the coefficients should be optimized automatically. However, in most of the existing DRL algorithms in Table \ref{T:DRL}, these coefficients are predetermined as hyper-parameters. To determine the values of the coefficients, constrained DRL can be applied \cite{liang2018accelerated}.

By formulating some long-term costs as statistic constraints in an MDP, we can apply constrained DRL for QoS guarantee in URLLC \cite{altman1999constrained}. To illustrate how to guarantee QoS in URLLC, $M_{\rm c}$ cost functions $C^{\rm MDP}_m(t)$, $m=1,...,M_{\rm c}$, are considered. The cost functions are defined as follows \cite{liang2018accelerated},
\begin{align}
C^{\rm MDP}_m(t) = {\mathbb{E}}\left[\sum_{i=t}^{\infty} \gamma_{\rm d}^{i-t}c^{\rm MDP}_m({\bf{s}}(i),{\bf{a}}^{\rm s}(i),{\bf{s}}(i+1))\right],\nonumber
\end{align}
where $c^{\rm MDP}_m({\bf{s}}(i),{\bf{a}}^{\rm s}(i),{\bf{s}}(i+1))$ is the instantaneous cost. A long-term cost could be the average queuing delay \cite{djonin2007q,ngo2010monotonicity}, the access cost for handovers \cite{sun2008constrained}, the visitation probability of some states \cite{marecki2013solution}, or the cost in the tail of a risk distribution \cite{chow2018risk}.

In URLLC, we aim to find the optimal policy that maximizes the long-term reward, such as resource utilization efficiency, subject to the QoS constraints, i.e.,
\begin{align}
{\mathop {\max }\limits_{\pi}}&\;{\mathbb{E}} [G(t)]\label{eq:CMDP}\\
\text{s.t.}\; & C^{\rm MDP}_m(t) \leq C_m^{\rm QoS}, m = 1,...,M_{\rm c}, \label{eq:longtermC}\tag{\theequation a}
\end{align}
where $C_m^{\rm QoS}$ is the QoS requirement. To solve problem \eqref{eq:CMDP}, one needs to evaluate the long-term reward as well as the long-term cost. Due to this fact, finding the solution of constrained DRL is more challenging than regular DRL. To solve constrained MDP, the primal-dual method, which is referred to as the Lagrangian approach in \cite{altman1999constrained}, can be applied \cite{liang2018accelerated}.

Similar to the constrained unsupervised deep learning, the Lagrangian of problem \eqref{eq:CMDP} can be expressed as
\begin{align}
&L_{\rm CMDP}(\pi, {\bm{\lambda}}^{\rm CMDP}) \nonumber\\
&= {\mathbb{E}} [G(t)]-\sum_m \lambda^{\rm CMDP}_m\left[C^{\rm MDP}_m(t) - C_m^{\rm QoS}\right],\label{eq:CLag}
\end{align}
where ${\bm{\lambda}}^{\rm CMDP} = (\lambda^{\rm CMDP}_1, ..., \lambda^{\rm CMDP}_{M_{\rm c}})$ is the Lagrangian multiplier. Then, problem \eqref{eq:CMDP} is converted to an unconstrained problem as follows,
\begin{align}
{\mathop {\min }\limits_{\lambda_m^{\rm CMDP}>0, m=1,...,M_{\rm c}}}{\mathop {\max }\limits_{\pi}} \;L_{\rm CMDP}(\pi, {\bm{\lambda}}^{\rm CMDP}).\nonumber
\end{align}
The details of the primal-dual method can be found in \cite{altman1999constrained,liang2018accelerated}. S. Bhatnagar \emph{et al.} proved the asymptotic almost sure convergence of the method to a locally optimal solution of problem \eqref{eq:CMDP} \cite{bhatnagar2012online}.

\begin{figure*}[ht]
        \centering
        \begin{minipage}[t]{0.7\textwidth}
        \includegraphics[width=1\textwidth]{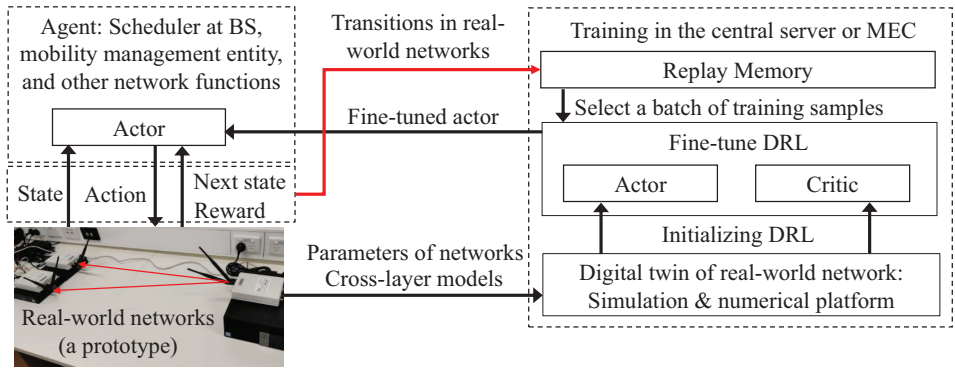}
        \end{minipage}
        \caption{A digital twin of a wireless network \cite{Dora2019Deep,Changyang2020Deep,Zhouyou2020Knowledge}.}
        \label{fig:DTmodel}
        \vspace{-0.2cm}
\end{figure*}

Compared with the DRL algorithms in Table \ref{T:DRL}, constrained DRL provides a more general framework for multi-objective optimization. Instead of setting some weighting coefficients among different KPIs manually, constrained DRL only includes one KPI in the objective function, while the other KPIs are considered in the constraints. By using the primal-dual method, the weighting coefficients, ${\bm{\lambda}}^{\rm CMDP}$ in \eqref{eq:CLag}, are optimized in the dual domain to achieve the target balance among different KPIs. For example, a weighted sum of the execution delay and the energy consumption is minimized by using DRL in \cite{li2018wcnc}. If constrained DRL is applied to this problem, we can minimize the energy consumption subject to a constraint on the execution delay.


\subsection{Off-line Initialization and Online Fine-tuning of DRL in URLLC}\label{subSec:DTforDRL}
In this subsection, we review an architecture that enables the implementation of DRL in URLLC. The basic idea is to initialize the DRL off-line in a simulation platform and fine-tune the pre-trained DRL in real-world networks.

\subsubsection{Digital Twins of Wireless Networks}
The digital twin of a wireless network can serve as a bridge between the model-based analysis and the data-driven DRL.  According to the definition in \cite{glaessgen2012digital}, ``a Digital Twin is an integrated multiphysics, multiscale, probabilistic simulation of an as-built vehicle or system that uses the best available physical models, sensor updates, fleet history, etc., to mirror the life of its corresponding real system". It has been considered as an important approach for realizing Industrie 4.0 in \cite{uhlemann2017digital,GE2018DT}.

As illustrated in Fig. \ref{fig:DTmodel}, a central server establishes a digital twin of a real-world network with the network topology, channel and queuing models, QoS requirements, and formulas in Section \ref{S:Tools}. Then, the DRL algorithm explores possible actions in the digital twin and initializes the actor and critic off-line. After off-line initialization, the actor will be sent to the control plane of the wireless network for real-time implementation.

\subsubsection{Exploration in a Digital Twin} By adopting the concept of the digital twin in DRL, the agent can explore in a simulation platform to avoid potential risks in real-world systems. Given an input, the central server explores possible actions based on the output of the policy (FNN in the actor). For each action, the constraints and the utility function are evaluated in the digital twin \cite{Dora2019Deep}. Then, all the actions that can satisfy the constraints will be used to update the FNN in the critic to obtain a better approximation of the value function. Among these actions, the best one that maximizes the value function is saved in the memory together with the input. Finally, some training samples are randomly selected from the memory to train the FNN in the actor.

\subsubsection{Transfer to Real-World Networks} Simulation environments are not exactly the same as real-world networks, which are dynamic. Thus, the actor initialized off-line may not be a good policy in non-stationary wireless networks. To handle the model mismatch problem, the system needs to update the actor and critic according to the transitions from the real-world networks. As discussed in Section \ref{subSec:transfer}, transfer learning can be used to fine-tune pre-trained FNNs. In this way, the initial performance of the pre-trained actor is much better than the actor initialized with random parameters, and the DRL converges faster in real-world networks with transfer learning \cite{Zhouyou2020Knowledge}. 


\subsection{{Summary of DRL}}
Domain knowledge can help establish a digital twin of a real-world network. By initializing DRL in the digital twin, the convergence time of DRL in the network can be reduced remarkably. In addition, exploration in the digital twin does not cause unexpected penalties in the network. Thus, exploration safety can be improved. To handle model mismatch, DRL keeps updating the actor and critic from the feedback of the network. In addition to the research challenges in Section \ref{subSec:DRLWN}, when applying DRL in URLLC, there are two open issues.
\begin{itemize}
\item Unlike unsupervised deep learning that has both long-term statistic constraints and instantaneous constraints, constrained DRL can only guarantee long-term constraints. How to include instantaneous constraints in DRL remains unclear.
\item When applying DRL in URLLC systems, the feedback of reward or penalty could be very sparse in the time horizon. This is because the packet loss probability in URLLC is extremely small. Improving the learning efficiency of DRL in URLLC systems is an urgent task.
\end{itemize}

\section{Case Studies}\label{Sec:cases}
In this section, we provide concrete examples to illustrate how to guarantee the QoS requirements of URLLC by integrating domain knowledge into deep learning.

\subsection{Deep Transfer Learning for URLLC}
\subsubsection{Example problem} We consider an example system in \cite{Changyang2019Optimizing}, where the transmit power of a BS serving multiple users is minimized by optimizing the transmit power and bandwidth allocation subject to the QoS requirements of different types of services. The problem is formulated as follows \cite{Changyang2019Optimizing},
\begin{align}
\mathop {\min }\limits_{{W_k},P_k^{\rm{t}}} \; &\sum\limits_{k = 1}^K {P_k^{\rm{t}}}\label{eq:objSuper}\\
\text{s.t.}\;& \sum\limits_{k = 1}^K {{W_k}}  \le {W_{\max }},\nonumber\\
&\text{QoS constraints,}\nonumber\\
& P_k^{\rm{t}} \geq 0, \;\text{and}\;{W_k}\geq0,\nonumber
\end{align}
where $P_k^{\rm{t}}$ and ${W_k}$ are the transmit power and the bandwidth allocated to the $k$-th user, respectively, and ${W_{\max }}$ is the total bandwidth of the system.

The QoS constraints depend on the requirements of different kinds of services. For delay-tolerant services, the average service rate of each user should be equal to or higher than the average arrival rate of the user, i.e.,
\begin{align}\label{eq:QoStoler}
{\mathbb{E}}_{g_k}\left\{W_k\log_2\left(1+\frac{\alpha_k g_k P_k^{\rm{t}}}{W_k N_0 N_{\rm T}}\right)\right\} \geq \bar{a}_k(t),
\end{align}
where $N_{\rm T}$ is the number of antennas at the BS, $N_0$ is the single-side noise spectral density, and $\alpha_k$ and $g_k$ are the large-scale and small-scale channel gains of the $k$-th user. In this system, each user is equipped with a single antenna. The instantaneous channel is available at each user, but the BS only knows $\alpha_k$ and the distribution of $g_k$.

For URLLC, the overall packet losses include decoding errors and queuing delay violations, and the E2E delay includes the transmission delay and the queuing delay. According to the conclusion in \cite{Cross2018she}, we can set the required decoding error probability and the required queuing delay violation probability as equal, i.e., $\varepsilon_{\rm tot}^{\max }/2$. The result in \cite{Joint2018She} shows that the optimal transmission delay is one frame, $D_{\rm{t}} = T_{\rm f}$, and the queuing delay bound should be $D_{\rm q}^{\max} = D_{\rm tot}^{\max} - T_{\rm f}$.

To guarantee the overall packet loss probability and E2E delay, the following constraint should be satisfied,
\begin{align}
&{\mathbb{E}}_{g_k}\left\{{f_{\rm{Q}}}\left( {\sqrt {{{{T_{\rm{f}}}W_k}}} \left[ {\ln \left( {1 + \frac{\alpha_k g_k P_k^{\rm{t}}}{W_k N_0 N_{\rm T}} } \right) - \frac{u_0{E_{B_k}\ln 2}}{{W_k}}} \right]} \right)\right\} \nonumber\\
&\leq \varepsilon_{\rm tot}^{\max}/2, \label{eq:const_decode}
\end{align}
where $u_0$ (bits/packet) is the packet size, $V \approx 1$ is applied, and the effective bandwidth of the $k$-th user, $E_{B_k}$ \cite{Cross2018she},
\begin{align}
E_{B_k} = \frac{\ln(2/\varepsilon_{\rm tot}^{\max})}{(D_{\rm tot}^{\max} - T_{\rm f}) \ln \left[\frac{T_{\rm f}\ln(2/\varepsilon_{\rm tot}^{\max})}{\lambda_k^{\rm a} (D_{\rm tot}^{\max} - T_{\rm f})}+1\right]}\; \text{(packets/s)}, \label{eq:UpdatedEB}
\end{align}
where $\lambda_k^{\rm a}$ is the average packet arrival rate of the $k$-th user.

The optimization algorithm to find labeled training samples and the supervised deep learning algorithm were developed in \cite{Changyang2019Optimizing} and \cite{Dora2020Deep}, respectively.

\subsubsection{Simulation Results} The setup for simulation can be found in \cite{Changyang2019Optimizing}. We use two FNNs with parameters $\Omega_0$ and $\Omega_1$ to approximate the resource allocation policies of delay-tolerant services and URLLC services, respectively. For both FNNs, there are four hidden layers and each of them has $800$ neurons. The dimensions of the input and output layers are $40$ and $20$, respectively.

Since there is no URLLC services in the existing cellular networks, we assume that it is much easier to obtain labeled training samples for delay-tolerant services than that for URLLC services. The parameters of the first FNN, $\Omega_0$, are trained with the labeled samples for delay-tolerant services in set $\mathcal{S}_0$. To reduce the number of labeled training samples for URLLC, we initialize the values of $\Omega_1$ with $\Omega_0$ and apply network-based deep transfer learning \cite{Dora2020Deep}. Specifically, we fix the first three hidden layers and fine-tune the weights and bias in the last hidden layer. To evaluate the performance, we define the normalized accuracy as follows,
\begin{align}
\eta_{NN} = 1 - \frac{{\sum\limits_{k = 1}^K {\hat P_k^{\rm{t}}}  - \sum\limits_{k = 1}^K {\tilde P_k^{\rm{t}}} }}{{\sum\limits_{k = 1}^K {\tilde P_k^{\rm{t}}} }}\label{eq:PerformanceNN},
\end{align}
where $\hat{P}_k^{\rm t}$ and $\tilde{P}_k^{\rm t}$ are the transmit powers obtained from the supervised deep learning algorithm and the optimization algorithm, respectively.

\begin{figure}[ht]
        \centering
        \begin{minipage}[t]{0.48\textwidth}
        \includegraphics[width=1\textwidth]{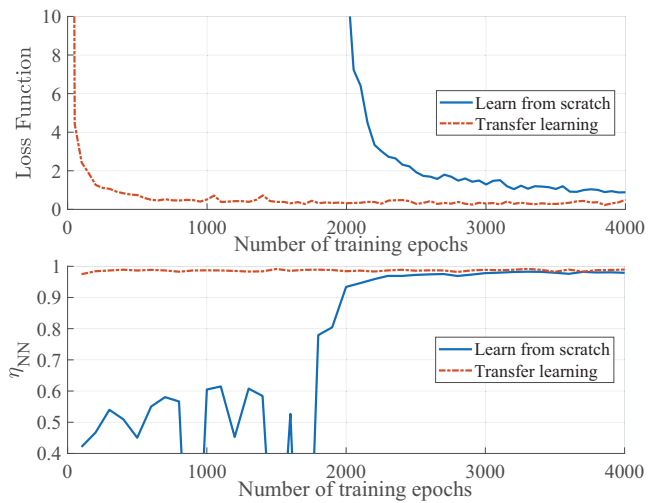}
        \end{minipage}
        \caption{Performance of deep transfer learning \cite{Changyang2019Optimizing,Dora2020Deep}.}
        \label{fig:TransferFNN}
        \vspace{-0.2cm}
\end{figure}

As shown in Fig. \ref{fig:TransferFNN}, if there are a large number of training samples, then the normalized accuracy approaches to one as the number of training epochs increases (The optimization algorithm generates one labeled training sample in each epoch). With the network-based deep transfer learning, the normalized accuracy reaches $97$\% after $100$ training epochs (with $100$ training samples). However, if the system learns from scratch, then it takes $2500$ epochs (with $2500$ training samples) to achieve the same performance. Therefore, by using deep transfer learning, we can obtain a good approximation of the optimal resource allocation policy for URLLC with $100$ training samples.

\subsection{Comparison of Supervised and Unsupervised Deep Learning}
\subsubsection{Example Problem} We take a bandwidth allocation problem in \cite{sun2019learning} to illustrate the difference between supervised and unsupervised deep learning. In this problem, the total bandwidth is minimized subject to the QoS requirement of URLLC services. The problem is formulated as follows,
\begin{align}
\mathop {\min }\limits_{{W_k}} \; &\sum\limits_{k = 1}^K {W_k}\label{eq:objW}\\
\text{s.t.}\;& \mathbb{E}_{g_k} \left\{e^{- \theta_k s_k}\right\} - e^{- \theta_k E_{{B}_k}} \leq 0, k=1,...,K,\label{eq:LineECEB}\\
& {W_k}\geq0,\nonumber
\end{align}
where $s_k$ is the achievable rate in the finite blocklength regime, the value of $\theta_k$ is determined by the delay bound and delay violation probability according to \eqref{eq:theta}, and the QoS constraints in \eqref{eq:LineECEB} is derived from \eqref{eq:ECEB}.

The labeled training samples for supervised deep learning are obtained from the optimization algorithm in \cite{sun2019learning}. With unsupervised deep learning, the primal-dual method is applied to solve the equivalent functional optimization problem, i.e.,
\begin{align}   \label{prob:RA_FuncOpt}
    \mathop \mathrm{min} \limits_{W_k(\alpha_k)} \quad& \mathbb{E}_{\alpha_k} \left\{W_k(\alpha_k)\right\} \\
    \text{s.t.} \quad & \mathbb{E}_{g_k} \left\{e^{- \theta_k s_k(\alpha_k)}\right\} - e^{- \theta_k E_{B_k}} \leq 0, \label{eq:EBECalpha}\tag{\theequation a} \\
    &W_k(\alpha_k) \geq 0, \nonumber
\end{align}
where $W_k(\alpha_k)$ is a mapping from the realization of the large-scale channel gain to the bandwidth allocation.

\subsubsection{Simulation results} To show whether supervised and unsupervised deep learning can guarantee the QoS requirements or not, we evaluate the CCDF of relative error of the QoS constraint. Specifically, the relative error is defined as follows,
\begin{align}
\nu &\triangleq \max\left\{\frac{\mathbb{E}_{g_k} \left\{e^{- \theta_k \hat{s}_k}\right\} - e^{- \theta_k E_{{B}_k}}}{e^{- \theta_k E_{{B}_k}}},0\right\}\nonumber\\
& = \max\left\{ \mathbb{E}_{g_k} \left\{e^{ \theta_k (E_{B_k} - \hat{s}_k)}\right\} - 1,0 \right\},\nonumber
\end{align}
where $\hat{s}_k$ is obtained by substituting the output of the DNN, $\hat{W}_k$, into the expression of the achievable rate in the finite blocklength regime.

To evaluate the performance of unsupervised deep learning, the bandwidth allocation policy is trained and tested in $100$ trails. In each trial, the policy is trained with $10000$ iterations by using the primal-dual method and is tested with $1000$ realizations of $\alpha_k$. The same realizations of $\alpha_k$ are used to train and test the bandwidth allocation policy in supervised deep learning, where the solutions obtained from the optimization algorithm, $W_k^*$, are used as the labels.

\begin{figure}[ht]
        \centering
        \begin{minipage}[t]{0.45\textwidth}
        \includegraphics[width=1\textwidth]{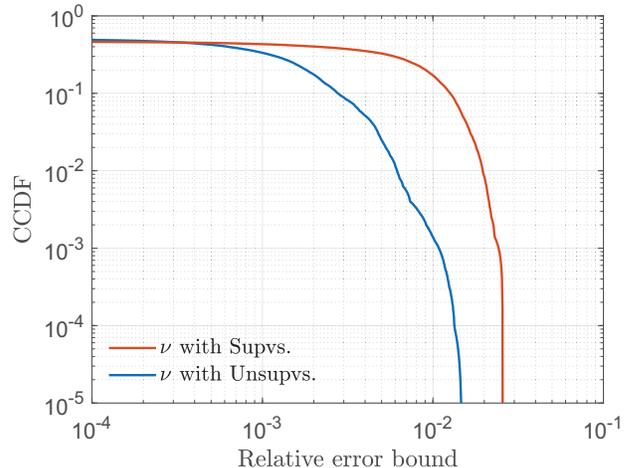}
        \end{minipage}
        \caption{Supervised deep learning versus unsupervised deep learning \cite{sun2019learning}.}
        \label{fig:SUS}
        \vspace{-0.2cm}
\end{figure}

\begin{figure}[ht]
        \centering
        \begin{minipage}[t]{0.48\textwidth}
        \includegraphics[width=1\textwidth]{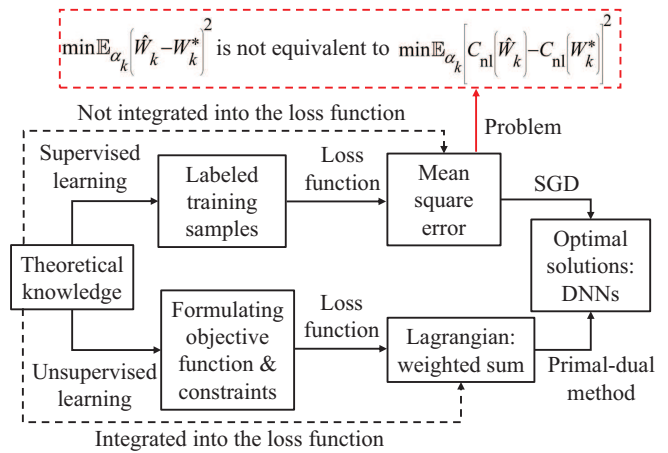}
        \end{minipage}
        \caption{Comparison of supervised and unsupervised deep learning \cite{sun2019learning}.}
        \label{fig:flowSUS}
        \vspace{-0.2cm}
\end{figure}

The results in Fig. \ref{fig:SUS} shows that with probability $1-10^{-5}$, the relative errors of supervised and unsupervised deep learning are less than $3$\% and $2$\%, respectively. By reserve a small portion of extra bandwidth (increasing effective capacity by $2\sim3$\%) to each user, the deep learning approaches can guarantee the QoS requirements with a probability of $1-10^{-5}$.

The results in Fig. \ref{fig:SUS} also indicate that unsupervised deep learning outperforms supervised deep learning in terms of the approximation error of the QoS constraint. The reason is illustrated in Fig. \ref{fig:flowSUS}, where $C_{\rm nl}$ represents the non-linear QoS constraint in \eqref{eq:EBECalpha}. In supervised learning, the domain knowledge is used to find labeled training samples but is not integrated into the loss function, ${\mathbb{E}}_{\alpha_k}\left(\hat{W}_k-W^*_k\right)^2$. However, to minimize the approximation error of the QoS constraint, we should minimize the following loss function ${\mathbb{E}}_{\alpha_k}\left[C_{\rm nl}(\hat{W}_k)-C_{\rm nl}(W^*_k)\right]^2$. Since the QoS constraint is non-linear, the two loss functions are not equivalent. Different from supervised deep learning, the loss function of unsupervised deep learning is a weighted sum of the objective function and the QoS constraint. The policy and the weighting coefficients are optimized in the primal and dual domains, respectively. In other words, the domain knowledge is integrated into the loss function. Therefore, unsupervised deep learning outperforms supervised deep learning.

\subsection{Knowledge-assisted DRL}
\begin{figure*}[ht]
        \centering
        \begin{minipage}[t]{0.6\textwidth}
        \includegraphics[width=1\textwidth]{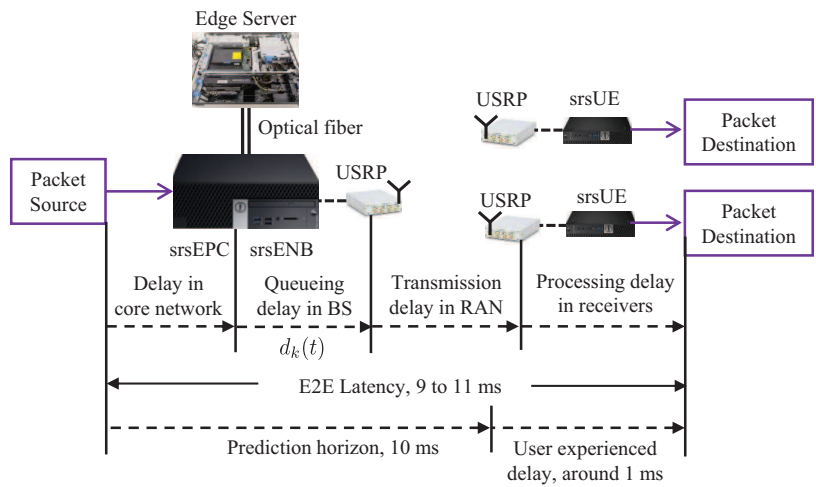}
        \end{minipage}
        \caption{Prototype for E2E delay evaluation \cite{Changyang2020Deep,Zhouyou2020Knowledge}.}
        \label{fig:proto}
        \vspace{-0.2cm}
\end{figure*}

\subsubsection{Example Problem} 
\begin{figure}[th]
        \centering
        \begin{minipage}[t]{0.40\textwidth}
        \includegraphics[width=1\textwidth]{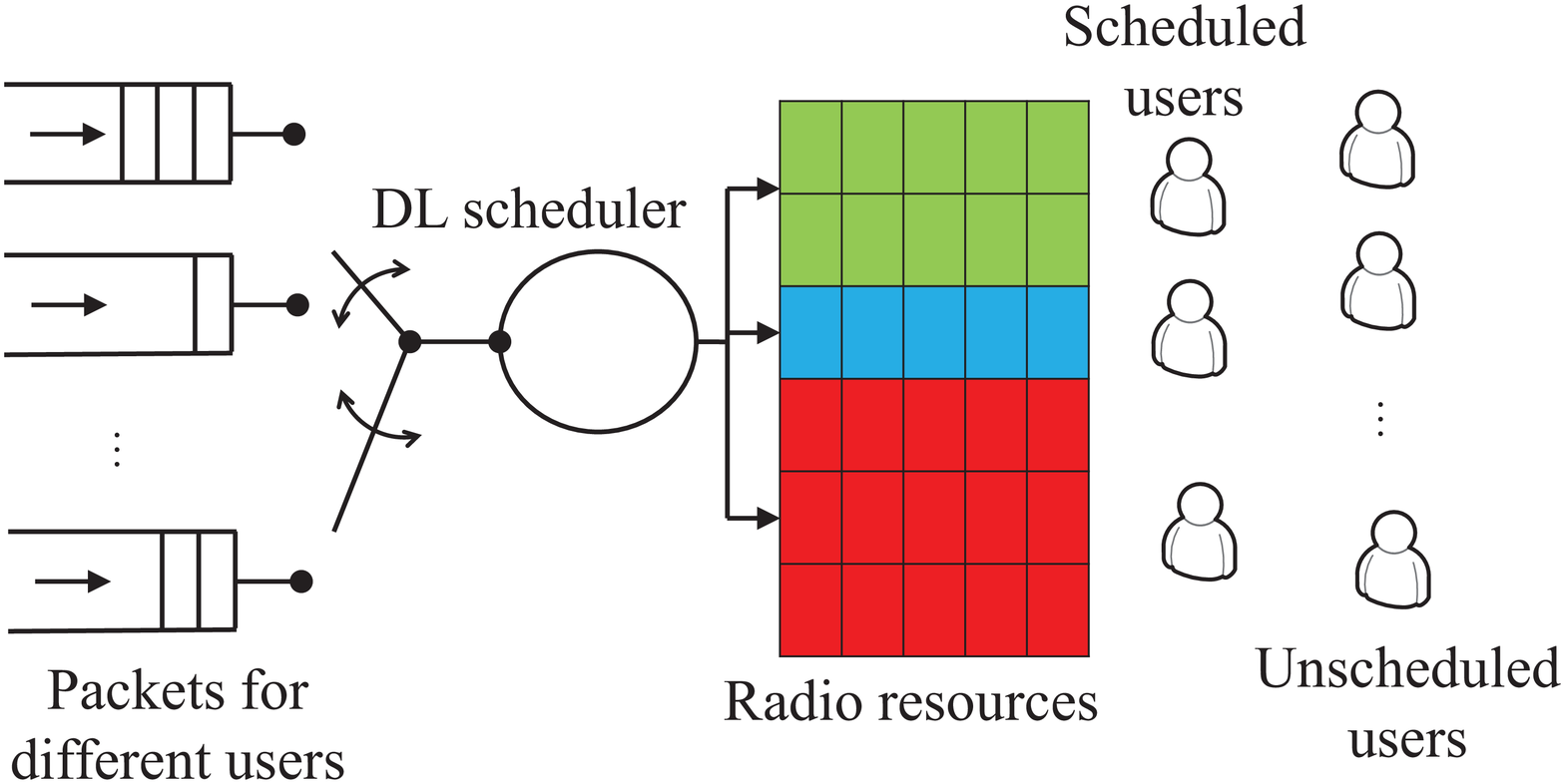}
        \end{minipage}
        \caption{DL scheduler \cite{Zhouyou2020Knowledge}.}
        \label{fig:Scheduler}
        \vspace{-0.2cm}
\end{figure}

To illustrate how to integrate domain knowledge into DRL by using the technologies in Section \ref{Sec:enhanceDRL}, we take the scheduler design problem in \cite{Zhouyou2020Knowledge} as an example. As shown in Fig. \ref{fig:Scheduler}, a DL scheduler at the BS determines which users should be scheduled and how many resource blocks should be allocated to the scheduled users. Let us denote the HoL delay of the $k$-th user in the $t$-th slot as $d_k(t)$. To meet the latency and jitter requirements of URLLC, the users should be scheduled when $d_k(t)$ lies in $[D_{\min},D_{\max}]$. As such, the jitter is $(D_{\max} - D_{\min})$. The goal is to minimize packet losses caused by decoding errors and delay bound violations. Specifically, in the $t$-th slot, we maximize the summation of the discounted rewards of all the users,
\begin{align}
\max_{{\bf{u}}(\cdot|\Omega_u)}\mathop{\mathbb{E}} \sum_{i=t}^{\infty} {\gamma_{\rm d}^{i-t}\sum_{k=1}^{K}r_k(i)}, \label{eq:returnDRL}
\end{align}
where $r_k(i)$ is the reward of $k$-th user in the $i$-th slot, and ${\bf{u}}(\cdot|\Omega_u)$ is the actor of DRL, where $\Omega_u$ represents the parameters of the actor.

\subsubsection{Simulation Results} We compare the performances achieved by three DRL algorithms with different definitions of state, action, and reward. The first one is a straightforward application of the deep deterministic policy gradient (DDPG) algorithm \cite{lillicrap2015continuous} (with legend ``DDPG"). The state includes the HoL delays and the channel quality indicators (CQIs) of users, the action is the numbers of resource blocks allocated to users, and the reward is given by
\begin{align}
{r}^{[1]}_k(t) =
    \begin{cases}
         \mathbf{1}_{D_\text{min} \leq d_k(t)\leq D_\text{max}} \cdot \mathbf{1}^{\text{dec}}_{k}(t) ,\ \text{if}\ x_k(t) =1\ , \\
        0\ ,\ \text{if}\ x_k(t) =0  \ ,
    \end{cases}
\end{align}
where $\mathbf{1}_{D_\text{min} \leq d_k(t)\leq D_\text{max}}$ and $ \mathbf{1}^{\text{dec}}_{k}(t)$ are two indicators. If a packet is scheduled when $d_k(t) \in [D_{\min},D_{\max}]$ and is successfully decoded by the receiver, then ${r}^{[1]}_k(t)= 1$. Otherwise, ${r}^{[1]}_k(t)= 1$.

With the second approach, DDPG is applied in a theoretical DRL framework (with legend ``DDPG-T-DRL"), where the expression in \eqref{eq:decode} is applied to compute the number of resource blocks that is required to achieve a target decoding error probability, denoted by $n_k(t)$. Since the value of $n_k(t)$ is determined by the CQI of the user, the state is re-defined as the HoL delays and $n_k(t), k=1,...,K$. Given the required numbers of resource blocks, $n_k(t), k=1,...,K$, the scheduler only needs to determine which users should be scheduled. With the help of the expression in \eqref{eq:decode}, the reward is given by
\begin{align}\label{eq:reward_ue_model_based}
{r}^{[2]}_k(t) =
    \begin{cases}
        \mathbf{1}_{D_\text{min} \leq d_k(t)\leq D_\text{max}}(1- \epsilon^{\rm c}_k(t)),\ \text{if}\ x_k(t) =1,\\
        0,\ \text{if}\ x_k(t) =0  \ .
    \end{cases}
\end{align}
where $\epsilon^{\rm c}_k(t)$ is the decoding error probability. Considering that \eqref{eq:reward_ue_model_based} is strictly smaller than $1$, the reward can be re-defined as $\hat{r}^{[2]}_k(t) = - \ln[1-r_k(t)]$ to further improve the learning efficiency \cite{Zhouyou2020Knowledge}.

To further improve the performance, a knowledge-assisted DDPG algorithm is applied in the theoretical DRL framework in \cite{Zhouyou2020Knowledge} (with legend ``KA-DDPG-T-DRL"). The basic idea is to exploit the knowledge of the rewards of multiple users, the target scheduling policy, and the approximation errors of the critic. To integrating the three kinds of knowledge into DDPG, we apply the techniques discussed in Section \ref{Sec:enhanceDRL}: multi-head critic, reward shaping, and importance sampling.

\begin{figure}[ht]
        \centering
        \begin{minipage}[t]{0.40\textwidth}
        \includegraphics[width=1\textwidth]{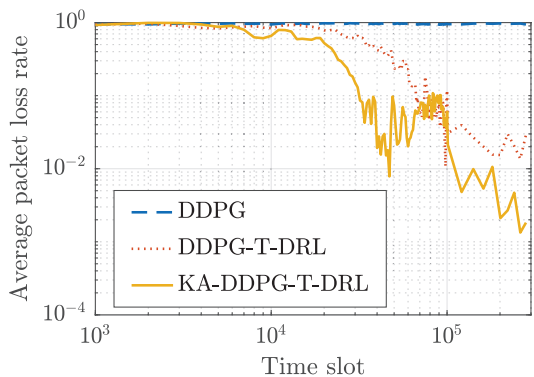}
        \end{minipage}
        \caption{Packet loss probability during the training phase, where the duration of each slot is $0.125$~ms \cite{Zhouyou2020Knowledge}.}
        \label{fig:PLPtrain}
        \vspace{-0.2cm}
\end{figure}

The details of the simulation setup of Fig. \ref{fig:PLPtrain} can be found in \cite{Zhouyou2020Knowledge}. To reduce the packet loss rate, we turn off the exploration after the $10^5$-th slot and keep fine-tuning the parameters of the actor and critic with a small learning rate, $10^{-4}$. In the first $10^5$ slots, the packet loss rate is evaluated every $10^3$ slots. After that, the packet loss rate is evaluated every $2\times10^4$ slots. The results in Fig. \ref{fig:PLPtrain} indicate that the straightforward implementation of DDPG does not converge. With the help of the theoretical DRL framework, DDPG converges slowly and the final packet loss rate is $10^{-2}$. If the knowledge is further exploited, the algorithm converges faster and the packet loss rate can be reduced by one order of magnitude.

\subsubsection{Experimental Results} To validate the architecture in Fig.~\ref{fig:DTmodel}, we built a prototype in \cite{Zhouyou2020Knowledge}. As illustrated in Fig.~\ref{fig:proto}, the prototype was built upon an open-source Long Term Evolution (LTE) software suit from Software Radio System (srs) Limited, a software company in wireless communications. The core network, the BS, and mobile devices are referred to as evolved packet core (srsEPC), eNodeB (srsENB), and user equipment (srsUE), respectively. The radio transceivers are Universal Software Radio Peripheral (USRP) B210. The E2E delay includes the delay in the core network, the queuing delay in the BS, the transmission delay in RAN, and the processing delay in the receiver.

\begin{figure}[ht]
        \centering
        \begin{minipage}[t]{0.40\textwidth}
        \includegraphics[width=1\textwidth]{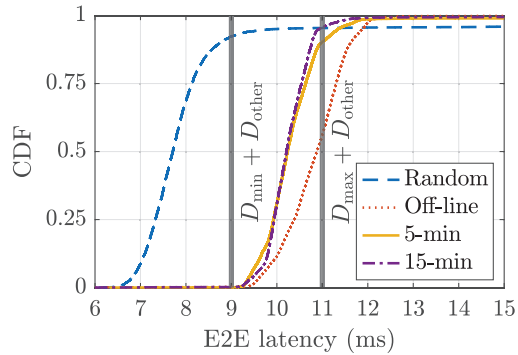}
        \end{minipage}
        \caption{Distribution of E2E delay \cite{Zhouyou2020Knowledge}.}
        \label{fig:latencyCDF}
        \vspace{-0.2cm}
\end{figure}

In this experiment, we aim to achieve $1$~ms user experienced delay and $\pm 1$~ms jitter in Tactile Internet, where the packet source is a tactile device \cite{Hou2019Prediction}. In the LTE software suit, the transmission delay is longer than $1$~ms. In addition, the propagation delay in the core network is larger than $1$~ms if the communication distance is longer than $300$~km. As a result, one can hardly achieve $1$~ms E2E delay. To handle this issue, the prediction and communication co-design method developed in \cite{Hou2019Prediction} is applied, where the prediction horizon is set to be $10$~ms. The transmitter predicts its future locations and sends them to the receivers $10$~ms in advance. According to the experimental results in \cite{Changyang2020Deep}, by using an FNN to predict the trajectory in the next $10$~ms, the prediction error probability is smaller than $10^{-5}$.

To achieve $1$~ms user experienced delay and $\pm 1$~ms jitter, the E2E delay in the communication system should lie in $[9,11]$~ms (given that the prediction horizon is $10$~ms). Since the scheduler can only control the queuing delay and cannot reduce the delay and jitter of other delay components, which cause $D_{\rm other} = 4$~ms delay in our prototype, we set $D_{\min}=5$~ms and $D_{\max}=7$~ms in scheduler design. The actor and the critic are first initialized in the digital twin that mimics the behavior of the real-world network. To handle the model-mismatch problem, we fine-tune the scheduling policy by using the transitions from the measurement in the prototype.

The results in Fig. \ref{fig:latencyCDF} shows that if we initialize the actor and the critic with random variables (with legend ``Random"), the initial E2E delay does not lie in $[9,11]$~ms. With the initialization in the digital twin, more than half of the packets can meet the delay and jitter requirements (with legend ``Off-line"). After $5$ minutes of fine-tuning in the prototype, around $90$~\% packets meet the requirements (with legend ``5-min"). If we keep fine-tuning the actor and the critic, the performance of the scheduler can be further improved. After $15$ minutes of fine-tuning, the delay bound violation probability is around $1$~\%. Such a delay bound violation probability is still too high for URLLC. This is because the scheduler can only control the jitters in buffers of the BS, and cannot reduce the jitters of the other delay components, which become the bottleneck of the system.

To further reduce the jitter, one possible way is to do the prediction at the receivers. Since the delay and the jitter experienced by each packet can be measured by the receiver, it is possible to adjust the prediction horizon to compensate variations of the delay and the jitter at the receiver. However, the historical trajectory at the receiver is not error-free since there are some packet losses in the communication system. As a result, the prediction accuracy at the receiver will be worse than that at the transmitter. Nevertheless, the comparison of these two approaches is not available in the existing literature.

\subsection{Summary of Case Studies}
The examples in this section imply that the application of deep learning in URLLC is not straightforward. We need to integrate the knowledge of communications and networking into different kinds of learning algorithms to reduce the convergence time, guarantee the QoS constraints, and improve the final performances in terms of E2E delay, reliability, and jitter. In addition, there are some open issues should be addressed.
\begin{itemize}
\item The QoS of URLLC is sensitive to a lot of hyper-parameters, such as the initial values of the parameters, the learning rate, and the structure of the neural network. These hyper-parameters are determined by trial-and-error in our case studies. How to find proper hyper-parameters without human efforts remains an important topic in deep learning.
\item The approximation errors of continuous variables are extremely small and the QoS constraints can be satisfied by reserving a small portion of extra resources. However, when the actions are discrete variables, such as the users to be scheduled, the reliability in our prototype is still unsatisfactory for URLLC. 
\item To fine-tune learning algorithms in non-stationary networks, the training phase should be fast enough to follow the variations of wireless networks. According to our results, it takes around $5$ to $15$ minutes to fine-tune the algorithms, which is fast enough for slow varying networks. It is still very challenging to update learning algorithms in highly dynamic networks.
\item In our case studies, FNNs are used to approximate the optimal policy or predict trajectories. FNNs work well in most of the small-scale problems. When the scales of the problems grow, novel structures of neural networks are needed.
\end{itemize}

\section{Future Directions}

%

\subsection{Extreme URLLC}
Due to the diverse QoS requirements in future application scenarios, the KPIs of URLLC, eMBB, and mMTC will be conflated in specific services. Such a new trend has been put forward in \cite{park2020extreme}, and is referred to as extreme URLLC.

\subsubsection{High Data Rate URLLC} As one of the killer applications in 5G networks, VR/AR applications require ultra-reliable and low-latency tactile feedback and high data rate ${360^{\rm o}}$ videos \cite{elbamby2018toward}. Meanwhile, as the size of devices shrinks, battery lifetime will become a bottleneck for enabling high data rate URLLC \cite{saad2019vision}. To implement VR/AR applications in future wireless networks, we need to investigate the fundamental tradeoffs among throughput, energy efficiency, reliability, and latency in communications, caching, and computing systems \cite{sun2019communications}, as well as enabling technologies such as touch user interface and haptic codecs \cite{de2018assessment,steinbach2018haptic}. With the sub-$6$GHz spectrum, it is difficult to meet the above requirements in 5G cellular networks. To overcome this difficulty, Terahertz communications were applied to support VR applications in \cite{chaccour2020can}. Since high-frequency communications rely on LoS paths, the blockage probability was analyzed in small cells in \cite{chaccour2020can}. To further reduce the blockage probability, novel network architectures that can provide multi-path diversity are much needed.

\subsubsection{Massive URLLC} Due to the explosive growth of the numbers of autonomous vehicles and mission-critical IoT devices \cite{Ericsson2019IoTforecast}, future wireless networks are expected to support massive URLLC. To support massive URLLC, novel communication and learning techniques are needed. With orthogonal multiple access technologies, the required bandwidth increases linearly with the number of devices. To achieve better tradeoffs among delay, reliability, and scalability, other multiple access technologies should be used, such as non-orthogonal multiple access and contention-based multiple access technologies  \cite{shirvanimoghaddam2017massive,Bikramjit2018Contention}. Meanwhile, we need to exploit the above $6$~GHz spectrum including mmWave \cite{hemadeh2017millimeter} and the Terahertz band \cite{xing2018propagation}.

\subsubsection{Secure URLLC} Future URLLC systems will suffer from different kinds of attacks that result in inefficient communications \cite{li20185g}. The widely used cryptography algorithms require high-complexity signal processing, and may not be suitable for URLLC, especially for IoT devices with low computing capacities. To defend against eavesdropping attacks in URLLC, physical layer security is a viable solution \cite{chen2019physical}. The maximal secret communication rate in the short blocklength regime over a wiretap channel was derived in \cite{yang2019wiretap}. The results show that there are tradeoffs among delay, reliability, and security. Based on this fundamental result, we can further investigate the technologies for improving physical-layer security.




\subsection{Advanced Learning Algorithms}
To meet the requirements of extreme URLLC, we need some advanced learning algorithms that were developed in recent years.

\subsubsection{Multi-agent DRL with Partial Observation} MEC is a promising network architecture for high data rate/massive URLLC \cite{Yuyi2017MEC,Mohammad2019Fog}, where the resource allocations in communication and computing systems are intertwined. When serving VR applications in an MEC system, the system can either project a 2D field of view into a 3D field of view at the MEC or the mobile VR device \cite{mangiante2017vr,sun2019communications}. In Industrial IoT, collecting the status of all the devices in the central server will bring considerable overheads to wireless networks, and hence the devices should be controlled in a distributed manner. To achieve this goal, multi-agent DRL with partial observation can be applied. With this framework, MEC servers, VR glasses, and IoT devices can take actions based on local observations. In this way, the overheads for exchanging control information and updating global status can be reduced remarkably \cite{sharma2019distributed}.

\subsubsection{GNNs for Flexible Network Management}
Since the dimensions of the input and the output of a FNN grows with the number of devices, we need to adjust the hyper-parameters of the FNN and retrain the parameters whenever the number of devices varies. Thus, FNNs are not flexible in managing dynamic networks. To address this issue, a promising approach is to use GNNs to represent the topology of wireless networks \cite{gama2019stability}. As indicated in \cite{wu2020comprehensive}, GNN is a very general structure that can be applied to solve large-scale problems with Non-Euclid data structure. Since the number of parameters of a GNN does not increase with the dimension of the input, GNNs are suitable for resource management in dynamic networks \cite{eisen2020optimal}.

\subsubsection{Learning with Limited Data Samples}
The real-world data samples from practical systems are limited and could be non-stationary in wireless networks. To train deep learning algorithms with limited data samples, we can use GANs to generate synthetic data, such as the inter-arrival time between packets and the channel gains \cite{kasgari2019experienced}. Another learning framework for fast adaptation is known as the few-shot learning \cite{sun2019meta}. The basic idea is to use meta learning to optimize hyper-parameters, including initial parameters, learning rates, and the structures of neural networks \cite{finn2017model,andrychowicz2016learning}. After offline training in existing task, the neural networks can be transferred to new tasks with limited data samples. Essentially, by using meta learning, we are trading online learning efficiency with offline computing resources. For example, in image recognition, hundreds of graph processing unites are used to train the meta learning algorithm over several days in \cite{zoph2018learning}.

\section{Conclusion}
In this paper, we first have summarized the KPIs and challenges of URLLC in four typical application scenarios: indoor large-scale scenarios, indoor wide-area scenarios, outdoor large-scale scenarios, and outdoor wide-area scenarios. It has been argued that existing optimization algorithms cannot meet these requirements and the application of deep learning is not straightforward. To address these issues, we have developed a road-map toward URLLC based on integrating domain knowledge of communications and networking into deep learning.

Following this road-map, we have reviewed promising 6G network architectures and discussed how to develop deep learning frameworks based on network architectures. Then, we have revisited domain knowledge in communications and networking including analytical tools and cross-layer optimization frameworks. To show how to integrate domain knowledge into deep learning, we have introduced different kinds of learning algorithms and provided concrete examples in case studies. These results indicate that the integration outperforms straightforward applications of learning algorithms in wireless networks. Finally, we have highlighted some future directions in this area.

\bibliographystyle{IEEEtran}
\bibliography{ref}

\end{document}